\documentclass[letterpaper,twocolumn,10pt,table,xcdraw]{article}

\usepackage{hegroup}
\usepackage{tikz}
\usepackage{amsfonts}
\usepackage{xspace} 
\usepackage{amsmath}
\usepackage{mathrsfs}
\usepackage{amssymb}
\usepackage{pifont}
\newcommand{\cmark}{\textcolor{green!70!blue}{\ding{51}}}%
\newcommand{\xmark}{\textcolor{red}{\ding{55}}}%
\usepackage{amsmath}
\usepackage{amsthm}
\usepackage{algorithmic}
\usepackage{algorithm}
\usepackage{multirow}
\usepackage{makecell}
\usepackage[sort&compress,numbers]{natbib}
\usepackage{caption}
\usepackage{subcaption}
\usepackage{enumitem}
\usepackage{float}
\usepackage{booktabs}
\usepackage{balance}
\usepackage[most]{tcolorbox}
\tcbset{width=0.49\textwidth,boxrule=1pt,colback=lightgray,arc=0pt,auto outer arc,left=0pt,right=0pt,top=0pt,bottom=0pt,boxsep=2pt}
\usepackage{hyperref}
\usepackage{cleveref}

\usepackage[hang,flushmargin]{footmisc}

\newcommand{\myta}[1]{%
    \begin{tcolorbox}[colback=black!3!white,enhanced jigsaw,breakable, width=\linewidth]
        \noindent \textbf{Takeaways.}
        #1
    \end{tcolorbox}
}

\newcommand{\Method}{\textit{FLPoison}\xspace}
\newcommand{\mypara}[1]{\noindent{\bf {#1}}\xspace}

\begin{document}

\date{}

\title{SoK: Benchmarking Poisoning Attacks and Defenses \\
in Federated Learning}
\author{
Heyi Zhang\textsuperscript{1}  \ \ \ 
Yule Liu\textsuperscript{2}  \ \ \ 
Xinlei He\textsuperscript{2}\thanks{Corresponding authors (\href{mailto:xinleihe@hkust-gz.edu.cn}{xinleihe@hkust-gz.edu.cn}, \href{mailto:jun.wu@ieee.org}{jun.wu@ieee.org}).} \ \ \ 
Jun Wu\textsuperscript{1}\textsuperscript{\textcolor{blue!60!green}{$\ast$}}\ \ \ 
Tianshuo Cong\textsuperscript{3} \ \ \
Xinyi Huang\textsuperscript{4}
\\
\\
\textsuperscript{1}\textit{Shanghai Jiao Tong University} \ \ \ 
\textsuperscript{2}\textit{Hong Kong University of Science and Technology (Guangzhou)} \ \ \ 
\\
\textsuperscript{3}\textit{Tsinghua University} \ \ \
\textsuperscript{4}\textit{Jinan University} \ \ \ 
}

\maketitle

\begin{abstract}

Federated learning (FL) enables collaborative model training while preserving data privacy, but its decentralized nature exposes it to client-side data poisoning attacks (DPAs) and model poisoning attacks (MPAs) that degrade global model performance.
While numerous proposed defenses claim substantial effectiveness, their evaluation is typically done in isolation with limited attack strategies, raising concerns about their validity. 
Additionally, existing studies overlook the mutual effectiveness of defenses against both DPAs and MPAs, causing fragmentation in this field.
This paper aims to provide a unified benchmark and analysis of defenses against DPAs and MPAs, clarifying the distinction between these two similar but slightly distinct domains.
We present a systematic taxonomy of poisoning attacks and defense strategies, outlining their design, strengths, and limitations.
Then, a unified comparative evaluation across FL algorithms and data heterogeneity is conducted to validate their individual and mutual effectiveness and derive key insights for design principles and future research.
Along with the analysis, we frame our work to a unified benchmark, \Method, with high modularity and scalability to evaluate 15 representative poisoning attacks and 17 defense strategies, facilitating future research in this domain. Code is available at \url{https://github.com/vio1etus/FLPoison}.

\end{abstract}

\section{Introduction}
\label{sec:introduction}

Federated learning (FL) is a distributed learning paradigm that enables multiple devices or organizations to collaboratively train a machine learning model without sharing their private data~\cite{fedsgdavg_AISTATS2017}.
It effectively leverages distributed data across different users without compromising their privacy, making it especially valuable for protecting individuals' sensitive information and complying with strict data privacy regulations in industries such as healthcare, mobile devices, Internet of Things, and finance~\cite{fl_healthcare_csur2022, fl_iot_cst2021, fl_mobile_cst2020}.
FL has been applied in real-world privacy-preserving distributed scenarios, such as next-word and emoji prediction in Google’s Gboard~\cite{flkeyboard_google2018,flemoji_google2019}, personalization of Apple’s Siri~\cite{fl_personalization_apple2022}, and credit rating for small enterprises at WeBank~\cite{fatecredit_fedai2020}.
The growth of data protection laws like GDPR~\cite{gdpr_2024} and CCPA~\cite{ccpa_2018} has further driven the adoption of FL, as it aligns with privacy-by-design principles, enabling companies to gain data insights while meeting regulatory requirements.

Nevertheless, the decentralized nature of federated learning introduces new attack surfaces due to its limited visibility of local data and model training.
One of the most critical security concerns is poisoning attacks, in which adversaries deliberately inject malicious data or manipulate local model training to degrade the global model’s performance or bias its predictions\cite{alie_nips2019, IPM_PMLR2020}.
Based on the attack vector, either data manipulation or model training manipulation, poisoning attacks are typically categorized as MPAs and DPAs.
According to the attack objective, they can further be classified as untargeted or targeted attacks.
These attacks can lead to severe consequences.
For instance, in a targeted attack, a self-driving car might misidentify stop signs with specific trigger stickers as speed limit signs, posing significant safety risks~\cite{badnets_NIPSWS2017}.

\begin{table*}[!t]
\centering
\caption{Comparison Between Our SoK and Previous Works.}
\label{tab:related_work}
\setlength{\tabcolsep}{3pt} 
\renewcommand{\arraystretch}{0.7} 
\resizebox{\textwidth}{!}{\begin{tabular}{ccccccccccc}
\toprule
\multicolumn{2}{c}{} & \textbf{Ours} & \begin{tabular}[c]{@{}c@{}}Tian et al.\\ \cite{poisonml_csur2022}-2022\end{tabular} & \begin{tabular}[c]{@{}c@{}}Lyu et al.\\ \cite{privacyrobustfl_tnnls2022}-2022\end{tabular} & \begin{tabular}[c]{@{}c@{}}Shejwalkar et al. \\ \cite{productionflpoison_sp2022}-2022\end{tabular} & \begin{tabular}[c]{@{}c@{}}Gong et al. \\ \cite{backdoorfl_wc2022}-2022\end{tabular} & \begin{tabular}[c]{@{}c@{}}Li et al. \\ \cite{blade_tbd2023}-2023\end{tabular} & \begin{tabular}[c]{@{}c@{}}Sharma et al. \\ \cite{sokflsecurity_acsac2023}-2023\end{tabular} & \begin{tabular}[c]{@{}c@{}}Wan et al.\\ \cite{poisonwflsurvey_cst2024}-2024\end{tabular} \\ \midrule
\multirow{5}{*}{\textbf{\begin{tabular}[c]{@{}c@{}}Unified \\ Evaluation\end{tabular}}} & Model Poisoning Evaluation & \cmark & \xmark & \xmark & \xmark & \xmark & \cmark & \xmark & \xmark \\ \cmidrule{2-10} 
& Data Poisoning Evaluation& \cmark & \xmark & \xmark & \cmark & \cmark & \xmark & \cmark & \cmark \\ \cmidrule{2-10} 
& Cross-FL Algorithm Evaluation & \cmark & \xmark & \xmark & \xmark & \xmark & \xmark & \xmark & \xmark \\ \cmidrule{2-10} 
& Unified Design Guideline & \cmark & \xmark & \xmark & \xmark & \xmark & \xmark & \xmark & \xmark \\ \cmidrule{2-10} 
& \begin{tabular}[c]{@{}c@{}}Unified Open-source Benchmark\end{tabular} & \cmark & \xmark & \xmark & \xmark & \xmark & \xmark & \xmark & \xmark \\ \midrule
\multirow{3}{*}{\textbf{Taxonomy}} &Model Poisoning Attacks & \cmark & \cmark & \cmark & \xmark & \xmark & \cmark & \xmark & \xmark \\ \cmidrule{2-10} 
& Data Poisoning Attacks & \cmark & \cmark & \cmark & \cmark & \cmark & \xmark & \cmark & \cmark \\ \cmidrule{2-10} 
& Fine-grained Categories & \cmark & \xmark & \cmark & \xmark & \xmark & \xmark & \xmark & \xmark \\
    \bottomrule
\end{tabular}
}
\end{table*}

Numerous defense strategies have been proposed to mitigate poisoning attacks.
Based on the attack vector they defend against, they are categorized into model poisoning and data poisoning defenses.
Model poisoning defenses~\cite{krum_nips2017,trimmedmean_ICML2018,rfa_arxiv2019,centeredclipping_icml2021} aim to protect the model's performance from degradation caused by MPAs, while data poisoning defenses~\cite{fltrust_ndss2020, foolsgold_raid2020, deepsight_ndss2022} focus on removing embedded targets or backdoors from the model while maintaining good performance.
These two types of poisoning defenses are often evaluated separately for their respective attacks.
Besides, several studies~\cite{backdoorfl_wc2022, sokflsecurity_acsac2023, poisonwflsurvey_cst2024} have theoretically analyzed poisoning attacks and defenses separately based on attack vectors and attack objectives, yielding a taxonomy that offers a degree of comprehensiveness.

However, most existing studies evaluate these attacks and defenses in isolation, often with inconsistent evaluation configurations and limited comparison strategies, raising concerns about validity.
Furthermore, most literature tends to focus separately on either model poisoning defenses or data poisoning defenses, without a unified consideration and evaluation perspective, leading to a fragmented understanding and situation in this field.

\mypara{Contributions.}
To bridge this gap and provide a quantitative analysis of existing poisoning attacks and defenses in FL, we develop an open-source benchmark featuring over 32 representative poisoning attacks and defenses, generally selected from highly cited and commonly compared works.
To the best of our knowledge, this is the first large-scale unified benchmark for poisoning defenses in the FL setting.
Based on the taxonomy, analysis, and unified evaluation of strategies, we clarify the connections and distinctions between model and data poisoning attacks and defenses, identify advanced attack and defense strategies, and establish design principles to benefit future research.
In summary, our contributions are:
\begin{itemize}
    \item We present a systematization of knowledge of state-of-the-art poisoning attacks and defenses in FL, illuminating the connections and distinctions among DPAs and MPAs by FL algorithms, methodology, strengths, and limitations.
    A fine-grained category based on their underlying principles is devised to better distinguish and understand the design and effectiveness of these attacks and defenses (\Cref{taxonomy:attack,taxonomy:defense}).

    \item We introduce the first unified benchmark with extensive quantitative comparisons of poisoning attacks and defenses across various FL algorithms and data heterogeneity levels.
    Our findings highlight their connections and distinctions, reveal strengths and limitations, identify advanced strategies, and provide insights into design principles and future work (\Cref{sec:experiments}).
    
    \item We present a highly modular and scalable evaluation benchmark that includes over 32 poisoning attack and defense strategies, compatible with 3 commonly used FL algorithm backbones. The source code will be released to advance research on poisoning attacks and defenses in federated learning ( \Cref{app:overview}).

\end{itemize}

\begin{figure*}[htbp]
    \centering
    \includegraphics [width=0.8\textwidth, keepaspectratio] {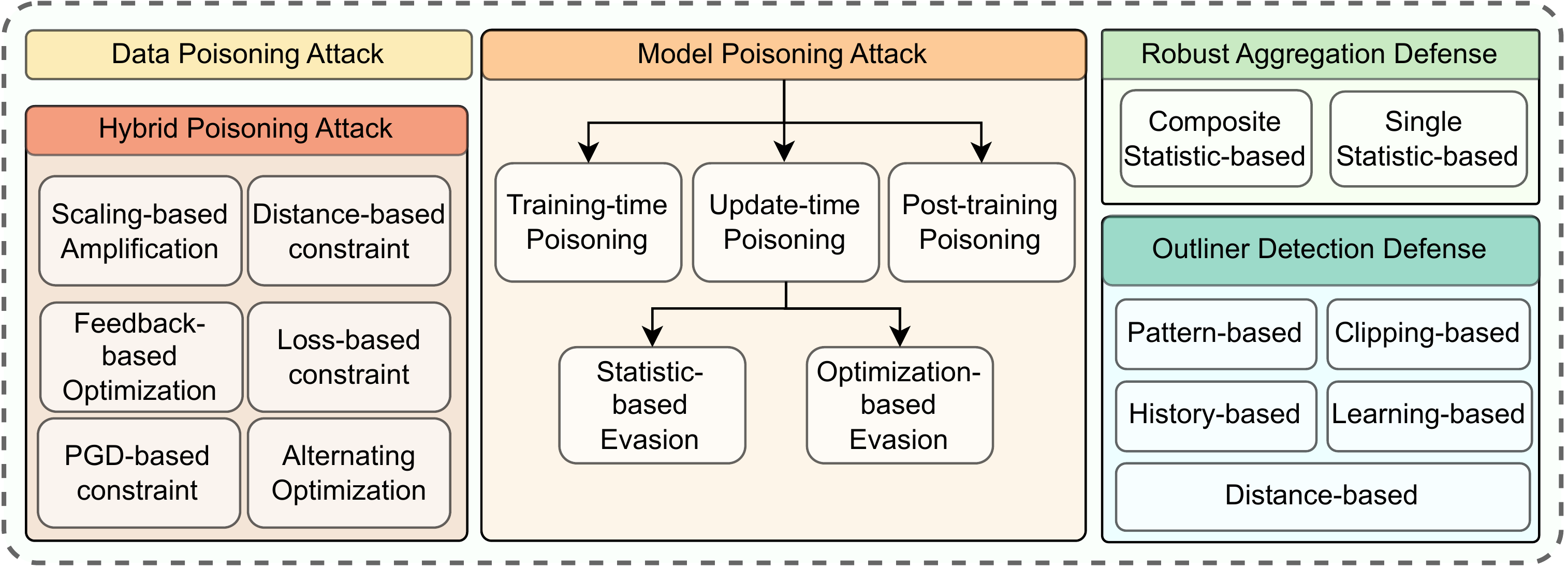}
    \caption{Taxonomies of Poisoning Attacks in Federated Learning}
    \label{pic:fp_taxonomy}
\end{figure*}

\section{Literature Review and Gap Analysis}
\label{sec:related_work}
\Cref{tab:related_work} presents a comparative summary of our SoK and the current literature on poisoning attacks and defenses in FL from 2022 to 2024. Detailed literature review and gap analysis are provided in~\Cref{sec:detailed_review}.

\section{Preliminaries and Problem Formulation}
\label{sec:preliminary}

In this section, we first introduce federated learning and three commonly used FL algorithms in the field of poisoning attacks to establish a system model in~\Cref{sec:preli_fl}
We then provide the principles of poisoning attacks in~\Cref{sec:preli_pa}  and develop a threat model in~\Cref{sec:preli_threat}.

\section{A Taxonomy of Poisoning Attacks}
\label{taxonomy:attack}

We categorize poisoning attacks into MPAs, DPAs, and hybrid poisoning attacks based on their attack point, following existing research~\cite{poisonml_csur2022, privacyrobustfl_tnnls2022}.
While some surveys distinguish between untargeted and targeted attacks, we focus on attack vectors to highlight their attack points, similarities, and differences.
Model poisoning attacks are mainly untargeted attacks, while data and hybrid poisoning attacks are typically targeted attacks.
Following \Cref{pic:fp_taxonomy}, we provide a systematic taxonomy analysis below, incorporating hybrid poisoning attacks into the data poisoning analysis due to their similar attack objective.
\Cref{tab:attack} provides a summary of poisoning attacks.

\begin{table*}[!t]
\centering
\caption{Summary of Poisoning Attacks}
\label{tab:attack}
\renewcommand{\arraystretch}{0.95} %
\setlength{\tabcolsep}{6pt} 
\resizebox{\textwidth}{!}{\begin{tabular}{cccccccc}\toprule
\multirow{2}{*}{\textbf{Category}} & \multirow{2}{*}{\textbf{Attacks}} & \multicolumn{5}{c}{\textbf{Attacker Knowledge}} & \multirow{2}{*}{\textbf{Algorithm Base}} \\ \cmidrule{3-7} 
 &  & Train Dataset & Training Process & Benign Updates & Malicious Updates & Defense Strategy \\ \midrule
\multirow{8}{*}{MPA} & Gaussian Random Attack~\cite{krum_nips2017} & \xmark & \xmark & \xmark & \xmark & \xmark & FedSGD \\ \cmidrule{2-7} 
 & Sign Flipping Attack~\cite{signflipping_icml2018} & \xmark & \xmark & \xmark & \xmark & \xmark & FedSGD \\ \cmidrule{2-7} 
 & ALIE Attack~\cite{alie_nips2019} & \xmark & \xmark & \cmark & \xmark & \xmark & FedSGD \\ \cmidrule{2-7} 
 & Inner Product Manipulation~\cite{IPM_PMLR2020} & \xmark & \xmark & \cmark & \xmark & \xmark & FedSGD \\ \cmidrule{2-7}
  & Fang Attack~\cite{fangattack_usenix_sec2020} & \xmark & \xmark & \cmark & \cmark & \cmark & FedAvg \\ \cmidrule{2-7} 
 & MinMax Attack~\cite{min_ndss2021} & \xmark & \xmark & \cmark & \xmark & \xmark & FedSGD \\ \cmidrule{2-7} 
 & MinSum Attack~\cite{min_ndss2021} & \xmark & \xmark & \cmark & \xmark & \xmark & FedSGD \\ \cmidrule{2-7} 
 & Mimic Attack~\cite{mimic_iclr2022} & \xmark & \xmark & \cmark & \xmark & \xmark & FedSGD \\ \midrule
\multirow{2}{*}{DPA} & Label Flipping Attacks~\cite{labelflipping_icml2012, foolsgold_raid2020, fangattack_usenix_sec2020} & \cmark & \xmark & \xmark & \xmark & \xmark & Centralized \\ \cmidrule{2-7} 
 & BadNet Attack~\cite{badnets_NIPSWS2017} & \cmark & \xmark & \xmark & \xmark & \xmark & Centralized \\ \midrule
\multirow{5}{*}{\begin{tabular}[c]{@{}c@{}}Hybrid \\ Poisoning\\ Attack\end{tabular}} & DBA Attack~\cite{dba_iclr2019} & \cmark & \xmark & \xmark & \xmark & \xmark & FedOpt \\ \cmidrule{2-7} 
 & Alternating Minimization Attack~\cite{altermin_icml2019} & \cmark & \xmark & \xmark & \xmark & \xmark & FedOpt \\ \cmidrule{2-7} 
 & Model Replacement Attack~\cite{modelreplacement_aistats2020} & \cmark & \xmark & \xmark & \xmark & \xmark & FedOpt \\ \cmidrule{2-7} 
 & Edge-case Backdoor~\cite{edgecase_nips2020} & \cmark & \xmark & \xmark & \xmark & \xmark & FedOpt \\ \cmidrule{2-7} 
 & Neurotoxin~\cite{neurotoxin_icml2022} & \cmark & \xmark & \xmark & \xmark & \xmark & FedOpt \\\bottomrule
\end{tabular}
}
\end{table*}

\subsection{Model Poisoning Attacks}

MPAs manipulate local training processes and are classified into three types based on the training stage: training-time, post-training, and update-time poisoning.
Most MPAs are untargeted, aiming to disrupt convergence (Byzantine attacks).
Below, we describe each category and summarize 8 representative attacks, including their designs, assumptions, limitations, behaviors, and key parameters.

\subsubsection{Training-time Poisoning Attacks}

In these attacks, the adversary can alter training configurations such as the number of epochs, learning rate, loss function, and other parameters.
These attacks often combine with data poisoning attacks, forming hybrid poisoning attacks.
We detail these attacks on hybrid poisoning attacks (see \Cref{sec:hybrid_attack}).

\subsubsection{Post-training Poisoning Attacks}

These attacks allow the adversary to manipulate the training results.
Since these attacks require only knowledge of the adversary itself, they are also referred to as non-omniscient attacks~\cite{krum_nips2017, bulyan_icml2018, alie_nips2019, IPM_PMLR2020, centeredclipping_icml2021}.
They are summarized below.

\mypara{Gaussian Random Attack}~\cite{krum_nips2017} tries to inject random noise in local gradient updates to disrupt the global model's training, particularly since the variance of benign updates becomes smaller as the model converges.
It involves each adversary submitting random vectors, sampled from a Gaussian distribution.
Two parameters control the attack's ability: the mean and variance of the distribution, with a larger variance leading to a stronger attack.
Specifically, the malicious gradient update can be formalized as:

\begin{equation}
    g_{mal} = n, \quad \text{where }n \sim \mathcal{N}(0, \sigma^2)
\end{equation}

\mypara{Sign Flipping Attack}~\cite{alie_nips2019} aims to deviate the global model's gradient in the wrong direction, hindering convergence and degrading the training performance.
The adversary disrupts training by flipping the signs of gradient updates.
Specifically, the malicious gradient update can be defined as:

\begin{equation}
    g_{mal} = - \lambda \dot g,
\end{equation}
where positive parameter $\lambda$ affects the attack impact.

\subsubsection{Update-time Poisoning Attacks}

These attacks occur when clients submit updates, enabling malicious clients to eavesdrop on benign ones or collude with others to launch attacks.
Known as omniscient attacks, they rely on information from others.
We first outline their main categories, followed by detailed explanations.
These attacks can be classified into two types based on how malicious updates are constructed: Statistic-based evasion~\cite{alie_nips2019, IPM_PMLR2020, mimic_iclr2022} and Optimization-based evasion~\cite{fangattack_usenix_sec2020, min_ndss2021}.  
The former primarily generates malicious updates resembling benign ones by eavesdropping on benign updates using statistical measures like mean and standard deviation.  
The latter optimizes updates to bypass server defenses, such as L2 distance or Krum, making them appear benign.  
Some attacks require knowledge of the aggregation rule or defense method (aggregator-specific).
In the absence of this knowledge, a random aggregation rule may be chosen instead~\cite{fangattack_usenix_sec2020, min_ndss2021}.

\mypara{A Little is Enough (ALIE) Attack}~\cite{alie_nips2019} identifies that instead of relying on large changes to parameters, carefully crafted small perturbations of a few clients are sufficient to undermine convergence performance and bypass defenses like Krum~\cite{krum_nips2017}, TrimmedMean~\cite{trimmedmean_ICML2018}, and Bulyan~\cite{bulyan_icml2018}.
The ALIE attack assumes that client model parameters are IID and follow a normal distribution.
Leveraging the empirical variance of benign weight updates following normal distribution, ALIE constrains the perturbation range of malicious updates to ensure stealth.
Therefore, it can mislead the median of overall weight updates through malicious updates, a key metric for some defenses' mitigation or detection, ensuring the negative impact is less likely to be eliminated.
Specifically, the adversary first calculates the mean $\mu_i$ and standard deviation $\sigma_i$ of benign weight updates for each parameter coordinate $i \in [d]$, where $d$ is the dimension of model weight update, then sets the malicious weight updates:

\begin{equation}
    w_{mal_i} \in (\mu_i-z^{max}\sigma_i, \mu_i-z^{max}\sigma_i),
\end{equation}
where $z^{\text{max}}$ controls the perturbation magnitude, which can be calculated via standard normal function $\phi(z)$:
\begin{equation}
    z^{max} = \text{max}_z \left(\phi(z) < \frac{n - \lfloor{\frac{n}{2}+1\rfloor}}{n-f}\right).
\end{equation}

\mypara{Inner Product Manipulation (IPM) Attack}~\cite{IPM_PMLR2020} points out that most robust aggregators, such as coordinate-wise median~\cite{trimmedmean_ICML2018} and Krum~\cite{krum_nips2017}, only guarantee an upper bound on the distance between the robust aggregated vector and the correct mean, but neglect the necessity of non-negative between the true gradient and the robust aggregated vector.
The IPM attack exploits the negation of their inner product to degrade gradient descent convergence.
Specifically, it assumes that the adversary knows the updates of benign clients so that it can calculate the mean value of the true benign gradient, and then set the malicious gradient update as:

\begin{equation}
    g_{mal} = - \epsilon \sum_{i=f+1}^n g_i,
\end{equation}
where $\epsilon$ is a positive parameter controlling the impact.

\mypara{Mimic Attack}~\cite{mimic_iclr2022} involves the adversary copying a specific benign update, amplifying its influence while under-representing others. Since the attack mimics legitimate updates, it is indistinguishable and hard to filter out. The targeted update can be any benign update.

\mypara{Fang Attack}~\cite{fangattack_usenix_sec2020} is an aggregator-specific attack involving collusive behavior, where each adversary is aware of the other adversaries' local training data, models, and gradient updates.
If the aggregator is unknown, a randomly assumed aggregator is used.
Here we present the Krum-version Fang attack, which deviates the update while optimizing it to be selected by Krum as if it were benign.
Specifically, in iteration $t$, it first estimates the changing direction $s_i$ of parameter $i$ using the current before-attack model of $f$ adversaries, where $s_i^{t} = sign(f_{\text{avg}}(w_{i \in (0, f)}))$.
The sign function returns 1, 0, or -1 for positive, zero, and negative values, respectively.
Then, it starts by crafting a malicious model:

\begin{equation}
w_i^t=w_{\text {global }}^{t-1}-\lambda s_i^t, i \in (0, f).
\end{equation}
After that, the before-attack and crafted models are submitted as candidate updates to the Krum aggregator, expecting to select the crafted model as one of the benign updates by (1) optimizing $\lambda$ value and (2) appending additional crafted models if the search fails. 

\mypara{MinMax Attack}~\cite{min_ndss2021} aims to find a malicious gradient, whose maximum distance from other benign gradient updates is smaller than the maximum distance between any two benign gradient updates via finding an optimal $\gamma$:

\begin{equation}
\begin{gathered}
\underset{\gamma}{\text{argmax}} \max _{i \in (0,f)}||g_i-g_j||_2 \leq \max _{j, k \in (f+1, n)}||g_j-g_k||_2, \\
g_i = \nabla^b + \gamma \nabla^p, \nabla^b = f_{\text{avg}}(g_{j \in (f+1, n)}),
\end{gathered}
\end{equation}
where $\nabla^p$ is the permutation vector.
$\nabla^p$ can be the inverse unit vector of $\nabla^b$, $-\frac{\nabla^b}{||\nabla^b||_2}$, the inverse sign vector, $-sign(\nabla^b)$, or simply the inverse standard deviation of the benign updates $-std(g_{j \in (f+1, n)})$.

\mypara{MinSum Attack}~\cite{min_ndss2021} seeks a malicious gradient whose sum of distances from other benign gradient updates is smaller than the sum of distances of any benign gradient updates from other benign updates via finding an optimal $\gamma$:

\begin{equation}
\begin{gathered}
\underset{\gamma}{\text{argmax}} \sum _{i \in (0,f)}||g_i-g_j||_2^2 \leq \sum _{j, k \in (f+1, n)}||g_j-g_k||_2^2, \\
g_i = \nabla^b + \gamma \nabla^p, \nabla^b = f_{\text{avg}}(g_{j \in (f+1, n)}).
\end{gathered}
\end{equation}

\subsection{Data Poisoning Attacks}

Data poisoning attacks primarily manipulate training data to achieve their objectives.
Some data poisoning attacks are often combined with training-time poisoning attacks, classifying them as hybrid poisoning attacks.
A specific subtype of these attacks, known as backdoor attacks, aims to embed backdoors into the model so that when samples with the corresponding triggers are fed into the model, attack-specific predictions are produced.
Below, we present two representative data poisoning attacks.

\mypara{Label Flipping Attacks}~\cite{labelflipping_icml2012,foolsgold_raid2020,fangattack_usenix_sec2020} alter training labels to attacker-specific labels with a label substitution model, which can produce random, inverse, or target labels.
In the random strategy, the source label is replaced with a random one; in the inverse strategy, label $l$ is flipped to $L-l-1$, where $L$ is the number of classes; and in the target strategy, a specific label is assigned to label $l$.
It supports both targeted and untargeted poisoning, and the targeted version is used in our paper.

\mypara{BadNets Attacks}~\cite{badnets_NIPSWS2017, edgecase_nips2020, crfl_icml2021, modelreplacement_aistats2020} first introduces backdoor attacks, where the goal is to train a model that performs normally on the user’s training and test data but misclassifies inputs containing a backdoor trigger to an attacker-specified target label.
It is achieved by injecting a certain ratio of malicious samples into the training set, created by adding triggers to images and altering their corresponding labels to the target label.

\subsection{Hybrid Poisoning Attacks}
\label{sec:hybrid_attack}

Hybrid poisoning attacks are conducted at both data poisoning and model poisoning attack points to enhance attack effectiveness and evade defenses.

\mypara{Distributed Backdoor Attack (DBA)}~\cite{dba_iclr2019}, unlike centralized backdoor attacks, BadNets where each adversary embeds the same global trigger, leverages the distributed nature of federated learning by breaking a global trigger pattern into distinct local patterns.
These local patterns are embedded into the training data of different adversarial participants.
Moreover, scale techniques~\cite{modelreplacement_aistats2020} are used to amplify malicious updates during training, while samples with a global trigger are used during inference to activate backdoor attacks.

\mypara{Alternating Minimization Attack (AlterMin)}~\cite{altermin_icml2019} assumes IID training and specific validation data samples, where the adversary controls a small number of non-colluding clients (typically one) without access to other clients' updates.
It jointly optimizes the model with the adversarial objective and the stealth objective (in an alternative manner).
The adversarial objective is to let the global model misclassify the crafted trigger-embedded input to an attacker-specified target label.
Meanwhile, the stealth objective is to make the malicious local updates deviate subtly from the benign updates, avoiding detection by distance-based defenses.

\mypara{Model Replacement Attack (ModelRep)}~\cite{modelreplacement_aistats2020}, also known as the Constrain-and-Scale Attack, demonstrates that the adversary can replace the global model with a malicious backdoored version by employing constrain-and-scale techniques.
It first constrains the training process by incorporating an anomaly detection term $\mathcal{L}_{\text{ano}}$ to loss function $\mathcal{L}_{\text{model}}$, based on various metrics, such as the p-norm distance between weight matrices or cosine distance (as used in the paper):
\begin{equation}
\mathcal{L}_{\text{model}}=\alpha \mathcal{L}_{\text {class}}+(1-\alpha) \mathcal{L}_{\text {ano}},
\end{equation}
where $\alpha$ controls the trade-off between evading anomaly detection and achieving the attack objectives.
Then, it scales the (pseudo) gradient update before submission by a factor of $\gamma = \frac{n}{\eta}$, where $n$ is the number of participants and $\eta$ is the global learning rate at the server.
Since clients typically lack knowledge of $n$ and $\eta$, the authors suggest that the attacker can empirically estimate $\gamma$.
    
\mypara{Edge-case Backdoor}~\cite{edgecase_nips2020} introduces a new type of backdoor attack where adversaries exploit edge-case samples for targeted attack, which are unlikely to appear in normal training or test data and reside in the tail of the input distribution.
The adversary can create edge-case datasets by mixing normal datasets with rare but similar ones, such as using the ARDIS dataset~\cite{ardis_nca2020} for the MNIST dataset~\cite{mnist_proceddingsIEEE}.
While both datasets involve handwritten digits, ARDIS exhibits greater variability in handwriting styles and more noise from historical sources.
Additionally, they proposed two attack strategies based on the edge-case backdoor.
The first involves adversaries using projected gradient descent (PGD) at each FL round to ensure the malicious model evades norm clipping defenses.
The second strategy combines PGD with a Model Replacement Attack~\cite{modelreplacement_aistats2020}, applying scaling before model submission.
    
\mypara{Neurotoxin}~\cite{neurotoxin_icml2022} introduces a durable backdoor attack strategy that involves updating coordinates infrequently used by benign clients to counter backdoor detection.
This approach is based on the observation that the sparsity of gradients in stochastic gradient descent (SGD) causes the majority of the L2 norm of aggregated benign gradients to be contained in a small number of parameters~\cite{sparsifiedSGD_nips2018, SGD_sketching_nips2019}.
Therefore, the backdoor can be maintained durably by only updating those parameters unlikely updated by benign clients.
Specifically, it first gets the top-k smallest absolute gradient values of the global model as the gradient mask, which is used to project the gradient to the infrequently updated parameters of the global model.
In addition, Neurotoxin applies gradient norm clipping to prevent the model from being updated too much.

\subsection{Summary of Attacks}

Poisoning attacks in FL progress from treating model updates as black boxes to understanding the importance of weight parameters.
Early MPAs, like Gaussian and Sign Flipping, are simple and low-cost, requiring minimal attacker knowledge while still causing significant disruption.
However, they can be effectively mitigated by defenses based on some robust statistical metrics, like Krum~\cite{krum_nips2017}, Multi-Krum~\cite{krum_nips2017}, Coordinate-wise Median~\cite{trimmedmean_ICML2018}, RFA~\cite{rfa_arxiv2019}.
Their impact remains notable against basic outlier detection defenses, such as Auror~\cite{auror_acsac2016}.
Then, it advances to statistical-based evasion like IPM and ALIE, which exploit the statistical properties of benign client updates for stealth, making them hard to detect but less damaging.
Advanced attacks such as FangAttack, MinSum, and MinMax are designed to evade common defense metrics.
FangAttack stands out by integrating metric evasion with statistical deviation techniques.
Besides, data heterogeneity also begins to be considered for deviation, such as Mimic attack.
Data poisoning attacks start with basic methods like Label Flipping and BadNets and advance to techniques such as ModelRep and AlterMin, which enhance attack performance but suffer from poor generalization due to many hyper-parameters.
More advanced strategies involve backdoor implantation, like DBA, which distributes backdoors across clients, and Edge-case backdoor, poisoning datasets with tail-distribution samples.
Fundamental attacks like BadNets, DBA, and Edge-case require minimal tuning to ensure stability.
As attacks evolve, model parameters are no longer being treated as black boxes, with approaches like Neurotoxin using key parameters for backdoor durability.
This broadens the attack surface but increases complexity, raising generalization concerns.

\myta{
Our analysis categorizes key poisoning attack methodologies to guide comprehensive defense development. These include: (1) Random Noise, adding noise to disrupt gradients (e.g., Gaussian Random Attack); (2) Sign Manipulation, altering gradient signs (e.g., Sign Flipping, IPM); (3) Statistic-based Evasion, using update statistics deviations (e.g., ALIE, FangAttack); (4) Optimization-Based Evasion, optimizing updates to avoid detection (e.g., FangAttack, MinMax); (5) Scaling-based Amplification, as in Model Replacement, DBA, AlterMin; (6) Distance-based Constraints, used in MinSum and MinMax; (7) Loss-based constraints, seen in AlterMin; (8) Alternating Optimization, as in AlterMin; (9) Adaptive Attacks, with stronger robustness; (10) Feedback-based Optimization, in emerging methods like 3DFed~\cite{3dfed_sp2023}.
}

\section{A Taxonomy of Poisoning Defenses}
\label{taxonomy:defense}

Poisoning Defenses, also known as aggregation strategies, are often considered separately for model and data poisoning attacks.
We discuss both perspectives, highlighting that these defenses can be categorized into robust aggregation and outlier detection classes.
Robust aggregation aims to generate reliable outputs based on robust statistics~\cite{robust_statistics_2011} even with malicious updates, while outlier detection identifies and removes abnormal updates that may harm performance.
Below, we outline the defense strategies for model and data poisoning attacks.
\Cref{tab:defense} provides a summary of defense strategies with their knowledge, assumptions, complexity, and categories.

\begin{table*}[]
\centering
\normalsize
\caption{Summary of Poisoning Defenses}
\label{tab:defense}
\setlength{\tabcolsep}{3pt} 
\renewcommand{\arraystretch}{0.8} 
\resizebox{\textwidth}{!}{%
\begin{tabular}{cccclcccccc}
\toprule
\multirow{2}{*}{\textbf{Defense/Aggregator}} & \multicolumn{3}{c}{\textbf{Aggregator Knowledge}} &  & \multicolumn{3}{c}{\textbf{Aggregator Assumption}} & \multirow{2}{*}{\textbf{Complexity}} & \multirow{2}{*}{\textbf{Algorithm Base}} & \multirow{2}{*}{\textbf{Category}} \\ \cmidrule{2-4} \cmidrule{6-8}
 & \begin{tabular}[c]{@{}c@{}}Bounded\\ Adversary\end{tabular} & \begin{tabular}[c]{@{}c@{}}Benign\\ Server\end{tabular} & History &  & \begin{tabular}[c]{@{}c@{}}Distribution\\ Assumption\end{tabular} & Momentum & \begin{tabular}[c]{@{}c@{}}\# \\ Adversaries\end{tabular} &  &  \\ \midrule
Krum~\cite{krum_nips2017} & \cmark & \xmark & \xmark &  & Balanced IID & \xmark & $2f+2 < n$ & $O(dn^2)$ & FedSGD & Single Statistic-based \\ \midrule
Multi-Krum~\cite{krum_nips2017} & \cmark & \xmark & \xmark &  & Balanced IID & \xmark & $2f+2 < n$ & $O(dn^2)$ & FedSGD & Composite Statistical-based \\ \midrule
Coordinate-wise Median~\cite{trimmedmean_ICML2018} & \xmark & \xmark & \xmark &  & Balanced IID & \xmark & $2f+1 < n$ & $O(dn)$ & FedSGD & Single Statistic-based \\ \midrule
TrimmedMean~\cite{trimmedmean_ICML2018} & \xmark & \xmark & \xmark &  & Balanced IID & \xmark & $2f+1 < n$ & $O(dn\log n)$ & FedSGD & Single Statistic-based \\ \midrule
Bulyan~\cite{bulyan_icml2018} & \cmark & \xmark & \xmark &  & Balanced IID & \xmark & $4f+3 \le n$ & $O(dn^2)$ & FedSGD & Composite Statistical-based \\ \midrule
RFA~\cite{rfa_arxiv2019} & \xmark & \xmark & \xmark &  & Balanced IID & \xmark & $2f+1 < n$ & $O(Tdn)$ & FedAvg & Single Statistic-based \\ \midrule
CenteredClipping~\cite{centeredclipping_icml2021} & \xmark & \xmark & \cmark &  & Balanced and Imbalanced IID & \cmark & $2f+1 < n$ & $O(dn)$ & FedSGD & History, Clipping-based \\ \midrule
FLTrust~\cite{fltrust_ndss2020} & \xmark & \cmark & \xmark &  & Balanced IID and non-IID & \xmark & - & $O(dn)$ & FedOpt & Learning-based \\ \midrule
DnC~\cite{min_ndss2021} & \xmark & \xmark & \xmark &  & Balanced IID & \xmark & $2f+1 < n$ & $O(d_{sub}^3)$ & FedSGD & Composite Statistical-based \\ \midrule
SignGuard~\cite{signguard_ICDCS22} & \xmark & \xmark & \xmark &  & Balanced IID and Non-IID & \xmark & $2f+1 < n$ & $O(dn + n \log n)$ & FedSGD & Pattern-based \\ \midrule
Bucketing~\cite{mimic_iclr2022} & \cmark & \xmark & \xmark &  & Balanced IID and Non-IID & \xmark & $2f+2 < n / S$ & $O(dn^2)$ & FedSGD & Composite Statistical-based \\ \midrule
Auror~\cite{auror_acsac2016} & \xmark & \xmark & \xmark &  & Balanced IID & \xmark & $2f+1 < n$ & $O(dn)$ & FedSGD & Similarity-based \\ \midrule
FoolsGold~\cite{foolsgold_raid2020} & \xmark & \xmark & \cmark &  & Balanced Non-IID & \xmark & $2f+1 < n$ & $O(sn^2)$ & FedSGD & Similarity-based \\ \midrule
NormClipping~\cite{normclipping_nips2019} & \xmark & \xmark & \xmark &  & Balanced IID and Non-IID & \xmark & $2f+1 < n$ & $O(dn)$ & FedOpt & Clipping-based \\ \midrule
CRFL~\cite{crfl_icml2021} & \xmark & \xmark & \xmark &  & Balanced IID and Non-IID & \xmark & $2f+1 < n$ & $O(dn)$ & FedOpt & Clipping-based \\ \midrule
DeepSight~\cite{deepsight_ndss2022} & \xmark & \xmark & \cmark &  & Balanced IID & \xmark & $2f+1 < n$ & $O(dn^2)$ & FedOpt & Pattern, Distance-based \\ \midrule
FLAME~\cite{flame_usenix_security2022} & \xmark & \xmark & \cmark &  & Balanced IID and Non-IID & \xmark & $2f+1 < n$ & $O(dn^2)$ & FedOpt & \begin{tabular}[c]{@{}c@{}}Distance, Clipping, \\ Single Statistic-based\end{tabular} \\ \bottomrule
\end{tabular}}
{\raggedright Note: Those who get - on the number of adversaries rely on other trust bases rather than a majority basis for defense.  \par}
\end{table*}

\subsection{Model Poisoning Defenses}

Model poisoning defenses, also referred to as Byzantine defenses, are designed to safeguard FL models against model poisoning attacks or Byzantine attacks.
Below, we discuss 11 representative defense strategies.

\mypara{Krum}~\cite{krum_nips2017} is the first algorithm proven to be Byzantine-resilient for distributed SGD.
It utilizes the majority and squared-distance principles to select one client's gradient update (vector) that is closest to its $n-f$ neighbors, thereby excluding the vectors that are too far away.
Specifically, following the majority assumption, $f < \frac{n}{2} - 1$, the selected vector, as the aggregation result, is:
\begin{equation}
\begin{aligned}
    & \text{Krum}(g_1, \dots, g_n) = g_k, \quad \text{where} \; k = \underset{i \in [n]}{\text{argmin}} \, s(i) \\
    & s(i) = \sum_{j \in [n], j \neq i} ||g_i - g_j||^2,
\end{aligned}
\end{equation}
where $s(i)$ is the score for $g_i$.
$i \rightarrow j$ indicates that $g_j$ belongs to the $n-f-2$ closest vectors to $g_i$ measured by squared Euclidean distance, which can be obtained by iterating over all $j \in [n]$, then sorting to get the $n-f-2$ smallest ones.
The time complexity of Krum is $O(n^2 \cdot d)$ for the $d$-dimensional vector and distances calculation.

\mypara{Multi-Krum (M.K.)}~\cite{krum_nips2017}, a variant of Krum, different from Krum selecting only one closest vector among the gradient update vectors, selects the $m$ closest ones, and computes their average.
Specifically, the aggregation results are:

\begin{equation}
\begin{aligned}
    &\text{Multi-Krum}(g_1,\cdots,g_n) = f_{\text{avg}} (g_k), \\
    &k \in \{i_1, i_2, \dots, i_m\} = \underset{i \in [n]}{\text{argmin}}^m\: s(i),\\
\end{aligned}
\end{equation}
where $s(i)$ is the same as in Krum.

\mypara{Coordinate-wise Median (Median)}~\cite{trimmedmean_ICML2018} computes the aggregation result as ${CM}(g_1,\cdot,g_n)$, where the $j$-th coordinate is:
\begin{equation}
CM(g_1,\cdots,g_n)[j] = \text{median}(g_1[j], \cdots, g_n[j])
\end{equation}
with  $j \in [d]$ and the median being the one-dimensional median.
Accordingly, the time complexity is $O(dn)$ to get the median of the d-dimensional vector.

\mypara{TrimmedMean (T.M.)}~\cite{trimmedmean_ICML2018} assumes the majority of clients are benign.
It first sorts each coordinate $j$ to get the coordinate-wise sorted updates, $g_{\sigma}$.
Then it averages the coordinate-wise sorted updates after trimming (excluding) a $\beta$ fraction of the largest and smallest coordinate values.
Specifically, 
\begin{equation}
\begin{aligned}
TM(g_1,\cdots,g_n)[j] = \frac{1}{(1-2\beta)n} \overset{n-\beta n}{\underset{i=\beta n}{\sum}}g_{\sigma (i)}[j].
\end{aligned}
\end{equation}
The primary time complexity arises from the d-dimensional sorting, $O(dn \log n)$.

\mypara{Bulyan}~\cite{bulyan_icml2018} acts as a meta-algorithm designed to enhance any norm-based aggregation rule by introducing a form of coordinate-wise robust aggregation, thereby improving its overall performance.
It selects Krum~\cite{krum_nips2017} to form a selection set $S$, consisting of the $m$ vectors with the smallest scores, and aggregates the coordinates closest to their median, which is,

\begin{equation}
\begin{aligned}
    &\text{Bulyan}(g_1,\cdots,g_n)[j] = \frac{1}{\beta} \sum_{j \in \mathcal{S}} S[I[j]], j \in [d],\\
    &S=\{i_1, i_2, \dots, i_m\} = \underset{i \in [n]}{\text{argmin}}^m\: s(i),\\
\end{aligned}
\end{equation}
where $I$ is a $(\beta, d)$ index vector that identifies the elements in the selection set $S$ that is closest to the median of updates coordinate-wisely.
In other words, each $i$-th coordinate of the aggregation result is the average of the $\beta$ closest $i$-th coordinates to the median of the $i$-th coordinate among the selected set $S$.
With $n \ge 4f + 3$, Bulyan guarantees the participation of at least $2f + 3$ workers for each use of the Krum.
Bulyan has $O(d n^2)$ time complexity for Krum and sort operations.

\mypara{Robust Federated Learning (RFA)}~\cite{rfa_arxiv2019} replaces the weighted mean aggregation with a weighted geometric median, approximated using a smoothed Weiszfeld algorithm.
This approach aims to produce an aggregated vector that is robust and closer to most updates, enhancing resilience to outliers.
The time complexity primarily stems from the approximation using the smoothed Weiszfeld algorithm, which requires $T$ iterations, leading to a $O(Tdn)$ complexity.

\begin{equation}
\operatorname{RFA}(w_1, \cdots, w_n)=\underset{v}{\text{argmin}} \sum_{i=1}^n ||v - w_i||_2
\end{equation}

\mypara{FLTrust}~\cite{fltrust_ndss2020} assumes that the server has access to a clean, small dataset, which is used to train the server model and bootstrap trust during the training process.
It characterizes model updates by their direction and magnitude.
First, it computes the cosine similarity between the server model and each client model update and clips them using ReLU to derive their trust scores.
Then, it normalizes each update by the norm of the server update and finally performs a weighted sum of the updates based on their trust scores as the aggregated updates.
Specifically, it gets the ReLU-clipped trust score, 

\begin{equation}
S_i = \operatorname{ReLU}(c_i), \quad \text{where}\; c_i = \cos \theta_i =\frac{\langle{g}_i, {g}_0\rangle}{|{g}_i| \cdot |{g}_0|}
\end{equation}
where $g_0$ is the server model update in that round, and $S_i$ is the trust score of update $i$.
Finally, the aggregated update can be derived as
\begin{equation}
\begin{aligned}
\text{FLTrust}(g_1,\cdots,g_n) &=\frac{1}{\sum_{j=1}^n S_j} \sum_{i=1}^n S_i \cdot \frac{|{g}_0|}{|{g}_i|} \cdot {g}_i.
\end{aligned}
\end{equation}
The time complexity $O(dn)$ mainly comes from the normalization and aggregation operation.

\mypara{CenteredClipping (C.C.)}~\cite{centeredclipping_icml2021} highlights that existing aggregation rules may not ensure convergence in quantity-imbalanced IID data distributions, even without adversaries.
This issue is further compounded by recent time-coupled attacks, such as ALIE~\cite{alie_nips2019} and IPM~\cite{IPM_PMLR2020}, which allow adversaries to coordinate their efforts across multiple rounds, stealthily compromising convergence.
The study demonstrates that incorporating historical data is essential for aggregation rules to effectively defend against these time-coupled attacks.
It proposes an iterative clipping rule that clips updates centered on the previous round's aggregated update, incorporating the assumption of worker momentum to reduce variance.
At round $l+1$, the aggregation results ${v}_{l+1}$ is,

\begin{equation}
{v}_{l+1}={v}_l+\frac{1}{n} \sum_{i=1}^n\left({x}_i-{v}_l\right) \min \left(1, \frac{\tau_l}{|{x}_i-{v}_l|}\right),
\end{equation}
where $x_i$ represents the uploaded momentum from client $i$ if momentum is used; otherwise, it represents the gradient update from client $i$.
It requires $O(dn)$ time complexity for clipping and assumes the use of momentum by workers.

\mypara{DnC}~\cite{min_ndss2021} identifies anomalous noise or deviations using singular value decomposition (SVD) based spectral methods for outlier filtering.
To mitigate the computational cost of applying SVD to high-dimensional gradients, DnC first randomly subsamples the gradients to reduce their dimensionality and centers the subsampled gradient updates by subtracting their coordinate-wise mean.
It then identifies the top right singular eigenvector and projects the centered updates onto this vector to compute outlier scores.
Finally, DnC selects the $n - \beta f$ indices corresponding to the vectors with the lowest outlier scores as benign updates.
The time complexity $O(d_{sub}^3)$ mainly depends on the subsample dimension $d_{sub}$.

\mypara{Bucketing}~\cite{min_ndss2021} addresses the challenge of FL on heterogeneous data.
It first shuffles the sequence of client updates and then divides them into several buckets of size $S$.
Finally, it aggregates the updates in each bucket using a specified aggregator.
Its time complexity primarily stems from the aggregator used.
Following the original paper, Krum is the default aggregator, resulting in $O(dn^2)$ complexity.

\mypara{SignGuard (S.G.)}~\cite{signguard_ICDCS22} detects outliers by leveraging sign statistics of gradients.
It first filters updates by applying lower and upper bounds, which are corresponding factors multiplied by the median of the clients' update norms.
Next, it randomly samples gradients, computes their sign statistics, and clusters them, selecting the majority cluster as benign.
Finally, the mean of the benign cluster is used as the aggregated update.
The time complexity is primarily due to the norm and sort operations, which is $O(dn + n \log n)$.

\subsection{Data Poisoning Defenses}

Data poisoning defenses aim to protect FL models from data poisoning attacks, with most of them focusing on backdoor defenses that aim to eliminate potential backdoors in the models.
We consider 6 representative defense algorithms.

\mypara{Auror}~\cite{auror_acsac2016} identifies that benign clients' indicative features follow a consistent distribution, while adversaries' features exhibit anomalies.
It selects indicative features by clustering each update coordinate into two clusters and filtering based on the distance between their cluster centers and a threshold.
Indicative features are then clustered, with the majority determining benign clients for aggregation.
Designed for IID data, its time complexity is $O(dn)$ for initial coordinate clustering and $O(sn)$ for clustering $s$ selected features.

\mypara{FoolsGold (F.G.)}~\cite{foolsgold_raid2020}, building on the intuition of Auror, observes that in a non-IID training data scenario, indicative features from benign clients exhibit greater diversity compared to those from adversaries, as adversaries tend to behave similarly in a Sybil poisoning setting.
Unlike Auror, it selects indicative features based on the magnitude of the model parameters in the output layer of the global model, as these directly influence the prediction probabilities.
It calculates the cosine similarity between the indicative features of historically accumulated multiple-round updates to improve client similarity estimation, then re-weights the similarity scores for weighted aggregation.
Its time complexity mainly stems from the pairwise cosine distance and the pardoning function, $O(sn^2)$, given the number of selected indicative features $s$.

\mypara{NormClipping (N.C.)}~\cite{normclipping_nips2019} demonstrates that norm clipping can effectively mitigate the backdoor task without compromising overall performance.
It handles the updates as follows,
\begin{equation}
\Delta w_{t+1}=\sum_{k \in S_t} \frac{\Delta w_{t+1}^k}{\max (1,||\Delta w_{t+1}^k||_2 / M)}.
\end{equation}
The updates are clipped based on a self-defined threshold, and their mean is then calculated to serve as the aggregated updates.
Its time complexity primarily arises from the norm calculation, which is $O(dn)$.
    
\mypara{CRFL}~\cite{crfl_icml2021} highlights the absence of robustness certification in existing backdoor defenses and introduces the first certifiably robust federated learning approach against backdoor attacks.
It uses a Markov kernel to quantify model closeness in each iteration and combines this with parameter smoothing to certify inference predictions, ensuring the global model is robust against backdoor attacks within a certified bound.
Training involves parameter clipping and perturbation, while inference uses randomized smoothing. The main complexity arises from norm clipping, $O(dn)$.

\mypara{DeepSight (D.S.)}~\cite{deepsight_ndss2022} addresses the issue of benign models being removed by defenses treating neural networks as black boxes, compromising convergence.
It filters adversarial updates by analyzing training data distribution and measuring fine-grained differences in neural network structures and outputs. 
DeepSight clusters similar updates using metrics: Division Differences (DDif), Normalized Update Energy (NEUP), and cosine distances.
DDif captures discrepancies in predicted scores, NEUP measures gradient magnitude, and cosine distance assesses bias update distances.
Suspicious models are flagged based on NEUP threshold exceedances, comparing them to median values, with benign models typically showing more exceedances.
Malicious models are filtered based on cluster-level suspicious counts.
Its complexity is dominated by pairwise cosine distance calculations, $O(dn^2)$.

\mypara{FLAME}~\cite{flame_usenix_security2022} offers an improved solution compared to existing differential privacy-based backdoor defenses by estimating the minimum amount of noise required to eliminate backdoors from the model.
It first clusters the model updates based on their cosine distance to identify benign updates with small angular deviations.
Then, it clips these benign updates using the median of L2 norms of the model updates, and finally, obtains the aggregated model.
With the above two methods, it adds Gaussian noise $N(0, \sigma^2)$ with adaptive standard deviation, $\sigma = \lambda * \text{median}(w_i), i \in (1, n)$, where $ \lambda=\frac{1}{\varepsilon} \cdot \sqrt{2 \ln \frac{1.25}{\delta}}$ is the factor of differential privacy.
The complexity of FLAME is dominated by the cosine clustering of HDBSCAN, $O(dn^2)$.

\subsection{Summary of Defenses}

For defenses against model poisoning attacks, early methods employ robust statistical metrics, such as Krum, M.K., T.M., and Median, for robust aggregation.
They benefit from the independence from specific pattern analysis or setting constraints, thus improving the robustness and generalization ability against noise and outliers. 
However, they fail to capture anomalies in update parameters when confronted with complex attacks.
Another line of defense is outlier detection, which relies on certain patterns in parameter updating.
For instance, sign statistics (SignGuard), cosine similarity (FLTrust), and cross-round historical momentum information (CenteredClipping) are all efficient features in defense.
However, outlier detection may weaken under minor perturbations or unsupported assumptions, limiting generalization.
Early data poisoning defenses, such as Auror and F.G., rely on pattern-based clustering with indicative update parameters.
However, these approaches often struggle with poor pattern generalization and are limited to either IID or non-IID scenarios.
Recent methods improve performance by integrating robust statistical metrics with pattern-based clustering.
For example, DeepSight employs median-based thresholding and pattern clustering, while FLAME combines cosine similarity clustering with median-of-norm clipping.
Nonetheless, excessive complexity in analysis and parameter configuration can hinder generalization and robustness.

\myta{
We identify key defense methods against poisoning attacks, including L2 Distance-Based (e.g., Krum, SignGuard), Median-Based (e.g., Median, FLAME), Coordinate-Wise (e.g., RFA, Neurotoxin), Sign Distribution Analysis (e.g., SignGuard), Cosine Similarity (e.g., FLTrust, FLAME), and PCA (e.g., DnC). These methods vary in robustness.
\textbf{Besides, current defenses primarily address label and quantity skew, with feature and quality skew underexplored}~\cite{heterogeneous_csur2023}. Further research is encouraged to close these gaps and improve real-world protection.}

\section{Evaluation}
\label{sec:experiments}

In this section, we first outline the experimental setup.
Next, we systematically evaluate the performance of various attacks and defenses across 2,040 unique settings (15 attacks $\times$ 17 defenses $\times$ 2 datasets $\times$ 2 algorithms $\times$ 2 levels of data heterogeneity).
Specifically, we first analyze their overall effectiveness, identify the most effective attack and defense strategies, and the factors contributing to their success or limitations, and highlight the distinctions between defenses against MPAs and DPAs.
We then compare and analyze these strategies in terms of FL algorithms and data heterogeneity separately, highlighting their adaptability.
Furthermore, we also present the time overhead for both attacks and defenses and analyze their underlying causes in \Cref{sec:timeoverhead}.
Finally, an ablation study on the impact of adversary ratio and the most effective strategies is provided in \Cref{sec:ablation}.
We provide actionable insights as takeaway messages to enhance defenses and guide future research.

\subsection{Experimental Setup}

\mypara{Federated Learning System.}
We simulate a FL system with a central server and fully participating clients.
For FL algorithms, we observe that FedSGD and FedOpt are widely utilized in poisoning attacks, as highlighted in prior studies ~\cite{dba_iclr2019, neurotoxin_icml2022, deepsight_ndss2022, altermin_icml2019, modelreplacement_aistats2020, flame_usenix_security2022, crfl_icml2021, normclipping_nips2019, deepsight_ndss2022}.
As shown in \Cref{tab:attack} and \Cref{tab:defense}, most model poisoning attacks and defenses are developed based on the FedSGD algorithm~\cite{fedsgdavg_AISTATS2017}, while the majority of data poisoning attacks and defenses are built upon the FedOpt algorithm~\cite{fedopt_iclr2021}.
Notably, while some data poisoning studies claim to use FedAvg, a closer analysis of their algorithm descriptions or code implementations reveals they actually employ FedOpt~\cite{fedopt_iclr2021} with its server learning rate set to 1.
Therefore, both of them are evaluated for a comprehensive and unified evaluation.
The general settings, including the number of epochs, adversary ratio, batch size, optimizer, learning rate, configurations of attacks and defenses, etc., are detailed in \Cref{tab:experiment_setting} and \Cref{sec:config_attdef}.

\mypara{Models and Datasets.}
This paper trains LeNet-5~\cite{MNISTLeNet5_Proc_IEEE1998} on MNIST~\cite{MNISTLeNet5_Proc_IEEE1998} and ResNet-18~\cite{resnet18_he2015} on CIFAR-10~\cite{cifar10_2009} dataset.
MNIST includes 60,000 training and 10,000 testing grayscale images (28x28 pixels) of handwritten digits across 10 classes.
CIFAR-10 consists of 50,000 training and 10,000 testing color images (32x32 pixels) across 10 classes.
These models and datasets are widely used in FL evaluations~\cite{alie_nips2019, centeredclipping_icml2021, IPM_PMLR2020, fltrust_ndss2020, signguard_ICDCS22}.

\mypara{Data Heterogeneity.}
For data heterogeneity, we consider both IID and non-IID scenarios for evaluation.
Specifically, IID data is quantity- and class-balanced.
For non-IID, we observe two main partition strategies: FangAttack's partition~\cite{fangattack_usenix_sec2020}, followed by~\cite{fltrust_ndss2020, fldetector_kdd22, min_ndss2021}, and the Dirichlet distribution-based partition, pioneered by~\cite{dirichletfl_google, bayesiannn_icml2019} and followed by~\cite{modelreplacement_aistats2020, flame_usenix_security2022, dba_iclr2019, 3dfed_sp2023}.
We choose the Dirichlet distribution-based strategy for its widespread use in both poisoning attacks and federated learning field~\cite{dirichletfl_google,bayesiannn_icml2019,FedMA_iclr2020,modelreplacement_aistats2020, flame_usenix_security2022, dba_iclr2019, 3dfed_sp2023,blade_tbd2023,productionflpoison_sp2022}.
Specifically, the local data is sampled using label ratios from a Dirichlet distribution $Dir(\alpha)$. $\alpha$ (0.5 in our evaluation) controls data heterogeneity, with smaller $\alpha$ indicating greater class imbalance.
\Cref{pic:heterogeneity} visualizes our balanced IID and non-IID partition results on 20 clients for labels on the CIFAR-10 dataset.

\mypara{Evaluation Metrics.}
Two commonly used metrics are used to assess the effectiveness of attacks and defenses, (1) Accuracy (ACC) indicates the ratio of the number of correctly predicted labels to the total number of test samples (2) Attack Success Rate (ASR) indicates the ratio of correctly misclassified targets to the total number of poisoned test samples.

\begin{table*}[!ht]
\centering
\caption{\textbf{Evaluation of Model Poisoning Attacks:} Average performance of model poisoning attacks over different defense strategies.
The results are reported using accuracy and smaller values with red colors indicate better attack performance.}
\label{tab:mpa}
\setlength{\tabcolsep}{5pt} 
\renewcommand{\arraystretch}{0.8} 
{
\begin{tabular}{cc|ccccccccc}\toprule
  &  & \multicolumn{9}{c}{\textbf{Average MPA Performance Over Defenses (\%) -- MNIST}} \\ \cmidrule{3-11} 
\multirow{-2}{*}{Alg} & \multirow{-2}{*}{IID} & N.A. & Gaussian & S.F. & ALIE & IPM & F.A. & MinMax & MinSum & Mimic \\ \midrule
 & \cmark & \cellcolor[HTML]{D4E7F1}96.9 & \cellcolor[HTML]{FEF7F3}58.0 & \cellcolor[HTML]{FEF2EB}51.5 & \cellcolor[HTML]{DFEDF5}88.7 & \cellcolor[HTML]{FEFFFF}67.3 & \cellcolor[HTML]{FEFEFF}67.6 & \cellcolor[HTML]{E2EFF6}87.2 & \cellcolor[HTML]{E1EEF5}87.9 & \cellcolor[HTML]{D5E8F2}95.9 \\
\multirow{-2}{*}{FedSGD} & \xmark & \cellcolor[HTML]{D8E9F3}93.7 & \cellcolor[HTML]{FFFFFF}66.2 & \cellcolor[HTML]{FDEADE}42.3 & \cellcolor[HTML]{FDE1D0}31.6 & \cellcolor[HTML]{FDEADE}42.1 & \cellcolor[HTML]{FEEDE3}45.9 & \cellcolor[HTML]{FDDDCB}27.6 & \cellcolor[HTML]{FDDCC8}26.0 & \cellcolor[HTML]{DCEBF4}91.2 \\ \midrule
 & \cmark & \cellcolor[HTML]{D1E5F0}98.5 & \cellcolor[HTML]{FEFDFD}65.0 & \cellcolor[HTML]{EAF3F8}81.4 & \cellcolor[HTML]{FEF3ED}53.1 & \cellcolor[HTML]{F8FBFD}71.2 & \cellcolor[HTML]{F2F8FB}76.0 & \cellcolor[HTML]{DCEBF4}91.4 & \cellcolor[HTML]{DCECF4}91.0 & \cellcolor[HTML]{D4E7F1}96.9 \\
\multirow{-2}{*}{FedOpt} & \xmark & \cellcolor[HTML]{DCEBF4}91.4 & \cellcolor[HTML]{F0F7FB}76.9 & \cellcolor[HTML]{FEF6F2}56.6 & \cellcolor[HTML]{FEF6F2}56.8 & \cellcolor[HTML]{FDECE2}45.0 & \cellcolor[HTML]{FEF4EE}54.3 & \cellcolor[HTML]{FDE4D5}35.2 & \cellcolor[HTML]{FDDBC7}24.5 & \cellcolor[HTML]{DCECF4}90.8 \\ \midrule
\multicolumn{2}{c|}{AVG} & \cellcolor[HTML]{D6E8F2}95.1 & \cellcolor[HTML]{FFFFFF}66.5 & \cellcolor[HTML]{FEF7F3}57.9 & \cellcolor[HTML]{FEF7F3}57.6 & \cellcolor[HTML]{FEF6F1}56.4 & \cellcolor[HTML]{FEFAF7}60.9 & \cellcolor[HTML]{FEF9F7}60.3 & \cellcolor[HTML]{FEF7F3}57.3 & \cellcolor[HTML]{D8E9F3}93.7 \\ \midrule
\multicolumn{2}{c|}{} & \multicolumn{9}{c}{\textbf{Average MPA Performance Over Defenses (\%) -- CIFAR-10}} \\ \midrule
 & \cmark & \cellcolor[HTML]{D3E6F1}74.7 & \cellcolor[HTML]{FEF1EA}52.1 & \cellcolor[HTML]{FEF7F3}57.8 & \cellcolor[HTML]{D4E7F1}74.4 & \cellcolor[HTML]{ECF5F9}69.5 & \cellcolor[HTML]{FEF2EC}53.3 & \cellcolor[HTML]{F8FBFD}67.1 & \cellcolor[HTML]{F9FCFD}66.8 & \cellcolor[HTML]{F6FAFC}67.5 \\
\multirow{-2}{*}{FedSGD} & \xmark & \cellcolor[HTML]{FEFCFB}63.0 & \cellcolor[HTML]{FDE8DC}42.8 & \cellcolor[HTML]{FDEADF}44.9 & \cellcolor[HTML]{FEF7F3}57.7 & \cellcolor[HTML]{FEF4EF}55.3 & \cellcolor[HTML]{FDDBC7}29.1 & \cellcolor[HTML]{FEEFE7}49.9 & \cellcolor[HTML]{FEF0E8}50.6 & \cellcolor[HTML]{FEF6F1}56.5 \\ \midrule
 & \cmark & \cellcolor[HTML]{D2E6F1}74.9 & \cellcolor[HTML]{FEF3ED}53.8 & \cellcolor[HTML]{FEF6F2}57.4 & \cellcolor[HTML]{D1E5F0}74.9 & \cellcolor[HTML]{E3EFF6}71.3 & \cellcolor[HTML]{FEFEFE}64.9 & \cellcolor[HTML]{D7E8F2}73.9 & \cellcolor[HTML]{D6E8F2}73.9 & \cellcolor[HTML]{F2F8FB}68.2 \\
\multirow{-2}{*}{FedOpt} & \xmark & \cellcolor[HTML]{DFEDF5}72.2 & \cellcolor[HTML]{FEF2EA}52.4 & \cellcolor[HTML]{FEF6F1}56.8 & \cellcolor[HTML]{D9EAF3}73.3 & \cellcolor[HTML]{FEFEFF}65.9 & \cellcolor[HTML]{FEF4EE}54.8 & \cellcolor[HTML]{E1EEF6}71.7 & \cellcolor[HTML]{E3EFF6}71.3 & \cellcolor[HTML]{EBF4F9}69.7 \\ \midrule
\multicolumn{2}{c|}{AVG} & \cellcolor[HTML]{E4F0F6}71.2 & \cellcolor[HTML]{FEEFE7}50.3 & \cellcolor[HTML]{FEF3ED}54.2 & \cellcolor[HTML]{E9F3F8}70.1 & \cellcolor[HTML]{FFFFFF}65.5 & \cellcolor[HTML]{FEF0E7}50.5 & \cellcolor[HTML]{FFFFFF}65.7 & \cellcolor[HTML]{FEFFFF}65.7 & \cellcolor[HTML]{FEFEFE}65.5 \\ \bottomrule\end{tabular}
}\end{table*}

\begin{table*}[!ht]
\centering
\caption{\textbf{Evaluation of Data Poisoning Attacks:} Average performance of data poisoning attacks over different defense strategies.
The results are reported using ACC and ASR.
For both metrics, higher values with blue colors indicate better attack performance.}
\label{tab:dpa}
\setlength{\tabcolsep}{6pt} 
\renewcommand{\arraystretch}{0.9} 
\resizebox{\textwidth}{!}{\begin{tabular}{cccccccccccccccc}\toprule
\multicolumn{1}{l}{} & \multicolumn{1}{l}{} & \multicolumn{14}{c}{\textbf{Average DPA Performance Over Defenses -- MNIST}} \\ \cmidrule{3-16} 
\multicolumn{1}{l}{} & \multicolumn{1}{l}{} & \multicolumn{2}{c|}{LabelFlipping} & \multicolumn{2}{c|}{BadNets} & \multicolumn{2}{c|}{DBA} & \multicolumn{2}{c|}{AlterMin} & \multicolumn{2}{c|}{M.P.} & \multicolumn{2}{c|}{EdgeCase} & \multicolumn{2}{c}{Neurotoxin} \\
\multicolumn{1}{c}{\multirow{-3}{*}{\textbf{Alg}}} & \multicolumn{1}{c}{\multirow{-3}{*}{\textbf{IID}}} & Acc & \multicolumn{1}{c|}{ASR} & Acc & \multicolumn{1}{c|}{ASR} & Acc & \multicolumn{1}{c|}{ASR} & Acc & \multicolumn{1}{c|}{ASR} & Acc & \multicolumn{1}{c|}{ASR} & Acc & \multicolumn{1}{c|}{ASR} & Acc & ASR \\ \midrule
 & \multicolumn{1}{c|}{\cmark} & \cellcolor[HTML]{E2EFF6}94.4 & \multicolumn{1}{c|}{\cellcolor[HTML]{FDE1D0}21.1} & \cellcolor[HTML]{D5E8F2}96.7 & \multicolumn{1}{c|}{\cellcolor[HTML]{FEFEFE}50.1} & \cellcolor[HTML]{D6E8F2}96.5 & \multicolumn{1}{c|}{\cellcolor[HTML]{FEF9F6}45.1} & \cellcolor[HTML]{FEFBFA}85.8 & \multicolumn{1}{c|}{\cellcolor[HTML]{FEF8F5}44.2} & \cellcolor[HTML]{FEFDFC}87.2 & \multicolumn{1}{c|}{\cellcolor[HTML]{FAFCFE}55.0} & \cellcolor[HTML]{F9FCFD}90.1 & \multicolumn{1}{c|}{\cellcolor[HTML]{E8F2F8}70.7} & \cellcolor[HTML]{FEF9F6}83.6 & \cellcolor[HTML]{FDDBC7}15.2 \\
\multirow{-2}{*}{FedSGD} & \multicolumn{1}{c|}{\xmark} & \cellcolor[HTML]{F3F9FC}91.1 & \multicolumn{1}{c|}{\cellcolor[HTML]{FEF3ED}39.2} & \cellcolor[HTML]{E3EFF6}94.1 & \multicolumn{1}{c|}{\cellcolor[HTML]{D2E6F1}89.8} & \cellcolor[HTML]{FEFEFF}89.2 & \multicolumn{1}{c|}{\cellcolor[HTML]{FDDECC}18.4} & \cellcolor[HTML]{F6FAFC}90.6 & \multicolumn{1}{c|}{\cellcolor[HTML]{FEF7F3}43.1} & \cellcolor[HTML]{FEF3ED}77.9 & \multicolumn{1}{c|}{\cellcolor[HTML]{F9FCFD}55.9} & \cellcolor[HTML]{FEF5EF}79.1 & \multicolumn{1}{c|}{\cellcolor[HTML]{FDECE1}31.9} & \cellcolor[HTML]{FDE7DA}65.7 & \cellcolor[HTML]{FEFEFD}49.4 \\ \midrule
 & \multicolumn{1}{c|}{\cmark} & \cellcolor[HTML]{F2F8FB}91.4 & \multicolumn{1}{c|}{\cellcolor[HTML]{FDE5D7}25.5} & \cellcolor[HTML]{E8F2F8}93.2 & \multicolumn{1}{c|}{\cellcolor[HTML]{F1F7FB}63.0} & \cellcolor[HTML]{D4E7F1}96.9 & \multicolumn{1}{c|}{\cellcolor[HTML]{F5FAFC}59.1} & \cellcolor[HTML]{EEF6FA}92.1 & \multicolumn{1}{c|}{\cellcolor[HTML]{FEF8F4}43.6} & \cellcolor[HTML]{FDE3D3}61.4 & \multicolumn{1}{c|}{\cellcolor[HTML]{F6FAFC}58.4} & \cellcolor[HTML]{FEFDFD}87.7 & \multicolumn{1}{c|}{\cellcolor[HTML]{F7FBFD}57.5} & \cellcolor[HTML]{D1E5F0}97.4 & \cellcolor[HTML]{E3F0F6}74.6 \\
\multirow{-2}{*}{FedOpt} & \multicolumn{1}{c|}{\xmark} & \cellcolor[HTML]{FEFFFF}89.2 & \multicolumn{1}{c|}{\cellcolor[HTML]{FDE0CF}20.6} & \cellcolor[HTML]{EBF4F9}92.7 & \multicolumn{1}{c|}{\cellcolor[HTML]{D1E5F0}90.2} & \cellcolor[HTML]{F6FAFC}90.7 & \multicolumn{1}{c|}{\cellcolor[HTML]{FEF0E7}35.7} & \cellcolor[HTML]{FEFBFA}85.8 & \multicolumn{1}{c|}{\cellcolor[HTML]{FEF5F0}41.3} & \cellcolor[HTML]{FDDBC7}53.2 & \multicolumn{1}{c|}{\cellcolor[HTML]{EFF6FA}64.5} & \cellcolor[HTML]{FEF7F2}81.0 & \multicolumn{1}{c|}{\cellcolor[HTML]{DEEDF5}78.9} & \cellcolor[HTML]{FEFAF8}84.8 & \cellcolor[HTML]{E9F3F8}69.7 \\ \midrule
\multicolumn{2}{c|}{AVG} & \cellcolor[HTML]{F1F7FB}91.5 & \multicolumn{1}{c|}{\cellcolor[HTML]{FDE6D9}26.6} & \cellcolor[HTML]{E3EFF6}94.2 & \multicolumn{1}{c|}{\cellcolor[HTML]{E5F1F7}73.3} & \cellcolor[HTML]{E7F2F8}93.3 & \multicolumn{1}{c|}{\cellcolor[HTML]{FEF4ED}39.6} & \cellcolor[HTML]{FEFEFE}88.6 & \multicolumn{1}{c|}{\cellcolor[HTML]{FEF7F3}43.0} & \cellcolor[HTML]{FDEBE1}69.9 & \multicolumn{1}{c|}{\cellcolor[HTML]{F6FAFC}58.4} & \cellcolor[HTML]{FEFAF8}84.5 & \multicolumn{1}{c|}{\cellcolor[HTML]{F5F9FC}59.7} & \cellcolor[HTML]{FEF8F5}82.9 & \cellcolor[HTML]{FDFEFF}52.2 \\ \midrule
\multicolumn{2}{l|}{} & \multicolumn{14}{c}{\textbf{Average DPA Performance Over Defenses -- CIFAR-10}} \\ \midrule
 & \multicolumn{1}{c|}{\cmark} & \cellcolor[HTML]{DCEAF1}72.8 & \multicolumn{1}{c|}{\cellcolor[HTML]{FDE5D7}12.2} & \cellcolor[HTML]{D5E7F1}73.9 & \multicolumn{1}{c|}{\cellcolor[HTML]{DAEAF3}71.5} & \cellcolor[HTML]{D8E8F1}73.4 & \multicolumn{1}{c|}{\cellcolor[HTML]{E6F1F7}62.0} & \cellcolor[HTML]{FCDECC}49.4 & \multicolumn{1}{c|}{\cellcolor[HTML]{FDDBC7}0.1} & \cellcolor[HTML]{E5EDF2}71.3 & \multicolumn{1}{c|}{\cellcolor[HTML]{EAF4F9}58.0} & \cellcolor[HTML]{F3F0EE}67.4 & \multicolumn{1}{c|}{\cellcolor[HTML]{F9FCFE}45.5} & \cellcolor[HTML]{E6EEF2}71.2 & \cellcolor[HTML]{FEF2EB}26.8 \\
\multirow{-2}{*}{FedSGD} & \multicolumn{1}{c|}{\xmark} & \cellcolor[HTML]{F6EAE4}62.0 & \multicolumn{1}{c|}{\cellcolor[HTML]{FDE1D1}7.4} & \cellcolor[HTML]{F8E7DE}58.7 & \multicolumn{1}{c|}{\cellcolor[HTML]{D8E9F3}73.6} & \cellcolor[HTML]{F8E6DC}57.9 & \multicolumn{1}{c|}{\cellcolor[HTML]{EFF6FA}54.4} & \cellcolor[HTML]{FDDBC8}46.9 & \multicolumn{1}{c|}{\cellcolor[HTML]{FDEADF}17.9} & \cellcolor[HTML]{FDDBC7}46.2 & \multicolumn{1}{c|}{\cellcolor[HTML]{EAF4F9}58.0} & \cellcolor[HTML]{FAE3D6}54.3 & \multicolumn{1}{c|}{\cellcolor[HTML]{F7FAFD}47.9} & \cellcolor[HTML]{F6EBE5}62.5 & \cellcolor[HTML]{FDECE1}19.4 \\ \cmidrule{2-16} 
 & \multicolumn{1}{c|}{\cmark} & \cellcolor[HTML]{D5E7F1}73.8 & \multicolumn{1}{c|}{\cellcolor[HTML]{FDE5D7}12.0} & \cellcolor[HTML]{D1E5F0}74.4 & \multicolumn{1}{c|}{\cellcolor[HTML]{D9EAF3}72.7} & \cellcolor[HTML]{D2E6F1}74.4 & \multicolumn{1}{c|}{\cellcolor[HTML]{E2EFF6}65.3} & \cellcolor[HTML]{F3F1F1}69.3 & \multicolumn{1}{c|}{\cellcolor[HTML]{FDDBC7}0.7} & \cellcolor[HTML]{F3F1F0}68.4 & \multicolumn{1}{c|}{\cellcolor[HTML]{F4F9FC}50.0} & \cellcolor[HTML]{DDEAF1}72.6 & \multicolumn{1}{c|}{\cellcolor[HTML]{FEF9F5}33.9} & \cellcolor[HTML]{DDEAF1}72.6 & \cellcolor[HTML]{FDEADE}17.2 \\
\multirow{-2}{*}{FedOpt} & \multicolumn{1}{c|}{\xmark} & \cellcolor[HTML]{E2ECF2}71.8 & \multicolumn{1}{c|}{\cellcolor[HTML]{FDDFCD}4.7} & \cellcolor[HTML]{E1ECF1}72.0 & \multicolumn{1}{c|}{\cellcolor[HTML]{D1E5F0}78.8} & \cellcolor[HTML]{E4EDF2}71.6 & \multicolumn{1}{c|}{\cellcolor[HTML]{DAEBF3}71.4} & \cellcolor[HTML]{F3EFEE}67.3 & \multicolumn{1}{c|}{\cellcolor[HTML]{FDDCC9}1.7} & \cellcolor[HTML]{F3F0EE}67.7 & \multicolumn{1}{c|}{\cellcolor[HTML]{E4F0F7}63.3} & \cellcolor[HTML]{EAEFF2}70.6 & \multicolumn{1}{c|}{\cellcolor[HTML]{FEF9F6}34.7} & \cellcolor[HTML]{EFF1F2}69.9 & \cellcolor[HTML]{FEFDFC}38.5 \\ \midrule
\multicolumn{2}{c|}{AVG} & \cellcolor[HTML]{EDF0F2}70.1 & \multicolumn{1}{c|}{\cellcolor[HTML]{FDE2D3}9.1} & \cellcolor[HTML]{F0F1F2}69.7 & \multicolumn{1}{c|}{\cellcolor[HTML]{D7E9F2}74.2} & \cellcolor[HTML]{F2F2F2}69.3 & \multicolumn{1}{c|}{\cellcolor[HTML]{E4F0F7}63.3} & \cellcolor[HTML]{F8E6DD}58.2 & \multicolumn{1}{c|}{\cellcolor[HTML]{FDDFCD}5.1} & \cellcolor[HTML]{F5ECE7}63.4 & \multicolumn{1}{c|}{\cellcolor[HTML]{EBF4F9}57.3} & \cellcolor[HTML]{F4EEEC}66.2 & \multicolumn{1}{c|}{\cellcolor[HTML]{FFFFFF}40.5} & \cellcolor[HTML]{F3F1F1}69.0 & \cellcolor[HTML]{FEF1EA}25.5 \\ \bottomrule\end{tabular}}\end{table*}

\begin{table*}[!ht]
\centering
\caption{\textbf{Evaluation of Defense Strategies Over Model Poisoning Attacks:} Average defense performance against model poisoning attacks.
The results are reported using accuracy and larger values with blue colors indicate better defense performance.
Note that those who receive ``-'' encounter running errors due to non-adaptive hyper-parameters.}

\setlength{\tabcolsep}{3pt} 
\renewcommand{\arraystretch}{1} 
\label{tab:mpa_defense}
\resizebox{\textwidth}{!}{
\begin{tabular}{cccccccccccccccccccc}\toprule
 &  & \multicolumn{18}{c}{\textbf{Average Defense Performance Over MPA -- MNIST}} \\ \cmidrule{3-20} 
\multirow{-2}{*}{Alg} & \multirow{-2}{*}{IID} & Mean & Krum & M.K. & T.M. & Median & Bulyan & RFA & FLTrust & C.C. & DnC & Bucketing & S.G. & Auror & F.G. & N.C. & CRFL & D.S. & FLAME \\ \midrule
 & \multicolumn{1}{c|}{\cmark} & \cellcolor[HTML]{FEFFFF}71.2 & \cellcolor[HTML]{FFFFFF}70.4 & \cellcolor[HTML]{FFFFFF}70.3 & \cellcolor[HTML]{FCFDFE}72.6 & \cellcolor[HTML]{DBEBF4}91.9 & \cellcolor[HTML]{FEFEFE}69.8 & \cellcolor[HTML]{FEFBF9}66.2 & \cellcolor[HTML]{D5E8F2}95.4 & \cellcolor[HTML]{FBFDFE}73.2 & \cellcolor[HTML]{D6E8F2}95.1 & \cellcolor[HTML]{FEF0E7}52.1 & \cellcolor[HTML]{FCFEFE}72.2 & \cellcolor[HTML]{E6F1F7}85.6 & \cellcolor[HTML]{E7F1F7}85.1 & \cellcolor[HTML]{EBF4F9}82.6 & \cellcolor[HTML]{F9FCFE}73.9 & \cellcolor[HTML]{EFF6FA}80.1 & \cellcolor[HTML]{D7E8F2}94.4 \\
\multirow{-2}{*}{FedSGD} & \multicolumn{1}{c|}{\xmark} & \cellcolor[HTML]{FDEBE1}47.0 & \cellcolor[HTML]{FDEADF}45.5 & \cellcolor[HTML]{FDEBDF}46.0 & \cellcolor[HTML]{FDE5D8}39.8 & \cellcolor[HTML]{FEF0E8}52.3 & \cellcolor[HTML]{FDE9DE}44.6 & \cellcolor[HTML]{FDEADF}45.3 & \cellcolor[HTML]{E1EEF6}88.3 & \cellcolor[HTML]{FDEBE1}46.8 & \cellcolor[HTML]{FEFBF9}66.1 & \cellcolor[HTML]{FDDBC7}26.4 & \cellcolor[HTML]{FEF2EB}54.7 & \cellcolor[HTML]{FEEFE6}50.8 & \cellcolor[HTML]{EDF5FA}81.1 & \cellcolor[HTML]{FEF8F4}61.8 & \cellcolor[HTML]{FEEFE6}51.3 & \cellcolor[HTML]{FDE2D2}35.8 & \cellcolor[HTML]{FEF2EB}55.0 \\ \midrule
 & \multicolumn{1}{c|}{\cmark} & \cellcolor[HTML]{F7FBFD}75.1 & \cellcolor[HTML]{FDFEFF}72.0 & \cellcolor[HTML]{FEFBF8}65.5 & \cellcolor[HTML]{F7FBFD}75.1 & \cellcolor[HTML]{E5F1F7}85.9 & \cellcolor[HTML]{FEFAF8}65.3 & \cellcolor[HTML]{ECF5F9}81.8 & \cellcolor[HTML]{D9EAF3}93.3 & \cellcolor[HTML]{E7F2F8}84.7 & \cellcolor[HTML]{D1E5F0}97.6 & \cellcolor[HTML]{FDEADF}45.9 & \cellcolor[HTML]{D5E7F2}95.5 & \cellcolor[HTML]{E1EEF6}88.2 & \cellcolor[HTML]{FAFDFE}73.4 & \cellcolor[HTML]{D7E8F2}94.5 & \cellcolor[HTML]{E4F0F6}86.8 & \cellcolor[HTML]{ECF4F9}82.0 & \cellcolor[HTML]{E5F1F7}85.9 \\
\multirow{-2}{*}{FedOpt} & \multicolumn{1}{c|}{\xmark} & \cellcolor[HTML]{FEF3ED}56.5 & \cellcolor[HTML]{FDEADF}45.8 & \cellcolor[HTML]{FDEBE1}47.0 & \cellcolor[HTML]{FEEDE3}48.8 & \cellcolor[HTML]{FEF6F1}59.7 & \cellcolor[HTML]{FDEADF}45.9 & \cellcolor[HTML]{FDEADE}44.9 & \cellcolor[HTML]{DDECF4}90.8 & \cellcolor[HTML]{FEF8F4}62.4 & \cellcolor[HTML]{F8FBFD}74.5 & \cellcolor[HTML]{FDDDCA}29.5 & \cellcolor[HTML]{F9FCFD}74.2 & \cellcolor[HTML]{FFFFFF}70.5 & \cellcolor[HTML]{FDE7DA}41.5 & \cellcolor[HTML]{FCFDFE}72.6 & \cellcolor[HTML]{FEF9F6}63.5 & \cellcolor[HTML]{FEFCFB}67.2 & \cellcolor[HTML]{FEFFFF}70.9 \\ \midrule
\multicolumn{2}{c|}{Acc AVG} & \cellcolor[HTML]{FEF8F4}62.4 & \cellcolor[HTML]{FEF5EF}58.4 & \cellcolor[HTML]{FEF4EE}57.2 & \cellcolor[HTML]{FEF5F0}59.1 & \cellcolor[HTML]{FCFDFE}72.5 & \cellcolor[HTML]{FEF3ED}56.4 & \cellcolor[HTML]{FEF6F1}59.5 & \cellcolor[HTML]{DBEBF4}92.0 & \cellcolor[HTML]{FEFCFA}66.8 & \cellcolor[HTML]{EAF3F8}83.3 & \cellcolor[HTML]{FDE4D6}38.5 & \cellcolor[HTML]{F9FCFD}74.2 & \cellcolor[HTML]{FAFCFE}73.8 & \cellcolor[HTML]{FEFEFE}70.3 & \cellcolor[HTML]{F3F8FB}77.9 & \cellcolor[HTML]{FEFDFD}68.9 & \cellcolor[HTML]{FEFBF9}66.3 & \cellcolor[HTML]{F5FAFC}76.6 \\ \midrule
\multicolumn{2}{c|}{} & \multicolumn{18}{c}{\textbf{Average Defense Performance Over MPA -- CIFAR10}} \\ \midrule
 & \multicolumn{1}{c|}{\cmark} & \cellcolor[HTML]{FEFBF9}57.4 & \cellcolor[HTML]{FAFCFE}65.8 & \cellcolor[HTML]{F5FAFC}67.1 & \cellcolor[HTML]{FEFEFD}62.6 & \cellcolor[HTML]{E5F0F7}71.4 & \cellcolor[HTML]{FFFFFF}64.5 & \cellcolor[HTML]{F7FAFD}66.7 & \cellcolor[HTML]{D6E8F2}75.4 & \cellcolor[HTML]{FEFDFD}62.0 & \cellcolor[HTML]{D6E8F2}75.2 & \cellcolor[HTML]{FEF2EA}41.1 & \cellcolor[HTML]{E3EFF6}71.8 & \cellcolor[HTML]{FAFCFE}65.9 & \cellcolor[HTML]{FFFFFF}- & \cellcolor[HTML]{FAFCFE}65.8 & \cellcolor[HTML]{FEF5F0}47.9 & \cellcolor[HTML]{FEFEFE}64.0 & \cellcolor[HTML]{D9E9F3}74.6 \\
\multirow{-2}{*}{FedSGD} & \multicolumn{1}{c|}{\xmark} & \cellcolor[HTML]{FEF7F3}51.1 & \cellcolor[HTML]{FEF3ED}43.9 & \cellcolor[HTML]{FEF4EF}46.0 & \cellcolor[HTML]{FEFAF7}56.1 & \cellcolor[HTML]{FEFDFC}61.0 & \cellcolor[HTML]{FEF3EC}43.3 & \cellcolor[HTML]{FEFEFE}63.4 & \cellcolor[HTML]{EDF5F9}69.3 & \cellcolor[HTML]{FEF9F6}54.3 & \cellcolor[HTML]{FEF8F4}52.3 & \cellcolor[HTML]{FDE8DC}24.2 & \cellcolor[HTML]{FEF3ED}44.5 & \cellcolor[HTML]{FDEBE0}28.7 & \cellcolor[HTML]{FFFFFF}- & \cellcolor[HTML]{FEFEFE}63.5 & \cellcolor[HTML]{FEF6F1}49.3 & \cellcolor[HTML]{FDE8DB}23.6 & \cellcolor[HTML]{FEF7F4}51.7 \\ \midrule
 & \multicolumn{1}{c|}{\cmark} & \cellcolor[HTML]{FEFEFE}63.5 & \cellcolor[HTML]{DCECF4}73.7 & \cellcolor[HTML]{DCECF4}73.6 & \cellcolor[HTML]{FEFEFE}64.1 & \cellcolor[HTML]{E4F0F6}71.7 & \cellcolor[HTML]{E0EEF5}72.7 & \cellcolor[HTML]{F3F8FB}67.6 & \cellcolor[HTML]{E0EEF5}72.7 & \cellcolor[HTML]{F4F9FC}67.3 & \cellcolor[HTML]{D1E5F0}76.5 & \cellcolor[HTML]{FEF9F6}54.3 & \cellcolor[HTML]{DDECF4}73.4 & \cellcolor[HTML]{F3F9FC}67.5 & \cellcolor[HTML]{FFFFFF}- & \cellcolor[HTML]{FDFEFF}64.9 & \cellcolor[HTML]{FEF6F2}49.9 & \cellcolor[HTML]{E6F1F7}71.2 & \cellcolor[HTML]{D6E8F2}75.3 \\
\multirow{-2}{*}{FedOpt} & \multicolumn{1}{c|}{\xmark} & \cellcolor[HTML]{FEFAF8}56.8 & \cellcolor[HTML]{FEFBF9}57.5 & \cellcolor[HTML]{E4F0F7}71.5 & \cellcolor[HTML]{FFFFFF}64.3 & \cellcolor[HTML]{E6F1F7}71.0 & \cellcolor[HTML]{E5F0F7}71.4 & \cellcolor[HTML]{EBF4F9}69.6 & \cellcolor[HTML]{E7F2F8}70.8 & \cellcolor[HTML]{FEFEFE}64.1 & \cellcolor[HTML]{D6E8F2}75.4 & \cellcolor[HTML]{FEF4ED}44.7 & \cellcolor[HTML]{E0EDF5}72.8 & \cellcolor[HTML]{FEFDFC}61.2 & \cellcolor[HTML]{FFFFFF}- & \cellcolor[HTML]{F3F9FB}67.6 & \cellcolor[HTML]{FEFBF9}58.5 & \cellcolor[HTML]{FFFFFF}- & \cellcolor[HTML]{F2F8FB}68.0 \\ \midrule
\multicolumn{2}{c|}{Acc AVG} & \cellcolor[HTML]{FEFBF8}57.2 & \cellcolor[HTML]{FEFCFB}60.2 & \cellcolor[HTML]{FFFFFF}64.6 & \cellcolor[HTML]{FEFDFC}61.8 & \cellcolor[HTML]{EFF6FA}68.8 & \cellcolor[HTML]{FEFEFD}63.0 & \cellcolor[HTML]{F6FAFC}66.8 & \cellcolor[HTML]{E2EFF6}72.0 & \cellcolor[HTML]{FEFDFC}61.9 & \cellcolor[HTML]{EBF4F9}69.8 & \cellcolor[HTML]{FEF2EA}41.1 & \cellcolor[HTML]{FAFDFE}65.6 & \cellcolor[HTML]{FEFAF7}55.8 & \cellcolor[HTML]{FFFFFF}- & \cellcolor[HTML]{FBFDFE}65.4 & \cellcolor[HTML]{FEF7F3}51.4 & \cellcolor[HTML]{FEF8F5}52.9 & \cellcolor[HTML]{F4F9FC}67.4\\\bottomrule
\end{tabular}
}\end{table*}

\begin{table*}[!t]
\centering
\caption{\textbf{Evaluation of Defense Strategies Over Data Poisoning Attacks - MNIST:} Average defense performance against data poisoning attacks.
For the upper table, the results are reported using accuracy and larger values with blue colors indicate better defense performance.
For the lower table, the results are reported using ASR and larger values with blue colors indicate better defense performance.}
\label{tab:dpa_defense_mnist}
\setlength{\tabcolsep}{3pt} 
\renewcommand{\arraystretch}{1} 
\resizebox{\textwidth}{!}{\begin{tabular}{cc|cccccccccccccccccc}\toprule
  &  & \multicolumn{18}{c}{\textbf{Average ACC Over DPA (\%) – MNIST}} \\ \cmidrule{3-20} 
\multirow{-2}{*}{Alg} & \multirow{-2}{*}{IID} & Mean & Krum & M.K. & T.M. & Median & Bulyan & RFA & FLTrust & C.C. & DnC & Bucketing & S.G. & Auror & F.G. & N.C. & CRFL & D.S. & FLAME \\ \midrule
 & \cmark & \cellcolor[HTML]{E8F2F8}95.2 & \cellcolor[HTML]{EDF5F9}94.5 & \cellcolor[HTML]{E1EEF5}96.4 & \cellcolor[HTML]{EBF4F9}94.8 & \cellcolor[HTML]{E7F2F7}95.4 & \cellcolor[HTML]{E2EFF6}96.2 & \cellcolor[HTML]{E7F2F8}95.3 & \cellcolor[HTML]{DAEBF3}97.4 & \cellcolor[HTML]{E9F3F8}95.1 & \cellcolor[HTML]{F3F8FB}93.6 & \cellcolor[HTML]{FDE6D8}45.3 & \cellcolor[HTML]{E5F0F7}95.7 & \cellcolor[HTML]{FEFBF9}84.7 & \cellcolor[HTML]{FEF3EC}69.6 & \cellcolor[HTML]{E6F1F7}95.5 & \cellcolor[HTML]{EFF6FA}94.1 & \cellcolor[HTML]{E3EFF6}96.1 & \cellcolor[HTML]{DDECF4}97.0 \\
\multirow{-2}{*}{FedSGD} & \xmark & \cellcolor[HTML]{ECF4F9}94.7 & \cellcolor[HTML]{FEF6F1}75.7 & \cellcolor[HTML]{FEF9F6}81.6 & \cellcolor[HTML]{F5FAFC}93.2 & \cellcolor[HTML]{FEF6F1}74.9 & \cellcolor[HTML]{FEF7F3}77.7 & \cellcolor[HTML]{F4F9FC}93.3 & \cellcolor[HTML]{FEFDFD}89.5 & \cellcolor[HTML]{FEFEFE}91.0 & \cellcolor[HTML]{FEF6F1}75.2 & \cellcolor[HTML]{FDE3D3}39.9 & \cellcolor[HTML]{F4F9FC}93.3 & \cellcolor[HTML]{ECF5F9}94.6 & \cellcolor[HTML]{FEEFE6}62.4 & \cellcolor[HTML]{EAF3F8}95.0 & \cellcolor[HTML]{F5F9FC}93.2 & \cellcolor[HTML]{F1F7FB}93.9 & \cellcolor[HTML]{FEFDFC}88.5 \\
 & \cmark & \cellcolor[HTML]{FEFBF9}85.0 & \cellcolor[HTML]{DEEDF5}96.8 & \cellcolor[HTML]{D6E8F2}98.1 & \cellcolor[HTML]{FEFBF9}84.9 & \cellcolor[HTML]{D6E8F2}98.1 & \cellcolor[HTML]{D9EAF3}97.6 & \cellcolor[HTML]{D7E9F2}97.9 & \cellcolor[HTML]{D7E9F2}97.9 & \cellcolor[HTML]{FEFBF9}85.0 & \cellcolor[HTML]{E6F1F7}95.6 & \cellcolor[HTML]{FDECE2}57.7 & \cellcolor[HTML]{D3E6F1}98.6 & \cellcolor[HTML]{D4E7F1}98.3 & \cellcolor[HTML]{FDE4D5}42.6 & \cellcolor[HTML]{FEFBF9}85.3 & \cellcolor[HTML]{FEFBF9}84.8 & \cellcolor[HTML]{D7E8F2}98.0 & \cellcolor[HTML]{D1E5F0}98.8 \\
\multirow{-2}{*}{FedOpt} & \xmark & \cellcolor[HTML]{FEFAF8}83.6 & \cellcolor[HTML]{FEFCFA}86.7 & \cellcolor[HTML]{EBF4F9}94.8 & \cellcolor[HTML]{FEFAF7}83.1 & \cellcolor[HTML]{FCFDFE}92.1 & \cellcolor[HTML]{FEF7F3}78.4 & \cellcolor[HTML]{F9FCFD}92.5 & \cellcolor[HTML]{FDFEFF}91.9 & \cellcolor[HTML]{FEFAF7}83.2 & \cellcolor[HTML]{FEFAF7}83.1 & \cellcolor[HTML]{FDE8DC}50.5 & \cellcolor[HTML]{DEECF5}96.9 & \cellcolor[HTML]{FEFBF8}84.2 & \cellcolor[HTML]{FDDBC7}24.8 & \cellcolor[HTML]{FEFEFD}90.3 & \cellcolor[HTML]{FEF9F6}81.3 & \cellcolor[HTML]{DFEDF5}96.6 & \cellcolor[HTML]{DEECF5}96.8 \\
\multicolumn{2}{c|}{Acc AVG} & \cellcolor[HTML]{FEFDFD}89.6 & \cellcolor[HTML]{FEFDFC}88.4 & \cellcolor[HTML]{F8FBFD}92.7 & \cellcolor[HTML]{FEFDFC}89.0 & \cellcolor[HTML]{FEFEFD}90.1 & \cellcolor[HTML]{FEFCFB}87.5 & \cellcolor[HTML]{EBF4F9}94.8 & \cellcolor[HTML]{EFF6FA}94.2 & \cellcolor[HTML]{FEFDFC}88.6 & \cellcolor[HTML]{FEFCFB}86.9 & \cellcolor[HTML]{FDE7DA}48.4 & \cellcolor[HTML]{E2EFF6}96.1 & \cellcolor[HTML]{FEFEFE}90.5 & \cellcolor[HTML]{FDE8DC}49.9 & \cellcolor[HTML]{FFFFFF}91.5 & \cellcolor[HTML]{FEFDFC}88.3 & \cellcolor[HTML]{E2EFF6}96.1 & \cellcolor[HTML]{E8F2F8}95.3 \\ \midrule
\multicolumn{2}{c|}{} & \multicolumn{18}{c}{\cellcolor[HTML]{FFFFFF}\textbf{Average ASR Over DPA (\%) – MNIST}} \\ \midrule
 & \cmark & \cellcolor[HTML]{E3EFF6}77.7 & \cellcolor[HTML]{FDDFCD}11.7 & \cellcolor[HTML]{FDDFCD}11.8 & \cellcolor[HTML]{E6F1F7}74.7 & \cellcolor[HTML]{FDE6D8}21.1 & \cellcolor[HTML]{FDDECC}10.6 & \cellcolor[HTML]{FDFEFF}55.5 & \cellcolor[HTML]{FDE5D7}20.1 & \cellcolor[HTML]{E4F0F6}76.9 & \cellcolor[HTML]{FEF3ED}38.9 & \cellcolor[HTML]{D6E8F2}88.1 & \cellcolor[HTML]{FDDDCB}10.0 & \cellcolor[HTML]{FDE0CF}13.4 & \cellcolor[HTML]{FAFCFE}58.4 & \cellcolor[HTML]{E2EFF6}78.0 & \cellcolor[HTML]{DEEDF5}81.5 & \cellcolor[HTML]{FDEADE}26.1 & \cellcolor[HTML]{FDDFCD}11.8 \\
\multirow{-2}{*}{FedSGD} & \xmark & \cellcolor[HTML]{F5F9FC}62.5 & \cellcolor[HTML]{FEF3ED}39.0 & \cellcolor[HTML]{FEF7F4}44.4 & \cellcolor[HTML]{FAFCFE}58.4 & \cellcolor[HTML]{FEF4EE}39.3 & \cellcolor[HTML]{FDE7D9}22.1 & \cellcolor[HTML]{FEF8F4}44.5 & \cellcolor[HTML]{FDE5D7}20.4 & \cellcolor[HTML]{FEFEFE}53.2 & \cellcolor[HTML]{F8FBFD}59.6 & \cellcolor[HTML]{F3F8FB}64.1 & \cellcolor[HTML]{FEFFFF}54.9 & \cellcolor[HTML]{EFF6FA}67.7 & \cellcolor[HTML]{FDE6D9}21.7 & \cellcolor[HTML]{F2F8FB}65.0 & \cellcolor[HTML]{FEFFFF}54.9 & \cellcolor[HTML]{FEF4ED}39.2 & \cellcolor[HTML]{FEEDE3}30.3 \\
 & \cmark & \cellcolor[HTML]{D3E6F1}90.9 & \cellcolor[HTML]{FDE5D7}20.0 & \cellcolor[HTML]{FDE4D6}19.0 & \cellcolor[HTML]{DDECF4}82.7 & \cellcolor[HTML]{F8FCFD}59.4 & \cellcolor[HTML]{FDE5D7}19.9 & \cellcolor[HTML]{D9EAF3}85.6 & \cellcolor[HTML]{FDE6D8}21.0 & \cellcolor[HTML]{D3E6F1}90.9 & \cellcolor[HTML]{FAFCFE}58.5 & \cellcolor[HTML]{DAEAF3}85.2 & \cellcolor[HTML]{FDE9DE}26.0 & \cellcolor[HTML]{FDE4D6}19.1 & \cellcolor[HTML]{E5F0F7}76.0 & \cellcolor[HTML]{D3E6F1}91.0 & \cellcolor[HTML]{D1E5F0}92.0 & \cellcolor[HTML]{FDE8DB}23.5 & \cellcolor[HTML]{FDDBC7}6.3 \\
\multirow{-2}{*}{FedOpt} & \xmark & \cellcolor[HTML]{E5F1F7}75.7 & \cellcolor[HTML]{FEF9F5}45.8 & \cellcolor[HTML]{FEFFFF}54.6 & \cellcolor[HTML]{F7FBFD}60.7 & \cellcolor[HTML]{FEFCFA}49.7 & \cellcolor[HTML]{F9FCFE}58.7 & \cellcolor[HTML]{F2F8FB}64.5 & \cellcolor[HTML]{FDE0CF}13.8 & \cellcolor[HTML]{E6F1F7}75.2 & \cellcolor[HTML]{FDFEFF}55.7 & \cellcolor[HTML]{D7E9F2}87.3 & \cellcolor[HTML]{FEF7F3}43.6 & \cellcolor[HTML]{EEF5FA}68.4 & \cellcolor[HTML]{FEF1E9}35.5 & \cellcolor[HTML]{EEF6FA}68.3 & \cellcolor[HTML]{E5F0F7}75.8 & \cellcolor[HTML]{FFFFFF}54.0 & \cellcolor[HTML]{FEF4EF}40.3 \\ \midrule
\multicolumn{2}{c|}{ASR AVG} & \cellcolor[HTML]{E4F0F6}76.7 & \cellcolor[HTML]{FDECE2}29.1 & \cellcolor[HTML]{FEEEE6}32.5 & \cellcolor[HTML]{EDF5F9}69.2 & \cellcolor[HTML]{FEF6F1}42.4 & \cellcolor[HTML]{FDEBE0}27.8 & \cellcolor[HTML]{F5F9FC}62.5 & \cellcolor[HTML]{FDE4D5}18.8 & \cellcolor[HTML]{E7F2F8}74.1 & \cellcolor[HTML]{FEFEFE}53.2 & \cellcolor[HTML]{DEEDF5}81.2 & \cellcolor[HTML]{FEEFE7}33.6 & \cellcolor[HTML]{FEF6F1}42.1 & \cellcolor[HTML]{FEFAF8}47.9 & \cellcolor[HTML]{E5F1F7}75.6 & \cellcolor[HTML]{E5F0F7}76.0 & \cellcolor[HTML]{FEF1E9}35.7 & \cellcolor[HTML]{FDE7D9}22.2 \\\bottomrule
\end{tabular}}\end{table*}

\subsection{Evaluation of Attacks}

\Cref{tab:mpa} presents the performance of model poisoning attacks over different defenses (smaller values indicate better performance).
We observe that under the MNIST dataset, ALIE, IPM, and MinSum achieve better performance than others, and under CIFAR-10 dataset, Gaussian, SignFlipping, and FangAttack achieve better performance than others, and in general, Gaussian, SignFlipping, and FangAttack achieve the best attack effectiveness than others.

For simple attacks, our analysis reveals that although attacks Gaussian and SignFlipping are relatively straightforward, they exhibit the highest effectiveness across various settings, especially under more heterogeneous datasets.
We believe the effectiveness of Gaussian attacks benefits from the fact that as model training converges, the distribution of model updates in machine learning approaches a Gaussian distribution, making Gaussian attacks increasingly challenging to identify at later stages.
Besides, both Gaussian attack and Sign Flipping attack can introduce large perturbations to defense algorithms that rely solely on outlier detection without robust statistical measures or other reliable references, such as~\cite{trimmedmean_ICML2018, mimic_iclr2022, auror_acsac2016, foolsgold_raid2020, crfl_icml2021}.
For advanced attacks, ALIE performs well on MNIST but poorly on CIFAR-10 due to the small attack hyper-parameter values from the $z_{max}$ estimation in \cite{alie_nips2019}.
This suggests that $z_{max}$ requires manual fine-tuning for stronger attack impact, highlighting the non-adaptiveness of ALIE across datasets.
As noted in the original paper, IPM indeed exhibits stronger attack effectiveness against some specific defenses such as Krum and coordinate-wise methods.
However, its overall impact on system compromise remains limited.
Furthermore, FangAttack demonstrates superior performance compared to attacks that rely solely on statistical deviation on benign updates, such as ALIE and IPM, or solely on distance-evasion methods like MinMax and MinSum, especially with a more heterogeneous CIFAR dataset.
FangAttack's effectiveness stems from (1) its targeted evasion of distance-based Krum defenses, which enhances its attack capability, and (2) its balance between attack impact and defense evasion through statistical deviation on benign updates.
Mimic shows improved performance on CIFAR-10, attributed to the advantages provided by its heterogeneity.

\Cref{tab:dpa} shows the performance of data poisoning attacks over different defenses.
We find that BadNets (centralized backdoor), DBA (distributed backdoor), and Edge-case attacks (Edge-case backdoor), represent the most powerful DPAs, in general.
We believe that their simplicity and high adaptivity are the key reasons for their effectiveness.
These attacks do not rely on complicated fine-tuning, which can be applied to various datasets and federated learning algorithms.
Conversely, other backdoor attacks that require adjusting multiple hyper-parameters based on different datasets and algorithms~\cite{altermin_icml2019, neurotoxin_icml2022, modelreplacement_aistats2020} and may degrade the backdoor performance as the model converges or result in training failure.

Data poisoning attacks, especially backdoor attacks, are influenced by various parameters such as data poisoning ratio, the number of malicious nodes, trigger size, and data distribution.
Methods like scaling and alternating optimization are increasingly used to enhance poisoning effects, with hyper-parameters like scaling factors and the number of training iterations for benign and poisoned datasets.
These parameters have different impacts depending on the dataset and algorithm.
Therefore, we derive that the adaptivity of hyper-parameters is crucial for the robustness of data poisoning attacks.

\myta{
We find that adaptive hyper-parameters enhance attack effectiveness and robustness across settings.
\textbf{When designing advanced attacks, we recommend focusing on both 1) deviation based on benign update statistics and 2) evasion of common detection metrics like Krum.}
While many advanced attacks aim to evade server-side defenses, excessive evasion constraints may reduce attack impact.
\textbf{To enhance defenses, integrating robust statistical measures into outlier detection is advised, enabling resistance to simple perturbations while filtering malicious updates.}
}
\subsubsection{Comparison of Attacks in Terms of FL Algorithms}

\mypara{MPAs.}
It is observed that all MPAs exhibit a greater attack impact under FedSGD compared to the FedOpt algorithm in~\Cref{pic:alg_mpa_cmp} and~\Cref{tab:mpa}.
Gaussian, Sign Flipping, and FangAttack exhibit stronger overall attack effectiveness under both two algorithms.
It is evident that, for attacks other than Mimic, there is a significant disparity in attack impact under FedSGD, whereas the difference is minimal under FedOpt.
We attribute this observation to the fact that FedOpt does one aggregation after several iterations, resulting in more stable statistical characteristics of convergence.
This stability diminishes the impact of the Gaussian, Sign Flipping, and FangAttack, which is more effective than others under FedSGD, making it comparable to other attacks in terms of performance.
Specifically, the MinMax and MinSum attacks, which share a similar underlying attack principle, exhibit comparable effects under FedSGD but demonstrate distinct impacts under FedOpt.
This highlights their attack adaptability to different scenarios.
This is also confirmed in previous work~\cite{min_ndss2021}.

\mypara{DPAs.} 
Taking both the main task accuracy and ASR, into account  we formulate a metric, named Targeted Attack Impact (TAI), for DPAs, which can be calculated by 
\begin{equation}
\text{TAI} = \frac{\text{Accuracy under Attack}}{\text{Accuracy without Attack}}  + \text{ASR},
\end{equation}
where a higher targeted attack impact indicates a more effective DPA.
It is observed in~\Cref{pic:alg_dpa_cmp} and~\Cref{tab:dpa} that most DPAs exhibit stronger targeted attack impacts under FedOpt compared to FedSGD.
This can be attributed to the multiple local training rounds in FedOpt, which enable attackers to stealthily amplify the targeted impact without significantly compromising main task accuracy, unlike FedSGD.
Notably, BadNets, DBA, ModelReplacement, and EdgeCase attacks demonstrate higher targeted attack impacts than others under both algorithms, with BadNets and DBA showing particularly pronounced effects.
Besides, we observe that, compared to other attacks, Neurotoxin demonstrates a greater attack improvement under FedOpt than FedSGD.
This is because multiple local rounds benefit its attack mechanism.

\myta{
MPAs are more effective under FedSGD due to their iterative aggregation, allowing malicious deviations to accumulate.
In contrast, most DPAs, especially targeted attacks, are stronger under FedOpt, as multiple local training rounds enable gradual embedding of the targeted effect.
\textbf{This explains why all MPAs are designed for FedSGD and most DPAs, particularly targeted attacks, are tailored for FedOpt.}
}

\subsubsection{Comparison of Attacks in terms of Data Heterogeneity}

\mypara{MPAs.}
As demonstrated in~\Cref{pic:heter_mpa_cmp} and~\Cref{tab:mpa}, almost all MPAs reveal better effectiveness under non-IID than IID.
Gaussian, Sign Flipping, and FangAttack are the best attacks under IID.
Specifically, Gaussian and Sign Flipping attacks are more effective under IID because the similar data distribution across clients allows the perturbations to uniformly impact the global model, whereas in non-IID settings, data distribution differences cause inconsistent attack effects, reducing their effectiveness.
FangAttack, MinMax, and MinSum are the best attacks under non-IID.
Specifically, MinMax and MinSum attacks are less effective under IID but perform better under non-IID settings.
This is due to greater variations and distance between benign updates introduced by heterogeneous data, making distance-based attacks harder to detect.
Additionally, FangAttack ranks at the top in both IID and non-IID settings, demonstrating the superiority of optimization-based evasion attacks.

\mypara{DPAs.}
It can be shown that, in~\Cref{pic:heter_dpa_cmp} and~\Cref{tab:dpa}, BadNets, DBA, ModelReplacement, EdgeCase ranking greater attacks than others under both IID and non-IID.
As for their attack effectiveness under IID and non-IID settings, there is no clear trend regarding which performs stronger.
It also demonstrates that the hyper-parameter settings of poisoning attacks significantly influence their effectiveness, highlighting the need for adaptive poisoning attacks with fewer hyper-parameters.

\myta{
Both MPAs and DPAs have attacks that demonstrate strong attack impact under IID.
Additionally, data heterogeneity has been explored to enhance MPA performance under non-IID.
However, the impact of data heterogeneity on DPAs remains unclear.}

\subsection{Evaluation of Defenses}

\Cref{tab:mpa_defense} demonstrates the defense performance against model poisoning attacks.
We find that Median, FLTrust, DnC, and FLAME are the best defenses and perform well across various experiment settings, in general.
For model poisoning defenses (Median, FLTrust, and DnC), Median demonstrates its effectiveness as a good robust statistic.
FLTrust benefits from a server-maintained model trained on the same data distribution as the client, guiding correct gradient updates and adapting to different datasets and algorithms.
DnC uses singular value decomposition and majority-based clustering, without relying on dataset or algorithm-specific hyper-parameters.
We believe that these two methods benefit from their strong robustness and adaptability, making them stable across variations in datasets, algorithms, and data heterogeneity.
As a backdoor defense, FLAME derives advantages from (1) dynamic capture differences with HDBSCAN, (2) robust statistics like the median of norms for updates clipping.
Some defenses like Krum~\cite{krum_nips2017} and S.G.~\cite{signguard_ICDCS22} fail for relying solely on outlier detection or robust statistical metrics, while others, like F.G.~\cite{foolsgold_raid2020} and D.S.~\cite{deepsight_ndss2022}, introducing data-dependent hyper-parameters degrade their performance.

\Cref{tab:dpa_defense_mnist} and \Cref{tab:dpa_defense_cifar10} show the defense performance against data poisoning attacks.
First, we find that Krum, M.K., Bulyan, FLTrust, DnC, and FLAME outperform other methods generally.
As model poisoning defenses, Krum, M.K., and Bulyan demonstrate that robust statistics-based methods can be effective for defending against data poisoning attacks.
As a combination of trimmed mean and coordinate-wise methodology, Bulyan captures update anomalies more effectively in backdoor attack scenarios, highlighting the advantage of composite statistics-based methods.
FLAME, which integrates outliner detection and robust statistical metrics, exhibits strong robustness across diverse settings.
Second, beyond Krum, M.K., Bulyan, and FLAME, which perform well on CIFAR-10, FLtrust, DnC, and Auror also exhibit superior robustness and adaptivity compared to other approaches.
Besides, we also observe that certain defenses like F.G.~\cite{foolsgold_raid2020} and D.S.~\cite{deepsight_ndss2022} lack generalizability due to non-adaptive hyper-parameters.

\myta{
Defenses designed for MPAs and DPAs can be effective against both types to some extent, as they share certain similarities.
However, not all defenses are universally applicable, and their effectiveness may vary depending on the specific characteristics of both the attack and the defense.
\textbf{Based on our experimental findings, we recommend including Krum, and M.K. when involving data poisoning, and FLAME when involving model poisoning.}
}

\subsubsection{Comparison of Defenses in Terms of FL Algorithms}

\mypara{Defense Against MPAs.}
As shown in~\Cref{pic:alg_def_mpa_cmp} and~\Cref{tab:mpa_defense}, most defenses achieve better performance against MPAs under FedOpt compared to FedSGD.
This improvement can be attributed to the multiple local rounds per aggregation in FedOpt, which help mitigate small deviations and amplify significant ones, enhancing detection effectiveness.
Under FedSGD, FLTrust, F.G., DnC, Median, and FLAME demonstrate strong defense capabilities, while under FedOpt, FLTrust, DnC, S.G., and FLAME perform effectively.
This indicates that FLTrust, DnC, and FLAME maintain consistent effectiveness across both algorithms, highlighting their adaptability, thus underscoring the effectiveness of leveraging cosine similarity, the SVD method, and update magnitude control in enhancing defense performance.

\mypara{Defense Against DPAs.}
Considering the defense goal against DPAs, we formulate a metric, named Targeted Defense Robustness (TDR), for a defense strategy,
\begin{equation}
\text{TDR} = \frac{\text{Accuracy under DPA}}{\text{Accuracy without Attack}} + 1 - \text{ASR}.
\end{equation}
A higher TDR indicates greater defense robustness.
As illustrated in~\Cref{pic:alg_def_dpa_cmp}, ~\Cref{tab:dpa_defense_mnist}, and~\Cref{tab:dpa_defense_cifar10},
there is no clear evidence that defenses against DPAs perform consistently better under one algorithm compared to another.
Specifically, under FedSGD, the top-performing defenses are FLAME, Bulyan, M.K., Krum, and S.G. Meanwhile, under FedOpt, the best-performing defenses are FLAME, M.K., Krum, Bulyan, and S.G.
Notably, except for FLAME, the other four defenses were originally proposed for MPA, suggesting that defenses designed for MPA can also be quite effective against DPA.
Furthermore, it is worth mentioning that only FLAME and M.K. achieve a performance of about 1.75, which is still far below the optimal metric of 2.
It indicates that there is significant room for improving defense robustness against DPAs.

\myta{Most existing defenses perform better against MPAs under FedOpt than FedSGD.
However, there is no clear evidence that existing defenses against DPAs consistently favor one algorithm over the other.
Future research can explore algorithm-specific factors to improve defense robustness across algorithm-specific scenarios.}

\subsubsection{Comparison of Defenses in terms of Data Heterogeneity}

\mypara{Defense Against MPAs.}
As illustrated in ~\Cref{pic:heter_def_mpa_cmp} and ~\Cref{tab:mpa_defense}, it is evident that all defense mechanisms exhibit better performance under IID conditions compared to non-IID conditions.
In the IID setting, methods such as DnC, FLTrust, FLAME, Median, and F.G. demonstrate superior performance, maintaining an average accuracy of approximately 80\% or higher.
However, the scenario in the non-IID setting is less favorable.
Among the defenses, only FLTrust achieves an average accuracy of approximately 80\%, while the second-tier defenses, such as DnC and NormClipping, manage to sustain an average accuracy of around 66\%.
This highlights a significant gap in the availability of effective defenses for non-IID settings.

\mypara{Defense Against DPAs.}
As presented in ~\Cref{pic:heter_def_dpa_cmp}, ~\Cref{tab:dpa_defense_mnist}, and~\Cref{tab:dpa_defense_cifar10}, under IID setting, FLAME, M.K., Krum, and Bulyan demonstrate the best performance, achieving targeted defense robustness of approximately 1.8.
However, under non-IID setting, the top-performing defenses, FLAME, FLTrust, M.K., and Bulyan only achieve a robustness of around 1.6.
This significant drop in performance raises concerns about the lack of effective defenses against DPAs in non-IID settings.

\myta{
Currently, there are some effective defense methods against MPA and DPA under the IID setting.
\textbf{However, the situation in the non-IID setting is less promising, with almost no robust defense methods available for either MPA or DPA, highlighting a significant gap.}
}

\section{Conclusion}
\label{sec:conclusion}

This paper offers a Systematization of Knowledge on benchmarking poisoning attacks and defenses in federated learning.
A comprehensive taxonomy and a large-scale, unified evaluation of 17 defense strategies against 15 representative poisoning attacks are provided.
Our extensive analysis highlights the strengths and limitations of these strategies, identifies the most advanced methods, and empirically reveals the key connections and distinctions between model poisoning and data poisoning.
Moreover, our discussion illuminates the current state of research both theoretically and empirically, providing design principles and highlighting future research directions from a unified perspective.

\clearpage
\section*{Ethics Considerations}
Federated learning has been widely applied in security-sensitive scenarios. However, its distributed training nature introduces vulnerabilities to poisoning attacks. 
Our SoK paper aims to comprehensively analyze the performance of both attacks and defenses against poisoning attacks in federated learning scenarios through a unified evaluation framework. 
We affirm the ethical compliance of this research and strive to advance the secure deployment of federated learning systems in the future.

\section*{Open Science}
To ensure compliance with availability, functionality, and reproducibility standards, we will release our source code and data to the public.

\bibliographystyle{plain}
\bibliography{ref}

\begin{thebibliography}{10}

\bibitem{ccpa_2018}
California consumer privacy act (ccpa), state of california - department of justice - office of the attorney general, oct 2018.
\newblock Accessed: 2024-10-03.

\bibitem{gdpr_2024}
General data protection regulation (gdpr), apr 2024.
\newblock Accessed: 2024-10-03.

\bibitem{modelreplacement_aistats2020}
Eugene Bagdasaryan, Andreas Veit, Yiqing Hua, Deborah Estrin, and Vitaly Shmatikov.
\newblock How to backdoor federated learning.
\newblock In {\em International conference on artificial intelligence and statistics}, pages 2938--2948. PMLR, 2020.

\bibitem{alie_nips2019}
Gilad Baruch, Moran Baruch, and Yoav Goldberg.
\newblock A little is enough: Circumventing defenses for distributed learning.
\newblock {\em Advances in Neural Information Processing Systems}, 32, 2019.

\bibitem{altermin_icml2019}
Arjun~Nitin Bhagoji, Supriyo Chakraborty, Prateek Mittal, and Seraphin Calo.
\newblock Analyzing federated learning through an adversarial lens.
\newblock In {\em International conference on machine learning}, pages 634--643. PMLR, 2019.

\bibitem{labelflipping_icml2012}
Battista Biggio, Blaine Nelson, and Pavel Laskov.
\newblock Poisoning attacks against support vector machines.
\newblock In {\em Proceedings of the 29th International Coference on International Conference on Machine Learning}, ICML'12, page 1467–1474, Madison, WI, USA, 2012. Omnipress.

\bibitem{krum_nips2017}
Peva Blanchard, El~Mahdi El~Mhamdi, Rachid Guerraoui, and Julien Stainer.
\newblock Machine learning with adversaries: Byzantine tolerant gradient descent.
\newblock {\em Advances in neural information processing systems}, 30, 2017.

\bibitem{fltrust_ndss2020}
Xiaoyu Cao, Minghong Fang, Jia Liu, and Neil~Zhenqiang Gong.
\newblock Fltrust: Byzantine-robust federated learning via trust bootstrapping.
\newblock {\em arXiv preprint arXiv:2012.13995}, 2020.

\bibitem{signflipping_icml2018}
Georgios Damaskinos, Rachid Guerraoui, Rhicheek Patra, Mahsa Taziki, et~al.
\newblock Asynchronous byzantine machine learning (the case of sgd).
\newblock In {\em International Conference on Machine Learning}, pages 1145--1154. PMLR, 2018.

\bibitem{fatecredit_fedai2020}
editor2fedai.
\newblock Utilization of fate in risk management of credit in small and micro enterprises, 2020.

\bibitem{fangattack_usenix_sec2020}
Minghong Fang, Xiaoyu Cao, Jinyuan Jia, and Neil Gong.
\newblock Local model poisoning attacks to $\{$Byzantine-Robust$\}$ federated learning.
\newblock In {\em 29th USENIX security symposium (USENIX Security 20)}, pages 1605--1622, 2020.

\bibitem{foolsgold_raid2020}
Clement Fung, Chris~JM Yoon, and Ivan Beschastnikh.
\newblock The limitations of federated learning in sybil settings.
\newblock In {\em 23rd International Symposium on Research in Attacks, Intrusions and Defenses (RAID 2020)}, pages 301--316, 2020.

\bibitem{backdoorfl_wc2022}
Xueluan Gong, Yanjiao Chen, Qian Wang, and Weihan Kong.
\newblock Backdoor attacks and defenses in federated learning: State-of-the-art, taxonomy, and future directions.
\newblock {\em IEEE Wireless Communications}, 30(2):114--121, 2022.

\bibitem{badnets_NIPSWS2017}
Tianyu Gu, Brendan Dolan-Gavitt, and Siddharth Garg.
\newblock Badnets: Identifying vulnerabilities in the machine learning model supply chain.
\newblock {\em arXiv preprint arXiv:1708.06733}, 2017.

\bibitem{bulyan_icml2018}
Rachid Guerraoui, S{\'e}bastien Rouault, et~al.
\newblock The hidden vulnerability of distributed learning in byzantium.
\newblock In {\em International Conference on Machine Learning}, pages 3521--3530. PMLR, 2018.

\bibitem{flkeyboard_google2018}
Andrew Hard, Kanishka Rao, Rajiv Mathews, Swaroop Ramaswamy, Fran{\c{c}}oise Beaufays, Sean Augenstein, Hubert Eichner, Chlo{\'e} Kiddon, and Daniel Ramage.
\newblock Federated learning for mobile keyboard prediction.
\newblock {\em arXiv preprint arXiv:1811.03604}, 2018.

\bibitem{resnet18_he2015}
Kaiming He, Xiangyu Zhang, Shaoqing Ren, and Jian Sun.
\newblock Deep residual learning for image recognition. arxiv e-prints.
\newblock {\em arXiv preprint arXiv:1512.03385}, 10, 2015.

\bibitem{dirichletfl_google}
Tzu-Ming~Harry Hsu, Hang Qi, and Matthew Brown.
\newblock Measuring the effects of non-identical data distribution for federated visual classification.
\newblock {\em arXiv preprint arXiv:1909.06335}, 2019.

\bibitem{robust_statistics_2011}
Peter~J Huber and Elvezio~M Ronchetti.
\newblock {\em Robust statistics}.
\newblock John Wiley \& Sons, 2011.

\bibitem{SGD_sketching_nips2019}
Nikita Ivkin, Daniel Rothchild, Enayat Ullah, Ion Stoica, Raman Arora, et~al.
\newblock Communication-efficient distributed sgd with sketching.
\newblock {\em Advances in Neural Information Processing Systems}, 32, 2019.

\bibitem{centeredclipping_icml2021}
Sai~Praneeth Karimireddy, Lie He, and Martin Jaggi.
\newblock Learning from history for byzantine robust optimization.
\newblock In {\em International Conference on Machine Learning}, pages 5311--5319. PMLR, 2021.

\bibitem{mimic_iclr2022}
Sai~Praneeth Karimireddy, Lie He, and Martin Jaggi.
\newblock Byzantine-robust learning on heterogeneous datasets via bucketing.
\newblock In {\em International Conference on Learning Representations}, 2022.

\bibitem{cifar10_2009}
Alex Krizhevsky, Geoffrey Hinton, et~al.
\newblock Learning multiple layers of features from tiny images.
\newblock 2009.

\bibitem{ardis_nca2020}
Huseyin Kusetogullari, Amir Yavariabdi, Abbas Cheddad, H{\aa}kan Grahn, and Johan Hall.
\newblock Ardis: a swedish historical handwritten digit dataset.
\newblock {\em Neural Computing and Applications}, 32(21):16505--16518, 2020.

\bibitem{mnist_proceddingsIEEE}
Yann LeCun, L{\'e}on Bottou, Yoshua Bengio, and Patrick Haffner.
\newblock Gradient-based learning applied to document recognition.
\newblock {\em Proceedings of the IEEE}, 86(11):2278--2324, 1998.

\bibitem{MNISTLeNet5_Proc_IEEE1998}
Yann LeCun, L{\'e}on Bottou, Yoshua Bengio, and Patrick Haffner.
\newblock Gradient-based learning applied to document recognition.
\newblock {\em Proceedings of the IEEE}, 86(11):2278--2324, 1998.

\bibitem{3dfed_sp2023}
Haoyang Li, Qingqing Ye, Haibo Hu, Jin Li, Leixia Wang, Chengfang Fang, and Jie Shi.
\newblock 3dfed: Adaptive and extensible framework for covert backdoor attack in federated learning.
\newblock In {\em 2023 IEEE Symposium on Security and Privacy (SP)}, pages 1893--1907. IEEE, 2023.

\bibitem{blade_tbd2023}
Shenghui Li, Edith C-H Ngai, and Thiemo Voigt.
\newblock An experimental study of byzantine-robust aggregation schemes in federated learning.
\newblock {\em IEEE Transactions on Big Data}, 2023.

\bibitem{fl_mobile_cst2020}
Wei Yang~Bryan Lim, Nguyen~Cong Luong, Dinh~Thai Hoang, Yutao Jiao, Ying-Chang Liang, Qiang Yang, Dusit Niyato, and Chunyan Miao.
\newblock Federated learning in mobile edge networks: A comprehensive survey.
\newblock {\em IEEE communications surveys \& tutorials}, 22(3):2031--2063, 2020.

\bibitem{privacyrobustfl_tnnls2022}
Lingjuan Lyu, Han Yu, Xingjun Ma, Chen Chen, Lichao Sun, Jun Zhao, Qiang Yang, and S~Yu Philip.
\newblock Privacy and robustness in federated learning: Attacks and defenses.
\newblock {\em IEEE transactions on neural networks and learning systems}, 2022.

\bibitem{fedsgdavg_AISTATS2017}
Brendan McMahan, Eider Moore, Daniel Ramage, Seth Hampson, and Blaise~Aguera y~Arcas.
\newblock Communication-efficient learning of deep networks from decentralized data.
\newblock In {\em Artificial intelligence and statistics}, pages 1273--1282. PMLR, 2017.

\bibitem{fl_iot_cst2021}
Dinh~C Nguyen, Ming Ding, Pubudu~N Pathirana, Aruna Seneviratne, Jun Li, and H~Vincent Poor.
\newblock Federated learning for internet of things: A comprehensive survey.
\newblock {\em IEEE Communications Surveys \& Tutorials}, 23(3):1622--1658, 2021.

\bibitem{fl_healthcare_csur2022}
Dinh~C Nguyen, Quoc-Viet Pham, Pubudu~N Pathirana, Ming Ding, Aruna Seneviratne, Zihuai Lin, Octavia Dobre, and Won-Joo Hwang.
\newblock Federated learning for smart healthcare: A survey.
\newblock {\em ACM Computing Surveys (Csur)}, 55(3):1--37, 2022.

\bibitem{flame_usenix_security2022}
Thien~Duc Nguyen, Phillip Rieger, Roberta De~Viti, Huili Chen, Bj{\"o}rn~B Brandenburg, Hossein Yalame, Helen M{\"o}llering, Hossein Fereidooni, Samuel Marchal, Markus Miettinen, et~al.
\newblock $\{$FLAME$\}$: Taming backdoors in federated learning.
\newblock In {\em 31st USENIX Security Symposium (USENIX Security 22)}, pages 1415--1432, 2022.

\bibitem{fl_personalization_apple2022}
Matthias Paulik, Matt Seigel, Henry Mason, Dominic Telaar, Joris Kluivers, Rogier van Dalen, Chi~Wai Lau, Luke Carlson, Filip Granqvist, Chris Vandevelde, Sudeep Agarwal, Julien Freudiger, Andrew Byde, Abhishek Bhowmick, Gaurav Kapoor, Si~Beaumont, Áine Cahill, Dominic Hughes, Omid Javidbakht, Fei Dong, Rehan Rishi, and Stanley Hung.
\newblock Federated evaluation and tuning for on-device personalization: System design \& applications, 2022.

\bibitem{rfa_arxiv2019}
Krishna Pillutla, Sham~M Kakade, and Zaid Harchaoui.
\newblock Robust aggregation for federated learning.
\newblock {\em IEEE Transactions on Signal Processing}, 70:1142--1154, 2022.

\bibitem{flemoji_google2019}
Swaroop Ramaswamy, Rajiv Mathews, Kanishka Rao, and Fran{\c{c}}oise Beaufays.
\newblock Federated learning for emoji prediction in a mobile keyboard.
\newblock {\em arXiv preprint arXiv:1906.04329}, 2019.

\bibitem{fedopt_iclr2021}
Sashank~J. Reddi, Zachary Charles, Manzil Zaheer, Zachary Garrett, Keith Rush, Jakub Kone{\v{c}}n{\'y}, Sanjiv Kumar, and Hugh~Brendan McMahan.
\newblock Adaptive federated optimization.
\newblock In {\em International Conference on Learning Representations}, 2021.

\bibitem{deepsight_ndss2022}
Phillip Rieger, Thien~Duc Nguyen, Markus Miettinen, and Ahmad-Reza Sadeghi.
\newblock Deepsight: Mitigating backdoor attacks in federated learning through deep model inspection.
\newblock {\em arXiv preprint arXiv:2201.00763}, 2022.

\bibitem{sokflsecurity_acsac2023}
Geetanjli Sharma, MAP Chamikara, Mohan~Baruwal Chhetri, and Yi-Ping~Phoebe Chen.
\newblock Sok: Systematizing attack studies in federated learning--from sparseness to completeness.
\newblock In {\em Proceedings of the 2023 ACM Asia Conference on Computer and Communications Security}, pages 579--592, 2023.

\bibitem{min_ndss2021}
Virat Shejwalkar and Amir Houmansadr.
\newblock Manipulating the byzantine: Optimizing model poisoning attacks and defenses for federated learning.
\newblock In {\em NDSS}, 2021.

\bibitem{productionflpoison_sp2022}
Virat Shejwalkar, Amir Houmansadr, Peter Kairouz, and Daniel Ramage.
\newblock Back to the drawing board: A critical evaluation of poisoning attacks on production federated learning.
\newblock In {\em 2022 IEEE Symposium on Security and Privacy (SP)}, pages 1354--1371. IEEE, 2022.

\bibitem{auror_acsac2016}
Shiqi Shen, Shruti Tople, and Prateek Saxena.
\newblock Auror: Defending against poisoning attacks in collaborative deep learning systems.
\newblock In {\em Proceedings of the 32nd annual conference on computer security applications}, pages 508--519, 2016.

\bibitem{sparsifiedSGD_nips2018}
Sebastian~U Stich, Jean-Baptiste Cordonnier, and Martin Jaggi.
\newblock Sparsified sgd with memory.
\newblock {\em Advances in neural information processing systems}, 31, 2018.

\bibitem{normclipping_nips2019}
Ziteng Sun, Peter Kairouz, Ananda~Theertha Suresh, and H~Brendan McMahan.
\newblock Can you really backdoor federated learning?
\newblock {\em arXiv preprint arXiv:1911.07963}, 2019.

\bibitem{poisonml_csur2022}
Zhiyi Tian, Lei Cui, Jie Liang, and Shui Yu.
\newblock A comprehensive survey on poisoning attacks and countermeasures in machine learning.
\newblock {\em ACM Computing Surveys}, 55(8):1--35, 2022.

\bibitem{poisonwflsurvey_cst2024}
Yichen Wan, Youyang Qu, Wei Ni, Yong Xiang, Longxiang Gao, and Ekram Hossain.
\newblock Data and model poisoning backdoor attacks on wireless federated learning, and the defense mechanisms: A comprehensive survey.
\newblock {\em IEEE Communications Surveys \& Tutorials}, 2024.

\bibitem{edgecase_nips2020}
Hongyi Wang, Kartik Sreenivasan, Shashank Rajput, Harit Vishwakarma, Saurabh Agarwal, Jy-yong Sohn, Kangwook Lee, and Dimitris Papailiopoulos.
\newblock Attack of the tails: Yes, you really can backdoor federated learning.
\newblock {\em Advances in Neural Information Processing Systems}, 33:16070--16084, 2020.

\bibitem{FedMA_iclr2020}
Hongyi Wang, Mikhail Yurochkin, Yuekai Sun, Dimitris Papailiopoulos, and Yasaman Khazaeni.
\newblock Federated learning with matched averaging.
\newblock {\em arXiv preprint arXiv:2002.06440}, 2020.

\bibitem{crfl_icml2021}
Chulin Xie, Minghao Chen, Pin-Yu Chen, and Bo~Li.
\newblock Crfl: Certifiably robust federated learning against backdoor attacks.
\newblock In {\em International Conference on Machine Learning}, pages 11372--11382. PMLR, 2021.

\bibitem{dba_iclr2019}
Chulin Xie, Keli Huang, Pin-Yu Chen, and Bo~Li.
\newblock Dba: Distributed backdoor attacks against federated learning.
\newblock In {\em International conference on learning representations}, 2019.

\bibitem{IPM_PMLR2020}
Cong Xie, Oluwasanmi Koyejo, and Indranil Gupta.
\newblock Fall of empires: Breaking byzantine-tolerant sgd by inner product manipulation.
\newblock In {\em Uncertainty in Artificial Intelligence}, pages 261--270. PMLR, 2020.

\bibitem{signguard_ICDCS22}
Jian Xu, Shao-Lun Huang, Linqi Song, and Tian Lan.
\newblock Signguard: Byzantine-robust federated learning through collaborative malicious gradient filtering.
\newblock {\em arXiv preprint arXiv:2109.05872}, 2021.

\bibitem{heterogeneous_csur2023}
Mang Ye, Xiuwen Fang, Bo~Du, Pong~C Yuen, and Dacheng Tao.
\newblock Heterogeneous federated learning: State-of-the-art and research challenges.
\newblock {\em ACM Computing Surveys}, 56(3):1--44, 2023.

\bibitem{yi2024jailbreak}
Sibo Yi, Yule Liu, Zhen Sun, Tianshuo Cong, Xinlei He, Jiaxing Song, Ke~Xu, and Qi~Li.
\newblock Jailbreak attacks and defenses against large language models: A survey.
\newblock {\em arXiv preprint arXiv:2407.04295}, 2024.

\bibitem{trimmedmean_ICML2018}
Dong Yin, Yudong Chen, Ramchandran Kannan, and Peter Bartlett.
\newblock Byzantine-robust distributed learning: Towards optimal statistical rates.
\newblock In {\em International conference on machine learning}, pages 5650--5659. Pmlr, 2018.

\bibitem{bayesiannn_icml2019}
Mikhail Yurochkin, Mayank Agarwal, Soumya Ghosh, Kristjan Greenewald, Nghia Hoang, and Yasaman Khazaeni.
\newblock Bayesian nonparametric federated learning of neural networks.
\newblock In {\em International conference on machine learning}, pages 7252--7261. PMLR, 2019.

\bibitem{fldetector_kdd22}
Zaixi Zhang, Xiaoyu Cao, Jinyuan Jia, and Neil~Zhenqiang Gong.
\newblock Fldetector: Defending federated learning against model poisoning attacks via detecting malicious clients.
\newblock In {\em Proceedings of the 28th ACM SIGKDD Conference on Knowledge Discovery and Data Mining}, pages 2545--2555, 2022.

\bibitem{neurotoxin_icml2022}
Zhengming Zhang, Ashwinee Panda, Linyue Song, Yaoqing Yang, Michael Mahoney, Prateek Mittal, Ramchandran Kannan, and Joseph Gonzalez.
\newblock Neurotoxin: Durable backdoors in federated learning.
\newblock In {\em International Conference on Machine Learning}, pages 26429--26446. PMLR, 2022.

\end{thebibliography}

\appendix
\section{Overview of Our \Method}
\label{app:overview}

\begin{figure*}[htbp]
    \centering
    \includegraphics [width=0.95\textwidth, keepaspectratio] {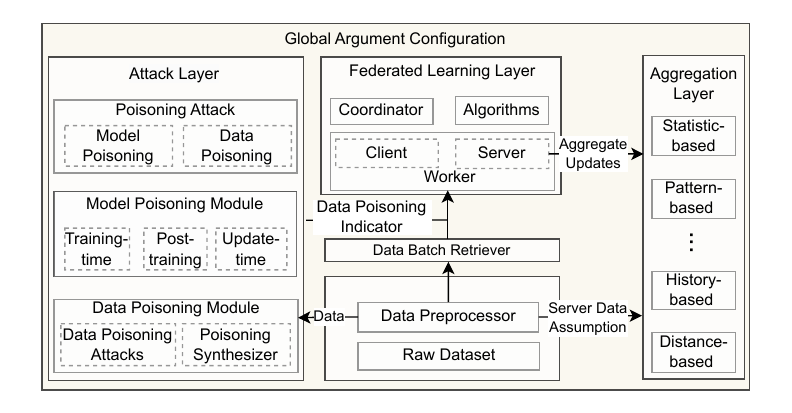}
    \caption{Overview of Our \Method}
    \label{pic:framework}
\end{figure*}

In this section, we provide a concise overview of our \Method framework.
As illustrated in Figure \ref{pic:framework}, our \Method is highly decoupled and modular, consisting of three layers: the federated learning layer, the attack layer, and the aggregation layer.
Each layer is configurable through a global argument.

In the federated learning layer, a Worker class is implemented with foundational training and inference functions, serving as a base for the Client and Server classes.
A Coordinator facilitates the federated learning workflow, including client initialization and federated learning algorithm configuration.
Supported algorithms, such as FedSGD, FedOpt, and FedAvg, are injected into clients as hooks to facilitate easy customization.
The attack layer comprises three main components.
The model and data poisoning modules enable various attack types by coupling with the client class.
An attack is defined by some of these four attributes: training-time poisoning, post-training poisoning, update-time poisoning, or data poisoning.
In the data poisoning module, a synthesizer takes preprocessed data and applies the specified poisoning strategies.
A data poisoning indicator determines whether a data batch will be poisoned, passing the modified batch for training or inference.
Finally, the aggregation layer collects updates from clients via the server and applies a chosen aggregation rule to produce the result returned to the server.
Some aggregation schemes assume the server has access to a dataset following the same distribution as the training data.
Each attack or defense is automatically registered through our implemented registry mechanism, ensuring high scalability.

\begin{algorithm}[htbp]
    \renewcommand{\algorithmicrequire}{\textbf{Input:}}
    \renewcommand{\algorithmicensure}{\textbf{Output:}}
    \caption{Algorithms: \colorbox{yellow}{FedSGD},\colorbox{pink}{FedAvg},and \colorbox{green}{FedOpt}}
    \label{alg:fl_algorithms}
    \begin{algorithmic}[1]
        \REQUIRE {Number of participants: $n$}\newline
        {Number of global epochs, local epochs: $T_g$, $T_l$}\newline
        {Learning rate of global and local training at epoch $t$: $\eta_g^t$, $\eta_l^t$}
        \ENSURE {Converged model: $w^{T_g}$}\newline
        
        \FOR{each global epoch $t \in [T_g]$}
            \STATE Distribute $w^{t-1}$ to participant $i \in [N]$ 
            \FOR[\textbf{in parallel}]{participant $i \in [N]$}
                \STATE $w_i^t \gets w^{t-1}$
                \STATE $T_l \gets 1$ (\colorbox{yellow}{FedSGD})
                \STATE Keep $T_l$ unchanged (\colorbox{pink}{FedAvg}, \colorbox{green}{FedOpt})
                \FOR{each local epoch $t \in [T_l]$}
                    \STATE Sample mini-batch $\xi_i$ from local data
                    \STATE $w_i^t \gets w_i^{t-1}-\eta_l^t \nabla F_k({w}_k^{t-1}, \xi_i)$
                \ENDFOR
                \STATE  ${\Delta}_i^t \gets {w}_i^t-{w}^{t-1}$
                \STATE Submits ${\Delta}_i^t$ back to server
                \STATE Submits $w_i^t$ back to server (\colorbox{pink}{FedAvg})
            \ENDFOR
            \STATE $\eta^t_g \gets 1$ (\colorbox{yellow}{FedSGD})
            \STATE ${w}^{t} \leftarrow {w}^{t-1}-\frac{\eta^t_g}{n} \left\{\Delta_i^{t-1}\right\}_{i \in[n]}$ (\colorbox{yellow}{FedSGD}, \colorbox{green}{FedOpt})
            \STATE ${w}^{t} \leftarrow \frac{1}{n}w^t_i$ (\colorbox{pink}{FedAvg})
        \ENDFOR
        \STATE return $w^{G}$
    \end{algorithmic}
\end{algorithm}

\section{Detailed Literature Review and Gap Analysis} \label{sec:detailed_review}

With increasing security concerns in FL and the emergence of new poisoning attacks, significant attention has been focused on this field.
\Cref{tab:related_work} presents a comparative summary of our SoK and the current literature on poisoning attacks and defenses in FL from 2022 to 2024.
A comprehensive evaluation and comparative analysis of these strategies from a unified perspective, as well as open-source benchmarks for consistent performance assessment, are lacking.

Most existing studies focus on surveys and analyses of either model or data poisoning attacks separately, often with limited scope.
Gong et al.~\cite{backdoorfl_wc2022} evaluate backdoor attacks as part of data poisoning attacks and their effectiveness against a few defenses.
Sharma et al.~\cite{sokflsecurity_acsac2023} and Wang et al.~\cite{poisonwflsurvey_cst2024} consider only partial data poisoning attacks and defense strategies without comprehensive analysis and evaluation.
Although a few studies~\cite{poisonml_csur2022, privacyrobustfl_tnnls2022} summarize and theoretically analyze both types of attacks, they lack experimental evaluations and fine-grained categories.
In summary, none provide a comprehensive and unified evaluation.
While a few experimental studies~\cite{productionflpoison_sp2022,blade_tbd2023} have been conducted, they remain limited in scope and perspective.
Shejwalkar et al.~\cite{productionflpoison_sp2022} focus exclusively on model poisoning attacks with a small number of malicious clients, leaving the effectiveness of model and data poisoning attacks involving larger groups unclear.
Li et al.~\cite{blade_tbd2023} evaluate model poisoning attacks in the context of Byzantine-robust federated learning but overlook data poisoning attacks and the intrinsic connections between these two types.
In summary, existing studies lack a comprehensive and unified analysis to clarify the distinctions and validate the mutual effectiveness of defenses against both DPAs and MPAs.

\subsection{Federated Learning} \label{sec:preli_fl}

Federal learning (FL) is a machine learning technique that focuses on training models on decentralized data while protecting the privacy of each local client~\cite{fedsgdavg_AISTATS2017}.
Typically, an FL algorithm comprises a global model hosted by a server and multiple local models hosted by different clients.
First, the server initializes the global model $w^0$ and distributes its weight to each client to set up their local models.
Clients with private data are randomly selected to participate in the training process, collaborating to train the global model on the server.
During each epoch, clients perform local training for a predefined number of local epochs and share their gradient updates or updated weights, and training information with the server.
The server then applies predefined aggregation rules, such as FedSGD~\cite{fedsgdavg_AISTATS2017}, FedAvg~\cite{fedsgdavg_AISTATS2017}, or FedOpt~\cite{fedopt_iclr2021}, to aggregate the updates and update the global model.
Finally, the updated weights are synchronized with the clients for the next epoch.
FL can be seen as a distributed optimization problem, defined as follows,
\begin{equation}
\min _{w \in R^d} F(w)=\frac{1}{n} \sum_{i=1}^n \mathbb{E}_{\xi_i \sim D_i}[F(w; \xi_i)],
\end{equation}
where $n$ is the number of participants, $D_i$ is $i$-clients' local data, and $F(w; \xi_i)$ is the loss function coordinated between server and clients, and calculated by model parameters $w$, and a batch of training data $\xi_i$.
Since the training data is distributed across clients, data heterogeneity exists.
Here, we consider independent and identically distributed (IID) or not (non-IID), following the previous work~\cite{fedsgdavg_AISTATS2017}.

For FL algorithms, McMahan et al.~\cite{fedsgdavg_AISTATS2017} develop FedSGD and FedAvg.
Reddi et al. propose FedOpt~\cite{fedopt_iclr2021}, a variant of FedAvg.
\Cref{alg:fl_algorithms} shows the details of these three algorithms.
In addition to the general procedures outlined in the previous section, there are notable distinctions among the algorithms.
In the FedSGD algorithm, the local epoch is set to 1, and the gradient, which is also the difference between the updated and received model weights, is submitted to the server.
The server then updates the global model using the aggregated updates.
FedAvg, designed as a communication-efficient version of FedSGD, typically uses multiple local epochs and transmits the updated local model weights directly to the server.
The server averages these updates based on the number of samples used in local training.
Following the setting of previous works~\cite{deepsight_ndss2022, flame_usenix_security2022}, we use equal weighting.
Building on the intuition of FedSGD and FedAvg, FedOpt theoretically validates that the negative of the average model difference can serve as a pseudo-gradient in general server optimizer updates.
Overall, FedOpt primarily differs from FedSGD in allowing multiple local epochs and adding a non-one global learning rate for server aggregation.
Consistent with prior studies~\cite{deepsight_ndss2022, flame_usenix_security2022}, we set the server learning rate of FedOpt to 1 for simplicity.

\subsection{Poisoning Attacks}\label{sec:preli_pa}

\begin{figure*}[htbp]
    \centering
    \includegraphics [width=0.9\textwidth, keepaspectratio] {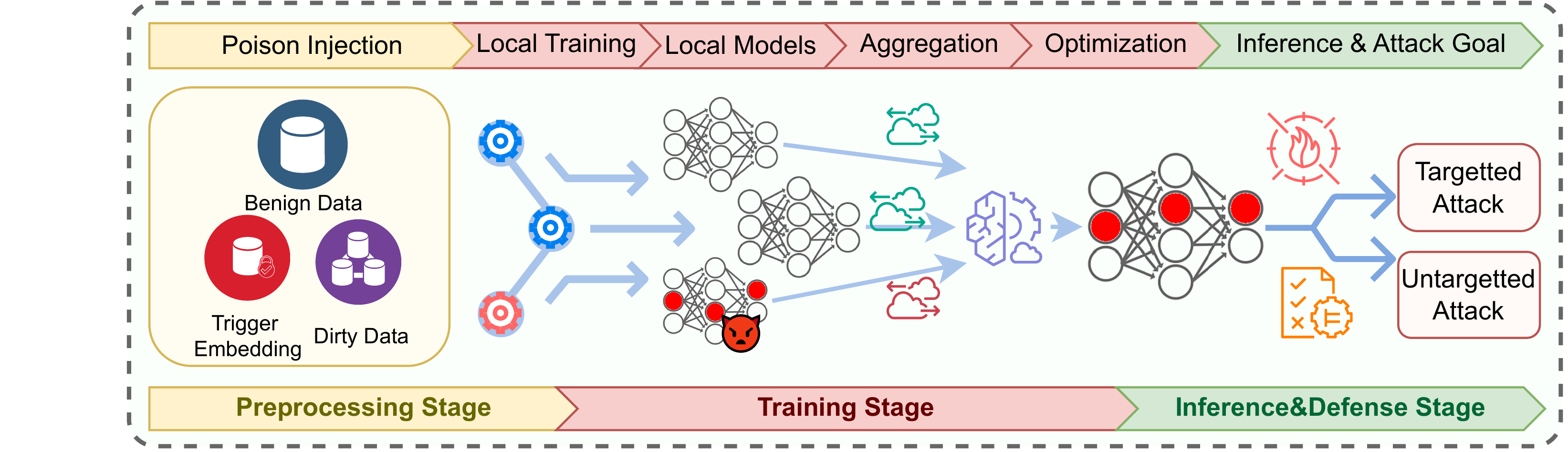}
    \caption{Illustration of Poisoning Attack Principles}
    \label{pic:fp_attackprocedure}
\end{figure*}

In the FL setting, poisoning attacks are generally classified into data poisoning attacks (DPAs) and model poisoning attacks (MPAs) based on the attack vector, and targeted or untargeted based on the attack goal.
As illustrated in \Cref{pic:fp_attackprocedure}, for data poisoning attacks, mostly targeted attacks, the adversary can control the training dataset during data preprocessing.
The adversary can then contaminate a clean dataset by adding dirty data~\cite{edgecase_nips2020,yi2024jailbreak} or performing malicious alterations, such as~\cite{labelflipping_icml2012, badnets_NIPSWS2017, dba_iclr2019, altermin_icml2019, modelreplacement_aistats2020, edgecase_nips2020, neurotoxin_icml2022}.
One common type of data poisoning attack is the backdoor attack, which aims to train a model that misclassifies any samples embedding a specified trigger as target labels $\tau_t$ rather than the original label $\tau_s$.
To execute this, the adversary first prepares a backdoored dataset with chosen target labels, then trains a backdoored model $w_{mal}^t$ using both benign and backdoored datasets $\mathcal{D} = \{\mathcal{D}_{be}, \mathcal{D}_{bd}\}$ to optimize for both the main task and the backdoor task.
Specifically, the attack objective in round $t$ for attacker $i$ is given by:
\begin{small} 
\begin{equation}
w_i^* = \arg \max _{w_i}\left(\sum_{j\in \mathcal{D}_{be}^i} P[F(w^t, \xi_j^i) = \tau_t] + \sum_{j \in \mathcal{D}_{bd}^i} P[F(w^t, \xi_j^i) = \tau_s]\right).
\end{equation}
\end{small}
Therefore, an effective data poisoning attack should demonstrate high accuracy in the main classification task while achieving a high success rate in poisoning the data.
Model poisoning attacks, mostly untargeted attacks, typically require knowledge of either the training process, benign, or malicious updates, to execute poisoning at various points such as training time, post-training, or update time.
Their attack objective is to compromise model convergence by maximizing the reduction in accuracy for the main classification task.

\subsection{Threat Model}\label{sec:preli_threat}

In an FL system, a central server coordinates global model training by first initializing the model and other auxiliary information and distributing them to $n$ clients for local training.
Among these clients, $f$ participants are either malicious or compromised by adversaries, potentially engaging in model poisoning attacks or data poisoning attacks as previously described.
For convenience and consistency with prior research~\cite{fangattack_usenix_sec2020}, we assume the first $f$ clients are either controlled or compromised by the adversary, which means client $i$, $i \in [0,f)$ is the malicious client, $i \in [f+1, n)$ is the benign client.
After clients complete their training and submit updates, the server processes these updates using a designated aggregator or defense strategy to generate a new global model for the next training round.
Although we implement all three FL algorithms mentioned above, we evaluate two commonly used algorithms, i.e., FedSGD and FedOpt, considering both IID and non-IID data distributions.
Notations used in this paper are provided in \Cref{tab:notation}.

\begin{table*}[htbp]
    \centering
\setlength{\tabcolsep}{4pt} 
\renewcommand{\arraystretch}{1.2} 
\caption{List of Abbreviations, Notations, and Their Definitions}
\label{tab:notation}
\resizebox{0.9\textwidth}{!}{%
\begin{tabular}{ccl}\toprule
Type & \textbf{Notations} & \textbf{Description} \\ \midrule
\multirow{5}{*}{\begin{tabular}[c]{@{}c@{}}Technical\\ Terms\end{tabular}} & SGD & Stochastic gradient descent \\ \cmidrule{2-3} 
 & IID & independent and identically distributed \\ \cmidrule{2-3} 
 & non-IID & Not independent and identically distributed \\ \cmidrule{2-3} 
 & ASR & Attack success rate \\ \cmidrule{2-3} 
 & Aggregator & A component or algorithm receives client updates and produces aggregated outputs. \\ \midrule
\multirow{11}{*}{\begin{tabular}[c]{@{}c@{}}Mathematical\\ Symbols\end{tabular}} & $n$ & The total number of participants or clients \\ \cmidrule{2-3} 
 & $f$ & The number of malicious clients \\ \cmidrule{2-3} 
 & $g$ & The benign gradient update in training, generally \\ \cmidrule{2-3} 
 & $g_{mal}$ & The malicious gradient update, generally \\ \cmidrule{2-3} 
 & $g_i$ & The gradient update of client $i$ \\ \cmidrule{2-3} 
 & $g_i, i\in(0, f)$ & The malicious gradient update of client $i$ \\ \cmidrule{2-3} 
 & $g_i, i\in(f+1, n)$ & The benign gradient update of client $i$ \\ \cmidrule{2-3} 
 & $w$ & The benign weights (update), generally \\ \cmidrule{2-3} 
 & $w_i$ & The weights (update) of client $i$ \\ \cmidrule{2-3} 
 & $w_{mal}$ & The malicious weights (update), generally \\ \cmidrule{2-3} 
 & $d$ & The dimension of model update \\ \bottomrule
\end{tabular}%
}
\end{table*}

\begin{table*}[htbp]
\centering
\caption{General Experimental Settings.
For data heterogeneity, we choose balanced IID and non-IID  (Dirichlet distribution with $\alpha$ = 0.5) partition strategies.
Settings for each attack and defense are in the $configs$ folder of our open-source codebase.
Some defenses under non-IID setting may require a smaller learning rate due to slower convergence, as specified in $batchrun.py$ of our codebase.}
\normalsize
\setlength{\tabcolsep}{3pt} 
\renewcommand{\arraystretch}{0.8} 
\label{tab:experiment_setting}
\begin{tabular}{ccclcc}
\toprule
\multirow{2}{*}{Setting} & \multicolumn{2}{c}{FedSGD} &  & \multicolumn{2}{c}{FedOpt} \\ \cmidrule{2-3} \cmidrule{5-6} 
 & \begin{tabular}[c]{@{}c@{}}MNIST \\ (LeNet-5)\end{tabular} & \begin{tabular}[c]{@{}c@{}}CIFAR-10 \\ (ResNet-18)\end{tabular} &  & \begin{tabular}[c]{@{}c@{}}MNIST \\ (LeNet-5)\end{tabular} & \begin{tabular}[c]{@{}c@{}}CIFAR-10 \\ (ResNet-18)\end{tabular} \\ \midrule
Epochs & 500 & 800 &  & 5 $\times$ 500 & 5 $\times$ 500 \\ \midrule
Learning Rate & 0.01 & 0.01 &  & 0.01 & 0.01 \\ \midrule
Batch Size & 64 & 64 &  & 64 & 64 \\ \midrule
\multicolumn{1}{l}{\# Adversaries / \# Participants} & \multicolumn{2}{c}{24/50} & \multicolumn{1}{l}{} & \multicolumn{2}{c}{4/20} \\ \midrule
Optimizer & \multicolumn{5}{c}{SGD optimizer with momentum 0.9, weight decay 5e-4} \\ \midrule
LR Scheduler & \multicolumn{5}{c}{MultiStepLR ([0.5, 0.8] $\times$ Epochs, gamma=0.01)} \\
\bottomrule
\end{tabular}
\end{table*}

\begin{figure*}[htbp]
    \centering
    \subfloat[Data Heterogeneity in CIFAR-10: Balanced IID ]{
        \centering
        \includegraphics[width=0.49\textwidth,keepaspectratio]{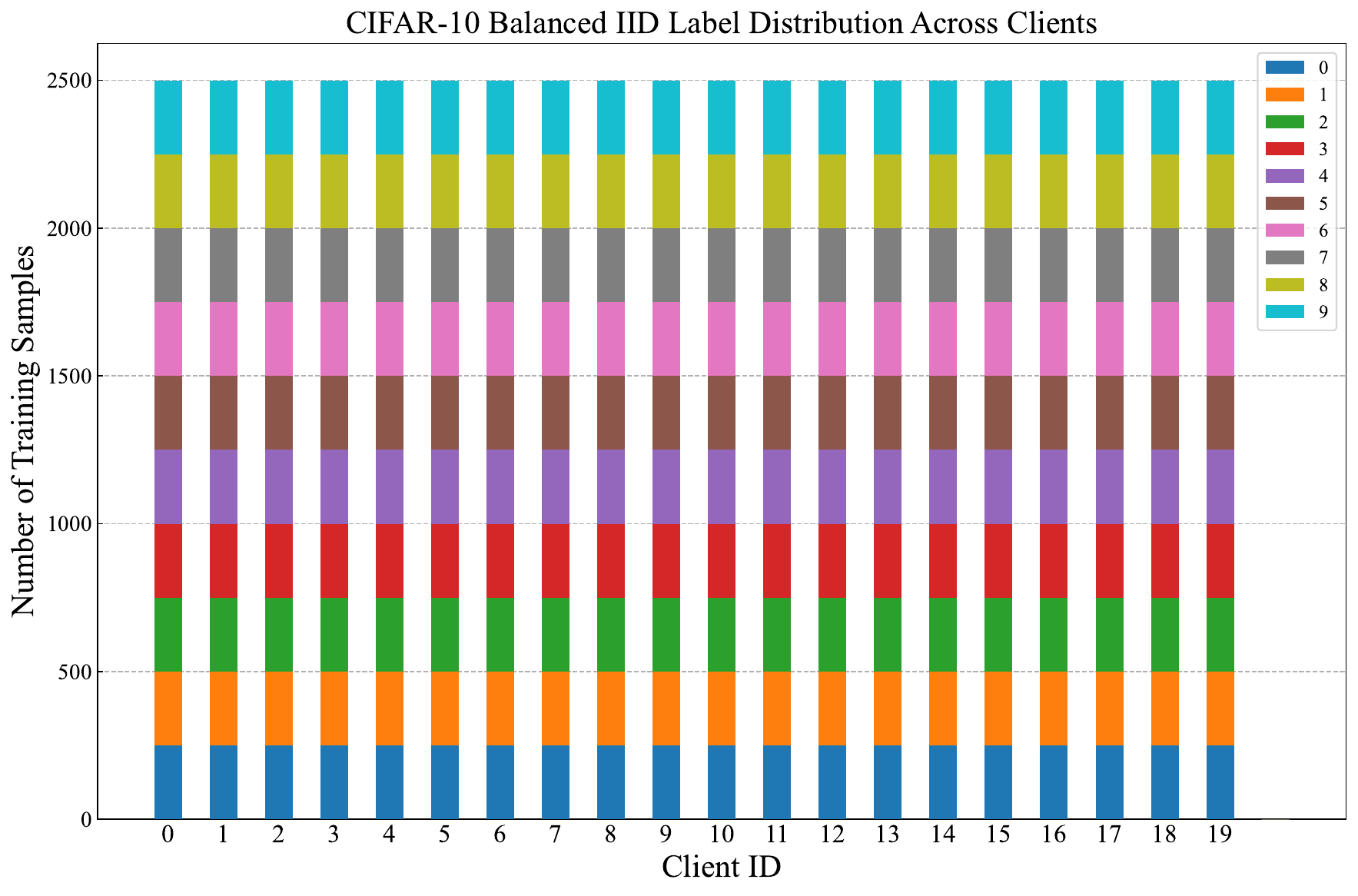}
    }
    \subfloat[Data Heterogeneity in CIFAR-10: Dirichlet-Based non-IID]{
        \centering
        \includegraphics[width=0.49\textwidth,keepaspectratio]{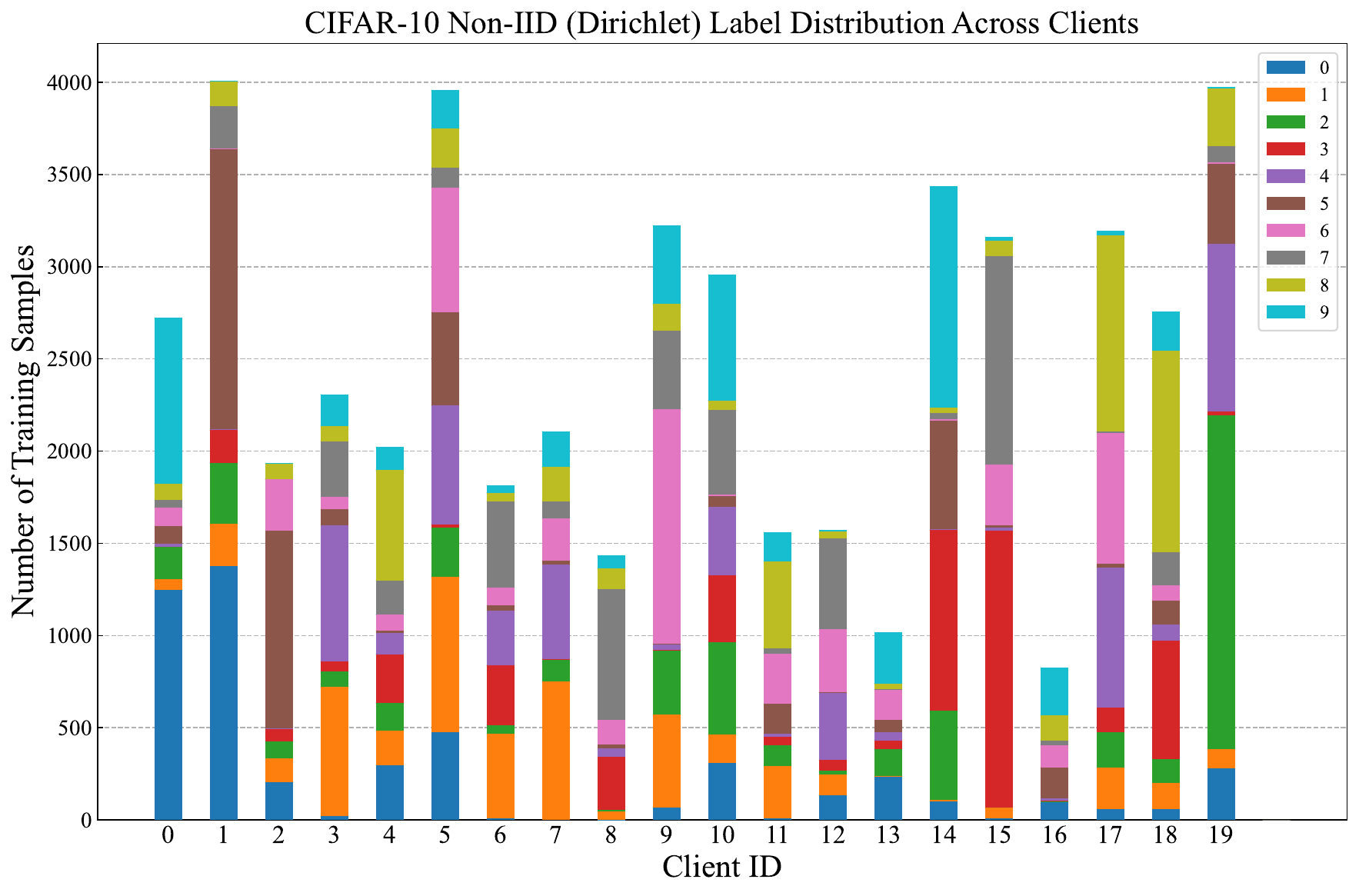}
    }
    \caption{Visualization of statistical heterogeneity across 20 clients for our IID and Non-IID partitions on CIFAR-10 dataset, where the x-axis indicates client IDs, the y-axis shows the number of training samples on that client, and colors represent label classes.}
    \label{pic:heterogeneity}
\end{figure*}

\begin{figure*}[htbp]
    \centering
    \subfloat[Accuracy under MPAs across FL algorithms]{
        \centering
        \label{pic:alg_mpa_cmp}
        \includegraphics[width=0.49\textwidth,keepaspectratio]{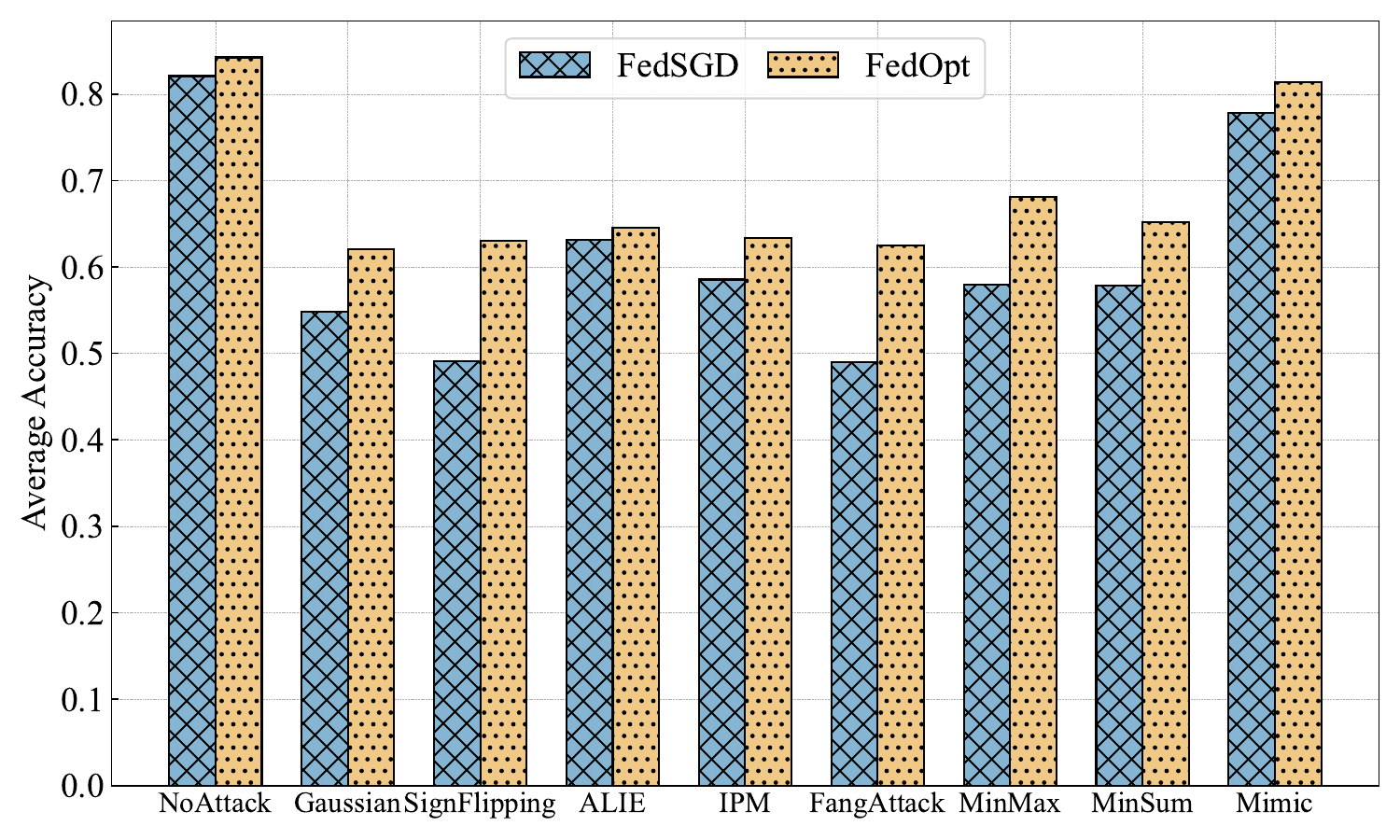}
    }
    \subfloat[Targeted Attack Impact of DPAs across FL algorithms]{
        \centering
        \label{pic:alg_dpa_cmp}
        \includegraphics[width=0.49\textwidth,keepaspectratio]{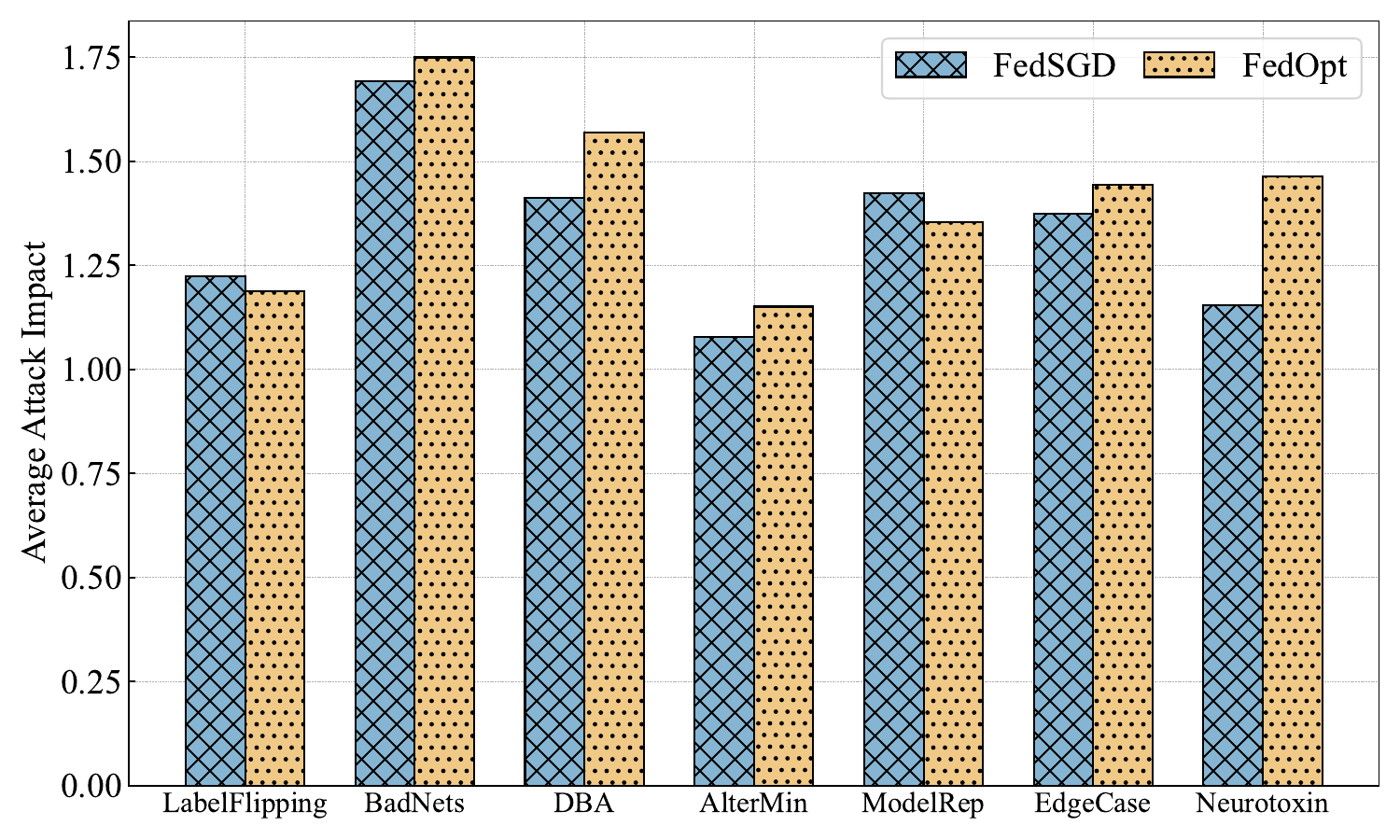}
    }
    \caption{Comparison of Attacks in terms of FL Algorithms}
    \label{pic:alg_cmp}
\end{figure*}

\begin{figure*}[htbp]
    \centering
    \subfloat[Accuracy under MPAs across Data Heterogeneity]{
        \centering
        \label{pic:heter_mpa_cmp}
        \includegraphics[width=0.49\textwidth,keepaspectratio]{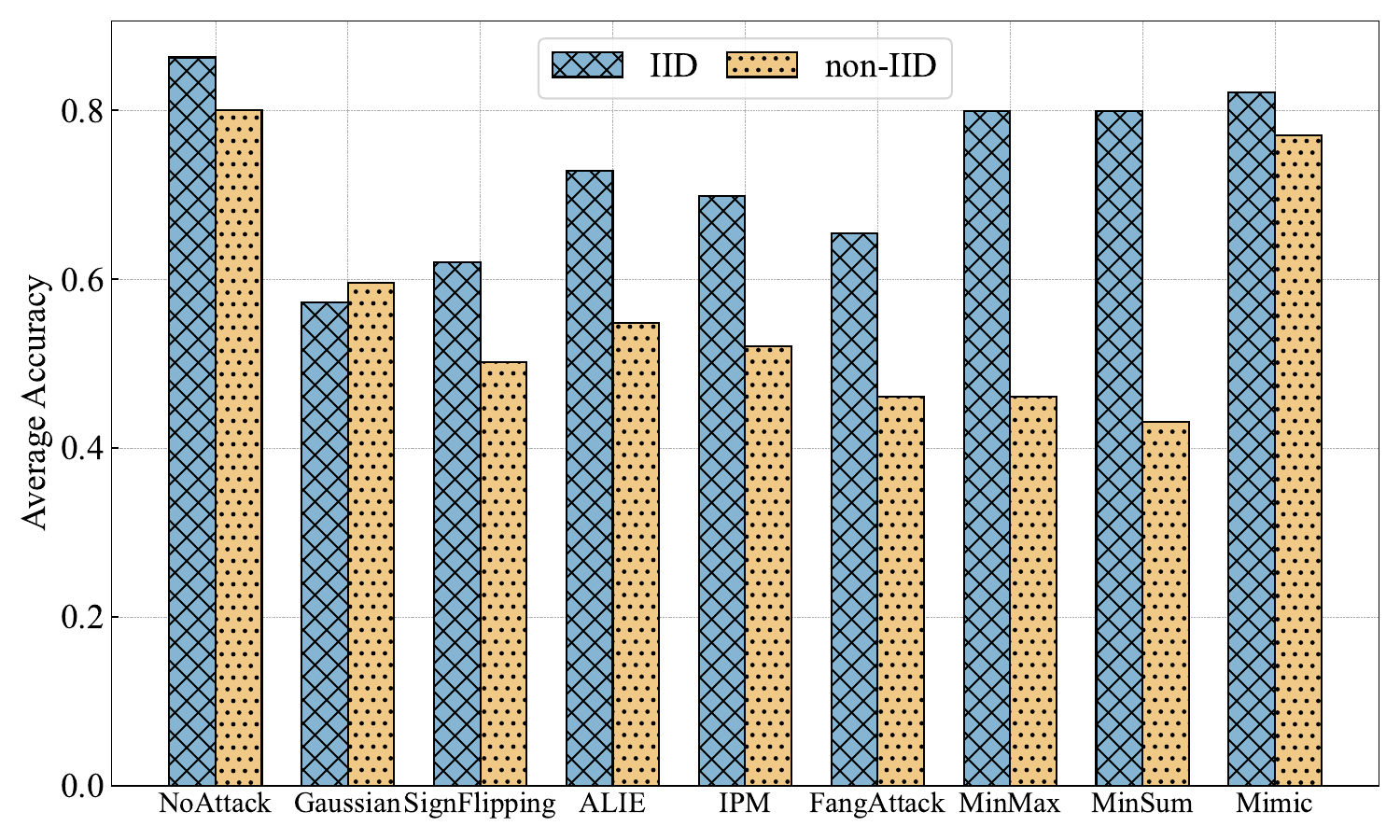}
    }
    \subfloat[Targeted Attack Impact of DPAs across Data Heterogeneity]{
        \centering
        \label{pic:heter_dpa_cmp}
        \includegraphics[width=0.49\textwidth,keepaspectratio]{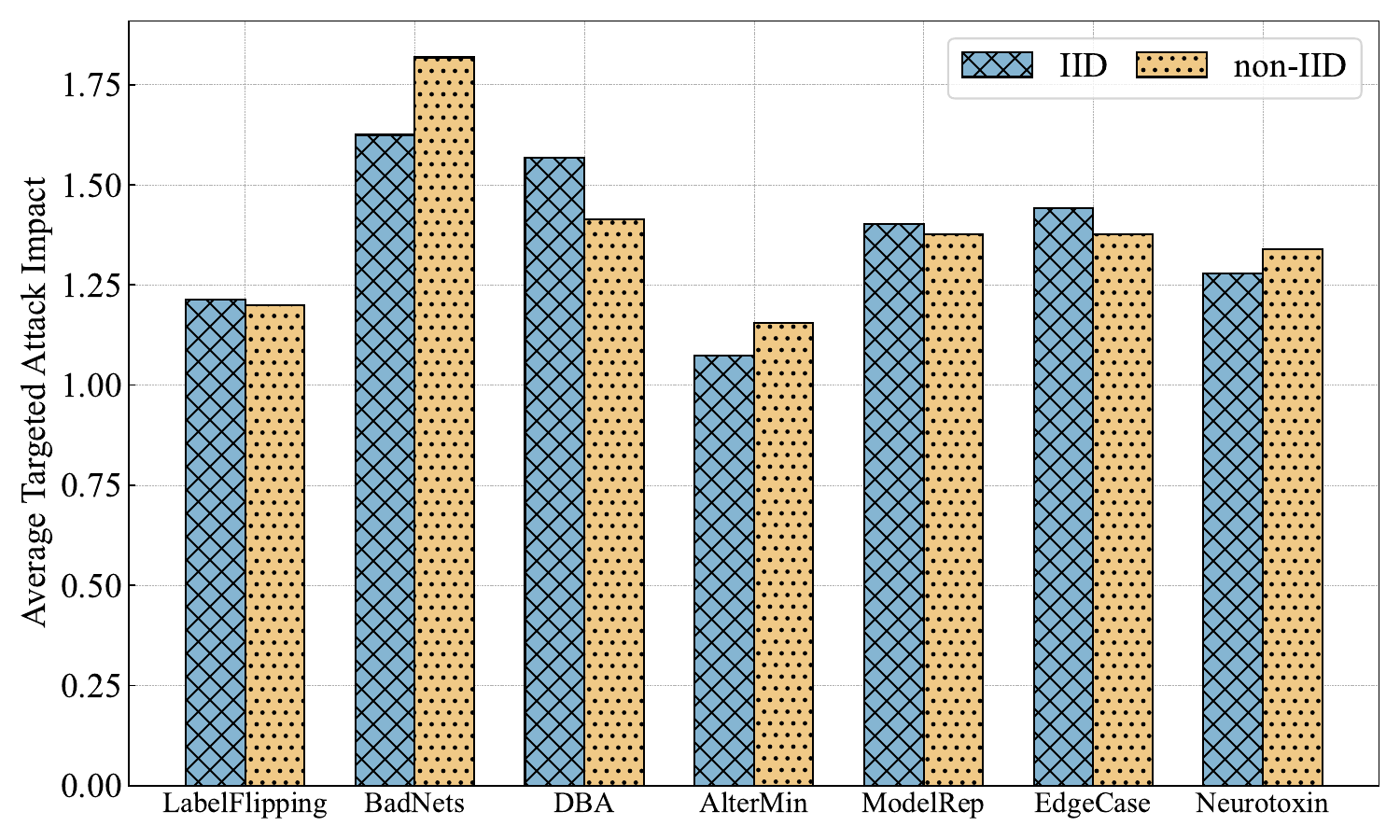}
    }
    \caption{Comparison of Attacks in terms of Data Heterogeneity}
    \label{pic:heter_cmp}
\end{figure*}

\begin{figure*}[!t]
    \centering
    \subfloat[Accuracy of Defenses under MPAs]{
        \centering
        \label{pic:alg_def_mpa_cmp}
        \includegraphics[width=0.49\textwidth,keepaspectratio]{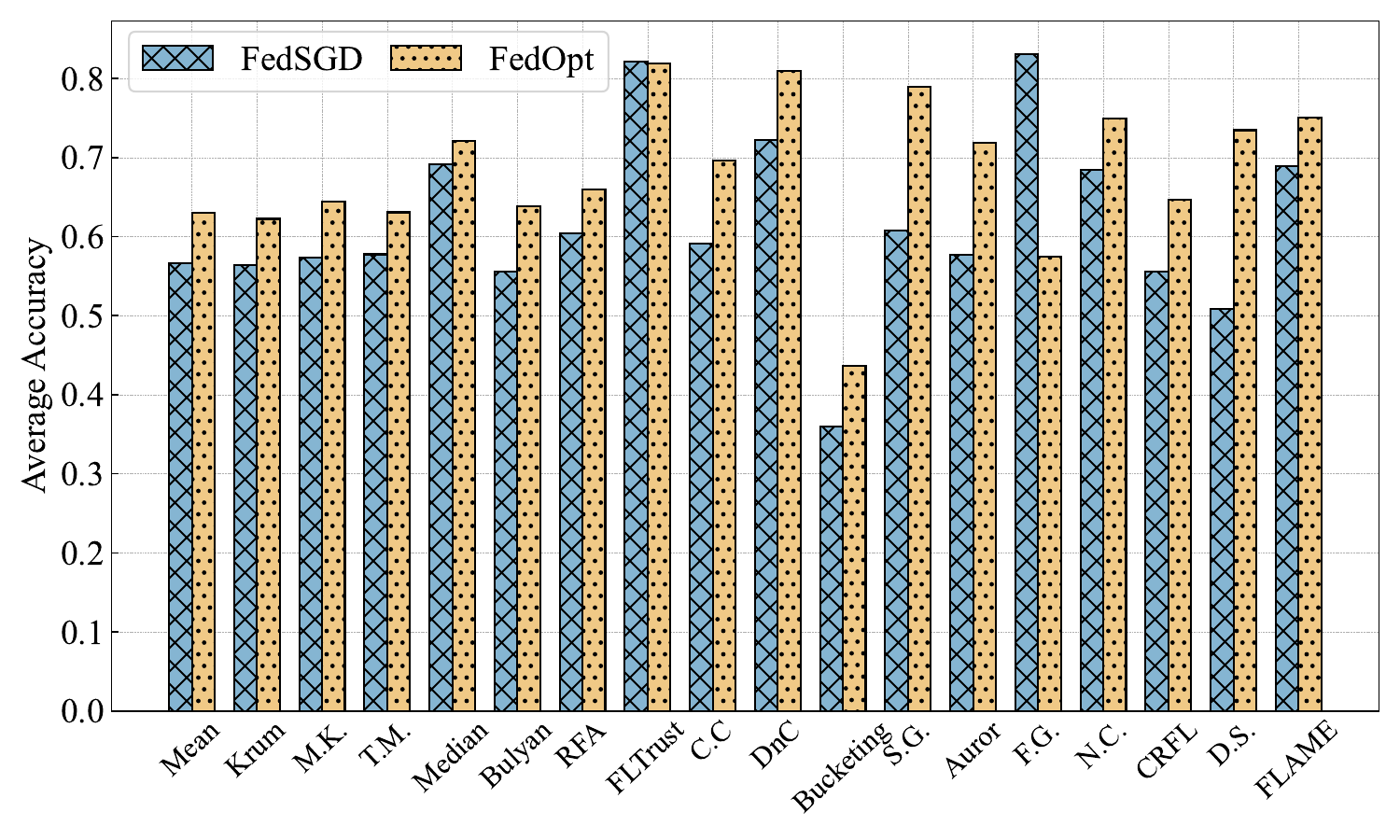}
    }
    \subfloat[Robustness of Defenses under DPAs]{
        \centering
        \label{pic:alg_def_dpa_cmp}
        \includegraphics[width=0.49\textwidth,keepaspectratio]{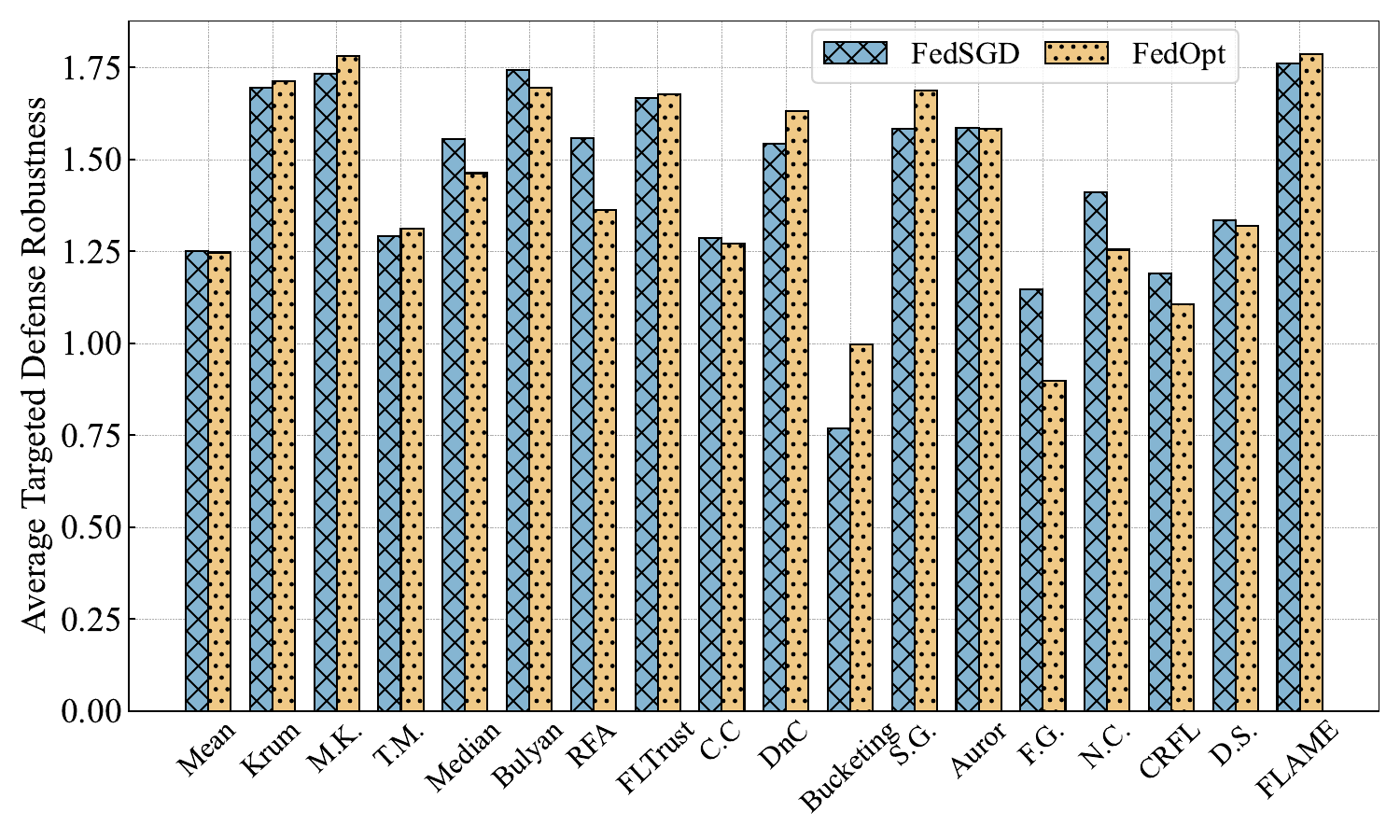}
    }
    \caption{Comparison of Defenses in terms of FL Algorithms}
    \label{pic:alg_def_cmp}
\end{figure*}

\begin{figure*}[!t]
    \centering
    \subfloat[Accuracy of Defenses under MPAs]{
        \centering
        \label{pic:heter_def_mpa_cmp}
        \includegraphics[width=0.49\textwidth,keepaspectratio]{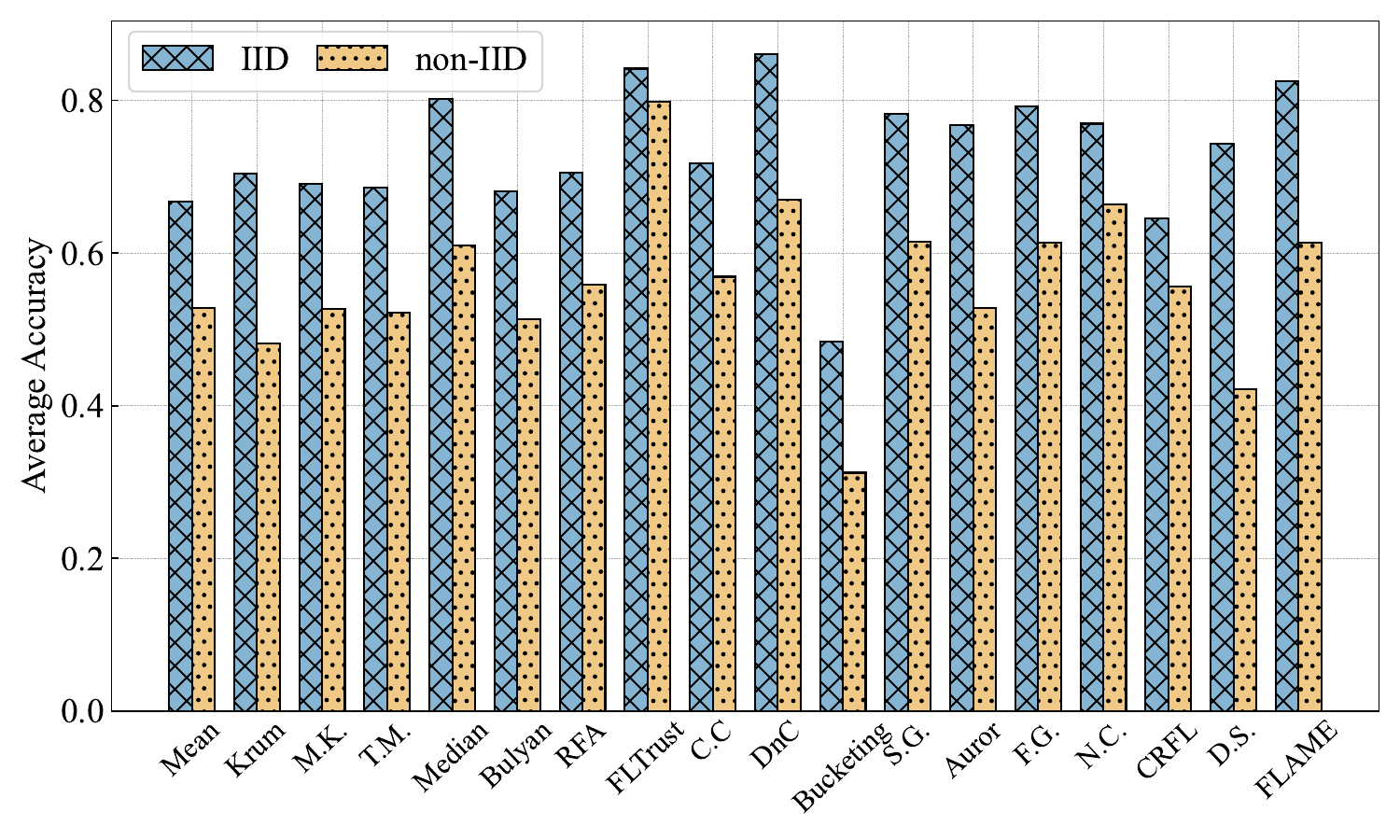}
    }
    \subfloat[Robustness of Defenses under DPAs]{
        \centering
        \label{pic:heter_def_dpa_cmp}
        \includegraphics[width=0.49\textwidth,keepaspectratio]{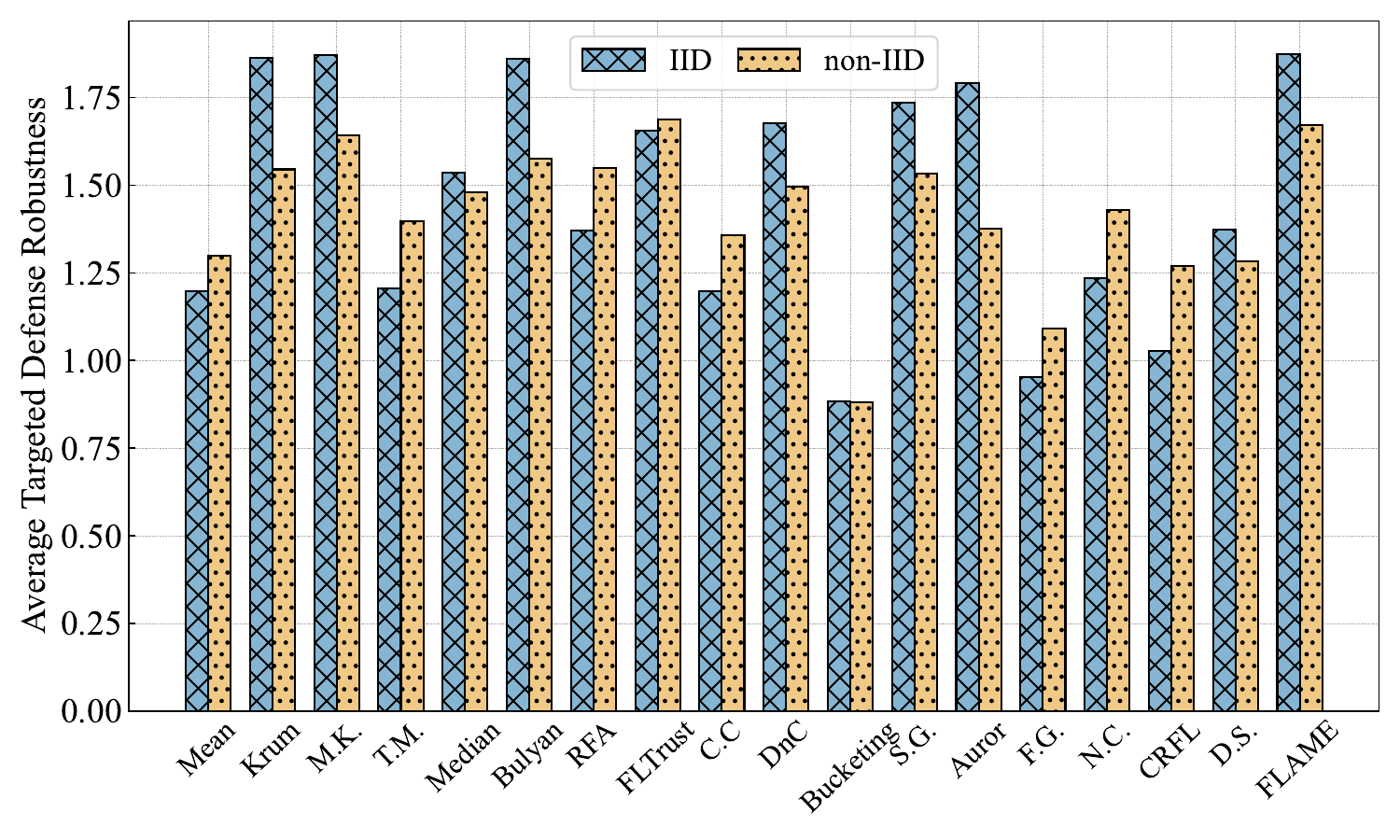}
    }
    \caption{Comparison of Defenses in terms of Data Heterogeneity}
    \label{pic:heter_def_cmp}
\end{figure*}

\begin{table*}[!ht]
\centering
\caption{\textbf{Evaluation of Defense Strategies Over Data Poisoning Attacks - CIFAR10:} Average defense performance against data poisoning attacks.
For the upper table, the results are reported using accuracy and larger values with blue colors indicate better defense performance.
For the lower table, the results are reported using ASR and larger values with blue colors indicate better defense performance. Note that those who receive ``-'' encounter running errors due to non-adaptive hyperparameters.}
\label{tab:dpa_defense_cifar10}
\setlength{\tabcolsep}{3pt} 
\renewcommand{\arraystretch}{1} 
\resizebox{\textwidth}{!}{
\begin{tabular}{cc|cccccccccccccccccc}\toprule
 &  & \multicolumn{18}{c}{\textbf{Average ACC Over DPA (\%) – CIFAR10}} \\ \cmidrule{3-20} 
\multirow{-2}{*}{Alg} & \multirow{-2}{*}{IID} & Mean & Krum & M.K. & T.M. & Median & Bulyan & RFA & FLTrust & C.C. & DnC & Bucketing & S.G. & Auror & F.G. & N.C. & CRFL & D.S. & FLAME \\ \midrule
 & \cmark & \cellcolor[HTML]{FEFBF9}65.9 & \cellcolor[HTML]{EDF5FA}72.6 & \cellcolor[HTML]{E7F2F7}73.7 & \cellcolor[HTML]{FEFBF9}66.1 & \cellcolor[HTML]{DEEDF5}75.1 & \cellcolor[HTML]{F0F7FB}72.0 & \cellcolor[HTML]{DFEDF5}74.9 & \cellcolor[HTML]{DEECF4}75.2 & \cellcolor[HTML]{FEFBF9}65.8 & \cellcolor[HTML]{E3F0F6}74.3 & \cellcolor[HTML]{FDE9DE}47.5 & \cellcolor[HTML]{E6F1F7}73.7 & \cellcolor[HTML]{FEFBF9}65.7 & \cellcolor[HTML]{FFFFFF}- & \cellcolor[HTML]{F7FBFD}71.0 & \cellcolor[HTML]{FDE9DE}47.7 & \cellcolor[HTML]{FFFFFF}- & \cellcolor[HTML]{E1EEF6}74.6 \\
\multirow{-2}{*}{FedSGD} & \xmark & \cellcolor[HTML]{FEF1E9}55.0 & \cellcolor[HTML]{FDE6D9}44.6 & \cellcolor[HTML]{FEEEE4}52.0 & \cellcolor[HTML]{FEF6F2}61.1 & \cellcolor[HTML]{FEFEFE}69.5 & \cellcolor[HTML]{FDE9DC}46.8 & \cellcolor[HTML]{DAEAF3}75.8 & \cellcolor[HTML]{FEFBF9}65.9 & \cellcolor[HTML]{FEF2EB}56.8 & \cellcolor[HTML]{FEEEE4}52.0 & \cellcolor[HTML]{FDDBC7}32.2 & \cellcolor[HTML]{FDECE2}50.5 & \cellcolor[HTML]{FDE6D9}44.3 & \cellcolor[HTML]{FFFFFF}- & \cellcolor[HTML]{E0EEF5}74.7 & \cellcolor[HTML]{FEF1E9}55.2 & \cellcolor[HTML]{FFFFFF}- & \cellcolor[HTML]{FEEDE4}51.6 \\
 & \cmark & \cellcolor[HTML]{FEFCFA}66.5 & \cellcolor[HTML]{DFEDF5}75.0 & \cellcolor[HTML]{E3EFF6}74.4 & \cellcolor[HTML]{D8E9F3}76.2 & \cellcolor[HTML]{D8E9F3}76.1 & \cellcolor[HTML]{E9F3F8}73.4 & \cellcolor[HTML]{D8E9F3}76.2 & \cellcolor[HTML]{EDF5FA}72.5 & \cellcolor[HTML]{D1E5F0}77.3 & \cellcolor[HTML]{D8E9F3}76.2 & \cellcolor[HTML]{FEF9F7}64.2 & \cellcolor[HTML]{D3E6F1}77.1 & \cellcolor[HTML]{DCEBF4}75.5 & \cellcolor[HTML]{FFFFFF}- & \cellcolor[HTML]{FEFCFB}67.4 & \cellcolor[HTML]{FEEDE3}51.1 & \cellcolor[HTML]{FFFFFF}- & \cellcolor[HTML]{D9EAF3}76.0 \\
\multirow{-2}{*}{FedOpt} & \xmark & \cellcolor[HTML]{FFFFFF}69.5 & \cellcolor[HTML]{FEF1E9}55.2 & \cellcolor[HTML]{F1F7FB}71.9 & \cellcolor[HTML]{DAEAF3}75.9 & \cellcolor[HTML]{E3EFF6}74.3 & \cellcolor[HTML]{FEFEFE}69.0 & \cellcolor[HTML]{DFEDF5}75.0 & \cellcolor[HTML]{F3F9FC}71.5 & \cellcolor[HTML]{DFEDF5}75.0 & \cellcolor[HTML]{DDECF4}75.3 & \cellcolor[HTML]{FEF3EC}57.4 & \cellcolor[HTML]{DBEBF4}75.6 & \cellcolor[HTML]{E4F0F6}74.2 & \cellcolor[HTML]{FFFFFF}- & \cellcolor[HTML]{FCFDFE}70.1 & \cellcolor[HTML]{FEF7F3}61.7 & \cellcolor[HTML]{FFFFFF}- & \cellcolor[HTML]{FEFEFE}68.9 \\
\multicolumn{2}{c|}{Acc AVG} & \cellcolor[HTML]{FEF9F7}64.2 & \cellcolor[HTML]{FEF7F3}61.8 & \cellcolor[HTML]{FEFDFC}68.0 & \cellcolor[HTML]{FEFEFF}69.8 & \cellcolor[HTML]{E6F1F7}73.7 & \cellcolor[HTML]{FEFAF8}65.3 & \cellcolor[HTML]{DCEBF4}75.5 & \cellcolor[HTML]{F5F9FC}71.3 & \cellcolor[HTML]{FEFEFD}68.7 & \cellcolor[HTML]{FEFEFE}69.5 & \cellcolor[HTML]{FDECE2}50.3 & \cellcolor[HTML]{FEFEFE}69.2 & \cellcolor[HTML]{FEFAF8}64.9 & \cellcolor[HTML]{FFFFFF}- & \cellcolor[HTML]{F8FBFD}70.8 & \cellcolor[HTML]{FEEFE7}53.9 & \cellcolor[HTML]{FFFFFF}- & \cellcolor[HTML]{FEFDFC}67.8 \\ \midrule
\multicolumn{2}{c|}{} & \multicolumn{18}{c}{\cellcolor[HTML]{FFFFFF}\textbf{Average ASR Over DPA (\%) – CIFAR10}} \\ \midrule
 & \cmark & \cellcolor[HTML]{DDECF4}72.2 & \cellcolor[HTML]{FDDBC7}8.2 & \cellcolor[HTML]{FDDCC8}9.1 & \cellcolor[HTML]{D1E5F0}80.5 & \cellcolor[HTML]{F7FBFD}53.2 & \cellcolor[HTML]{FDDCC9}9.4 & \cellcolor[HTML]{F2F8FB}57.0 & \cellcolor[HTML]{FEFFFF}48.5 & \cellcolor[HTML]{DDECF4}72.4 & \cellcolor[HTML]{FDE2D3}16.6 & \cellcolor[HTML]{D9EAF3}75.4 & \cellcolor[HTML]{FEF5F0}36.8 & \cellcolor[HTML]{FDDBC8}8.7 & \cellcolor[HTML]{FFFFFF}- & \cellcolor[HTML]{F6FAFC}54.0 & \cellcolor[HTML]{DEEDF5}71.2 & \cellcolor[HTML]{FFFFFF}- & \cellcolor[HTML]{FDDFCD}12.3 \\
\multirow{-2}{*}{FedSGD} & \xmark & \cellcolor[HTML]{EBF4F9}62.0 & \cellcolor[HTML]{FDDCC9}9.4 & \cellcolor[HTML]{FDDBC8}8.8 & \cellcolor[HTML]{F8FBFD}52.6 & \cellcolor[HTML]{F6FAFD}54.0 & \cellcolor[HTML]{FDE0CF}13.8 & \cellcolor[HTML]{FEF7F2}38.9 & \cellcolor[HTML]{FEFDFC}45.9 & \cellcolor[HTML]{F3F8FB}56.6 & \cellcolor[HTML]{FDEBE1}26.4 & \cellcolor[HTML]{E2EFF6}68.9 & \cellcolor[HTML]{FEFAF8}42.6 & \cellcolor[HTML]{FDE8DC}22.8 & \cellcolor[HTML]{FFFFFF}- & \cellcolor[HTML]{F8FBFD}52.6 & \cellcolor[HTML]{E8F2F8}64.2 & \cellcolor[HTML]{FFFFFF}- & \cellcolor[HTML]{FDE3D4}17.7 \\
 & \cmark & \cellcolor[HTML]{FEF9F6}41.6 & \cellcolor[HTML]{FDDBC7}7.9 & \cellcolor[HTML]{FDDBC7}8.3 & \cellcolor[HTML]{F6FAFD}54.0 & \cellcolor[HTML]{FAFCFE}51.6 & \cellcolor[HTML]{FDDBC8}8.8 & \cellcolor[HTML]{F7FBFD}53.4 & \cellcolor[HTML]{FEFDFC}45.7 & \cellcolor[HTML]{F3F8FB}56.4 & \cellcolor[HTML]{FDDCC9}9.8 & \cellcolor[HTML]{F9FCFD}52.4 & \cellcolor[HTML]{FEF2EB}33.8 & \cellcolor[HTML]{FDE4D5}17.9 & \cellcolor[HTML]{FFFFFF}- & \cellcolor[HTML]{F7FBFD}53.2 & \cellcolor[HTML]{EFF6FA}59.6 & \cellcolor[HTML]{FFFFFF}- & \cellcolor[HTML]{FDE7DA}21.4 \\
\multirow{-2}{*}{FedOpt} & \xmark & \cellcolor[HTML]{F4F9FC}55.8 & \cellcolor[HTML]{FDDDCA}10.5 & \cellcolor[HTML]{FDDBC7}8.2 & \cellcolor[HTML]{ECF4F9}61.7 & \cellcolor[HTML]{EFF6FA}59.2 & \cellcolor[HTML]{FDE0CE}13.4 & \cellcolor[HTML]{F1F7FB}57.6 & \cellcolor[HTML]{FEFCFB}45.1 & \cellcolor[HTML]{F6FAFD}53.9 & \cellcolor[HTML]{FDE3D4}17.7 & \cellcolor[HTML]{F4F9FC}55.6 & \cellcolor[HTML]{FEF4EE}35.9 & \cellcolor[HTML]{F3F8FB}56.7 & \cellcolor[HTML]{FFFFFF}- & \cellcolor[HTML]{F0F7FB}58.3 & \cellcolor[HTML]{F0F7FA}58.7 & \cellcolor[HTML]{FFFFFF}- & \cellcolor[HTML]{FDE6D8}20.3 \\ \midrule
\multicolumn{2}{c|}{ASR AVG} & \cellcolor[HTML]{F1F7FB}57.9 & \cellcolor[HTML]{FDDCC8}9.0 & \cellcolor[HTML]{FDDBC8}8.6 & \cellcolor[HTML]{EBF4F9}62.2 & \cellcolor[HTML]{F6FAFC}54.5 & \cellcolor[HTML]{FDDECB}11.3 & \cellcolor[HTML]{F9FCFE}51.7 & \cellcolor[HTML]{FEFDFD}46.3 & \cellcolor[HTML]{EEF6FA}59.8 & \cellcolor[HTML]{FDE3D4}17.6 & \cellcolor[HTML]{EAF3F8}63.1 & \cellcolor[HTML]{FEF5F0}37.3 & \cellcolor[HTML]{FDEBE1}26.5 & \cellcolor[HTML]{FFFFFF}- & \cellcolor[HTML]{F6FAFC}54.5 & \cellcolor[HTML]{E9F3F8}63.4 & \cellcolor[HTML]{FFFFFF}- & \cellcolor[HTML]{FDE4D5}17.9 \\ \bottomrule
\end{tabular}
}
\end{table*}

\begin{figure*}[!t]
    \centering
    \subfloat[Average Time Overhead of Poisoning Attacks]{
        \centering
        \label{pic:attack_time}
        \includegraphics[width=0.485\textwidth,keepaspectratio]{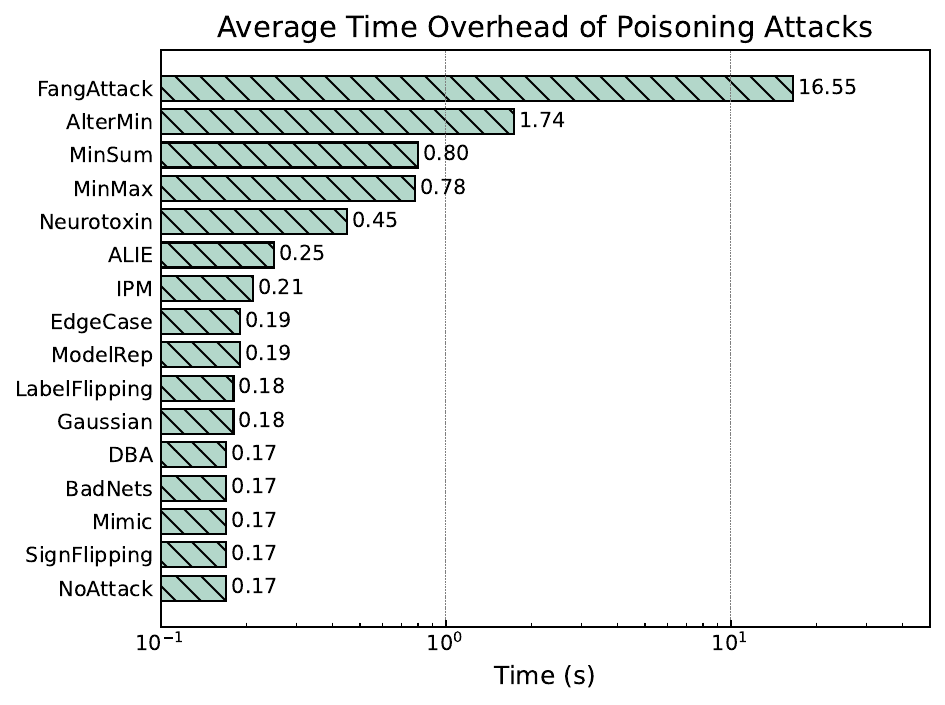}
    }
    \subfloat[Average Time Overhead of Poisoning Defenses]{
        \centering
        \label{pic:defense_time}
        \includegraphics[width=0.485\textwidth,keepaspectratio]{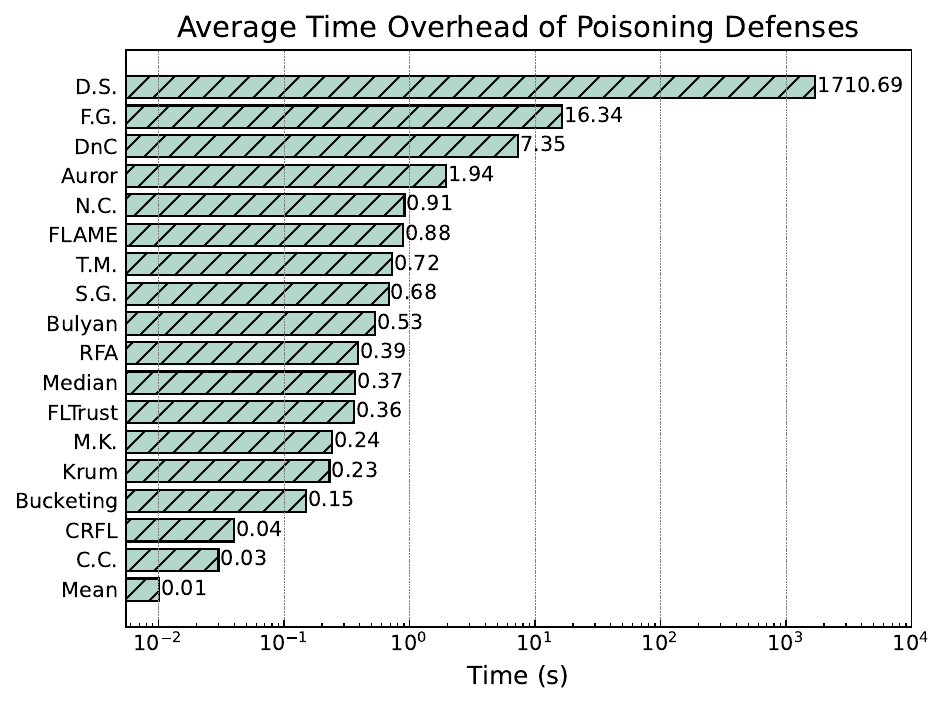}
    }
    \caption{Comparison of Average Time Overhead Per Epoch}
    \label{pic:time}
\end{figure*}

\subsection{Configurations of Attacks and Defenses.}\label{sec:config_attdef}
Since both data poisoning and hybrid poisoning attacks in our paper are targeted attacks, we classify them collectively as data poisoning for targeted evaluation purposes.
For data poisoning attacks, we set the poisoning ratio to 0.32 (20 images are poisoned per batch of size 64) following~\cite{badnets_NIPSWS2017, modelreplacement_aistats2020, dba_iclr2019}, unless otherwise stated.
For the specific configurations of attacks and defenses, we strive to follow the settings of the original work as closely as possible.
However, due to variations in algorithms, datasets, models, and space limitations, we have included these details we used in the ``configs'' folder of our codebase.
It is worth noting that some defenses converge slowly due to strict outlier filtering under non-IID settings, requiring a lower learning rate, as specified in ``batchrun.py'' of our codebase.

\subsection{Comparison of Time Overhead}\label{sec:timeoverhead}

In this section, we evaluate the empirical time overhead of poisoning attacks and defenses.
Specifically, we train a LeNet-5 model on the MNIST dataset in the IID setting under the FedSGD algorithm, as mentioned previously, for 300 rounds, reporting the average time overhead cost by each attack and defense strategy per epoch.

In~\Cref{pic:attack_time}, advanced optimization-based evasion strategies, such as FangAttack, AlterMin, MinSum, and MinMax, incur the most significant time overhead.
Among these, FangAttack incurs the highest cost, taking 97$\times$ times longer than NoAttack's normal training.
This significant overhead is attributed to bypassing Krum through repeated execution of its complex algorithm to optimize malicious updates.
AlterMin's high cost arises from its design, which involves running multiple epochs for each local training round to alternately optimize for stealth and malicious objectives, unlike other FedSGD clients that run only a single round.
MinSum and MinMax, similar to FangAttack, employ optimization-based evasion, though they focus exclusively on distance-based bypass, a task that is less complex than Krum.
Besides, Neurotoxin's overhead is due to the need to identify infrequently updated gradient coordinates before each local training round, followed by projecting the gradient onto these coordinates after training.
Statistic-based evasions, such as ALIE, IPM, or even Gaussian, SignFlipping, incur less time.
ALIE's cost stems from computing $z^{max}$ via the Inverse Cumulative Distribution Function, unlike the remaining attacks relying on simpler operations like trigger embedding, vector scaling.

\myta{\textbf{For attack design guidelines concerning time overhead, it's important to recognize that large search spaces in optimization-based bypass methods can result in significant time consumption.}
In contrast, statistical evasion methods, being less computationally expensive, can reduce time overhead while remaining effective in attacks.}

In \Cref{pic:defense_time}, time-consuming methods are primarily anomaly detection techniques due to their need to extract complex patterns, calculate metrics, filter anomaly, and often perform norm clipping, all of which significantly increase overhead.
The most time-intensive defenses, D.S., F.G., DnC, and Auror, consume 1,710,000$\times$, 1,634$\times$, 735$\times$, and 194$\times$ times of the Mean's cost, respectively.
Among these, D.S., F.G., and Auror focus on feature-level analysis.
D.S. is especially costly, extracting three complex patterns (DDif, NEUP, and Cosine Distance), performing ensemble clustering, and testing the model on up to 20,000 random data points to derive training probabilities distribution.
F.G. and Auror also incur high costs in feature-level pattern extraction, while DnC’s overhead arises mainly from SVD decomposition.
In contrast, FLAME and FLTrust are two of the most effective defenses with shorter costs.
CRFL and C.C. indicate that clipping techniques are time-efficient.
Most robust statistic-based methods, such as Bulyan, RFA, Median, M.K., and Krum, are less time-intensive with only 15$\times$ to 53$\times$ times the Mean's cost. 

\myta{
For defense design, developers should prioritize minimizing computational costs, especially in feature-level anomaly detection.
Avoid excessive pattern extraction, complex metrics, and feature-level clustering, as they increase time consumption.
Statistic-based robust aggregation methods offer a better balance between efficiency and defense effectiveness.}

\subsection{Ablation Study: Impact of Adversary Ratio}\label{sec:ablation}

\begin{figure*}[!t]
    \centering
    \subfloat[Average ACC under MPAs]{
        \centering
        \label{pic:mpa_ab}
        \includegraphics[width=0.231\textwidth,keepaspectratio]{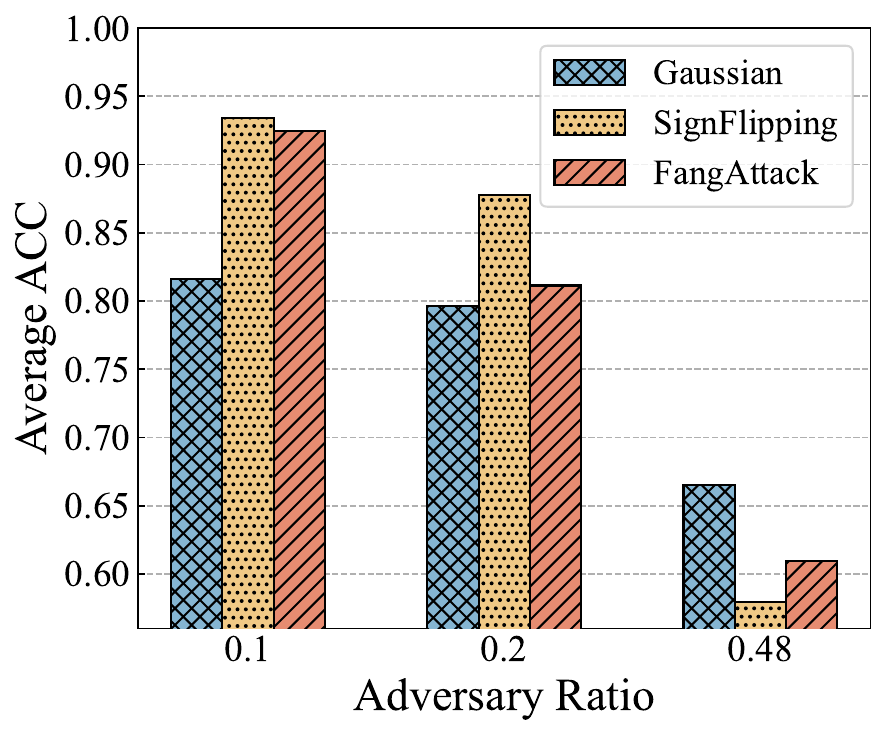}
    }
    \subfloat[Average ACC under DPAs]{
        \centering
        \label{pic:dpa_ab}
        \includegraphics[width=0.231\textwidth,keepaspectratio]{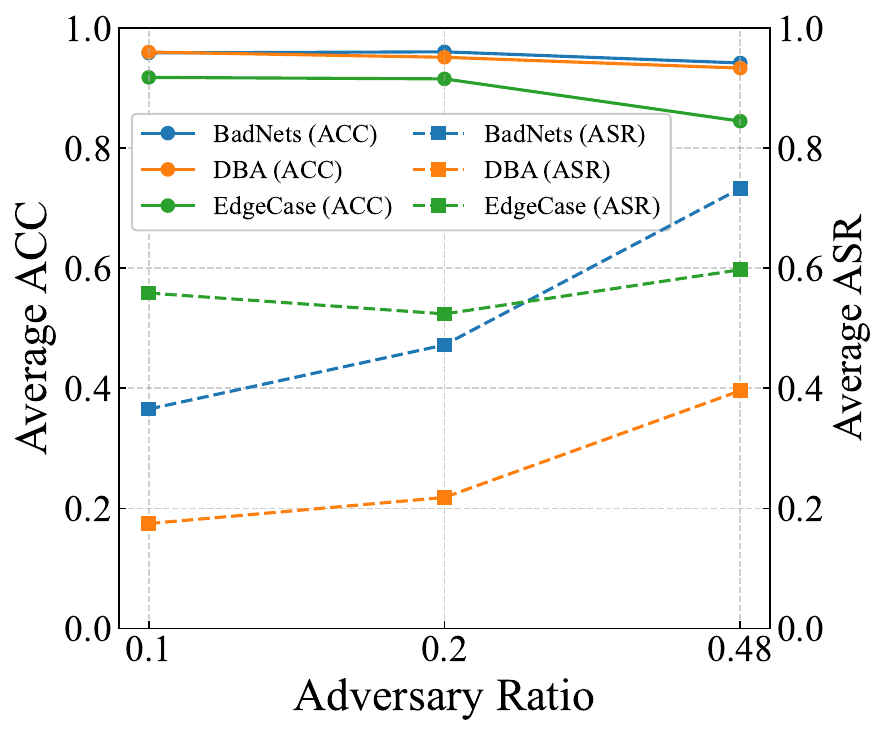}
    }
    \subfloat[Defenses Against MPAs]{
        \centering
        \label{pic:def_mpa_ab}
        \includegraphics[width=0.231\textwidth,keepaspectratio]{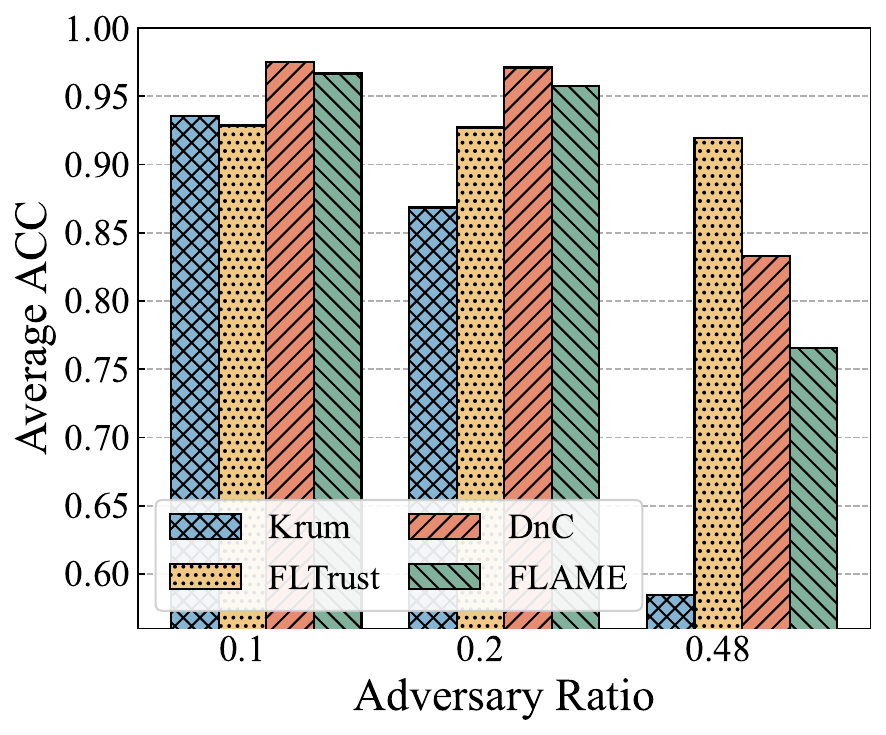}
    }
    \subfloat[Defenses Against DPAs]{
        \centering
        \label{pic:def_dpa_ab}
        \includegraphics[width=0.231\textwidth,keepaspectratio]{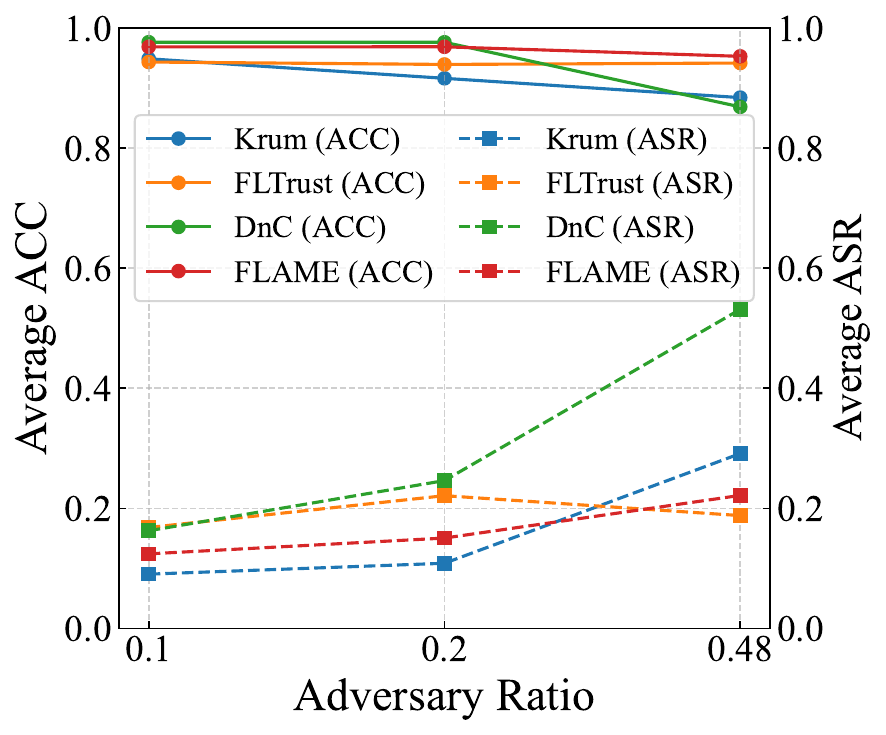}
    }
    \caption{Comparison of Average ACC/ASR Under Different Adversary Ratios}
    \label{pic:ablation}
\end{figure*}

In this section, we evaluate the robustness of advanced attacks (Gaussian Random, Sign Flipping, FangAttack, BadNets, DBA, EdgeCase) and defenses (Krum, FLTrust, DnC, FLAME).
The evaluation is conducted under adversary ratios of 0.1, 0.2, and 0.48 (5, 10, and 24 adversaries out of 50 participants) on the MNIST dataset, LeNet-5 model.
\Cref{pic:ablation} presents the average metrics (ACC or ASR) obtained under four different settings: the FedSGD and FedOpt algorithms, and the IID and non-IID settings.

\mypara{Advanced MPAs.}
\Cref{pic:mpa_ab} demonstrates that, as the proportion of malicious clients increases from 0.1 to 0.48, the impacts of MPAs progressively intensify, leading to a gradual decline in the accuracy of the global model.
However, as the adversary ratio increases from 0.1 to 0.48, it can be seen that the accuracy of the global model under the Gaussian attack decreases by $15\%$, while the effects of the SignFlipping and Fang attacks are more significant, with reductions of 35\% and 31\%, respectively.
{Advanced DPAs.}~\Cref{pic:dpa_ab} depicts that as th,e adversary ratio increases from 0.1 to 0.48, the ACCs of DPAs remain almost unchanged.
Its ASR increases overall, by around 4\%-37\%.
Specifically, BadNet and DBA increase more, whereas EdgeCase does not increase much.
Notably, the main task accuracy and ASR of BadNet are higher than that of DBA and EdgeCase, rendering it the most robust targeted attack.
s against MPAs.
For poisoning defenses, we select two advanced defenses against MPAs, FLTrust and DnC, and two advanced defenses against DPAs, Krum and FLAME.
In \Cref{pic:def_mpa_ab}, it is evident that as the adversary ratio increases, the average accuracies of model poisoning defenses exhibit an overall downward trend.
Specifically, Krum shows the largest drop, approximately 35\%, highlighting its inability to effectively mitigate excessive malicious deviations.
In contrast, advanced model poisoning defense mechanisms such as FLTrust and DnC exhibit minimal declines of about 1\%, underscoring their effectiveness and robustness across varying adversary ratios.
Furthermore, FLAME experiences a decrease of approximately 20\%.
Overall, FLTrust and DnC rank as the most effective model poisoning defense due to their high effectiveness and stability.

\mypara{Advanced Defenses Against DPAs.}
\Cref{pic:def_dpa_ab} illustrates that as the adversary ratio increases, the ACCs of data poisoning defenses decrease, while the ASRs simultaneously rise.
Specifically, DnC and FLTrust, two advanced model poisoning defenses, demonstrate relatively limited effectiveness in mitigating DPAs, as evidenced by the greatest decrease in ACC and the highest increase in ASR.
In contrast, FLAME outperforms Krum by maintaining the lowest ASR and the highest ACC, making it the most effective approach for DPAs.

\myta{
Overall, Gaussian, Sign Flipping, and FangAttack are identified as the most effective MPAs, while BadNets stands out as the most effective DPA.
Among the defenses, FLTrust and DnC demonstrate the highest efficacy against MPAs, whereas FLAME proves to be the most effective defense against DPAs.}

\end{document}